\documentclass[3p,10pt,letterpaper,number,times]{elsarticle}
\usepackage[numbers]{natbib}

\usepackage[T1]{fontenc}
\usepackage[latin9]{inputenc}
\usepackage{amsmath}
\usepackage{amssymb}
\usepackage{amsfonts}
\usepackage{graphicx}
\usepackage{esint}
\usepackage{microtype}
\usepackage{subfig}
\usepackage{booktabs} 
\usepackage{todonotes}
\usepackage{arydshln}
\usepackage{lineno}   
\usepackage{setspace} 
\usepackage{enumitem} 
\usepackage{accents}  
\usepackage{tcolorbox} 
\usepackage{mathrsfs} 
\usepackage{xcolor} 
\usepackage{hyperref}
\hypersetup{colorlinks=true, linkcolor=blue, citecolor=blue, urlcolor=blue}
\usepackage{xurl}
\renewcommand*{\dot}[1]{\accentset{\mbox{\normalsize .}}{#1}}
\newlength{\argwd}  \newlength{\arght}
\newcommand{\overharp}[3]{
  \settowidth{\argwd}{#2}\settoheight{\arght}{#2}%
  \raisebox{#3\arght}{
    \makebox[0pt][l]{
      \resizebox{\argwd}{0.7\arght}{$#1$}%
    }%
  }%
#2}
\newcommand{\overleftharp}[1]{\overharp{\leftharpoonup}{#1}{0.9}}
\newcommand{\overrightharp}[1]{\overharp{\rightharpoonup}{#1}{0.9}}
\makeatletter
\newsavebox{\@brx}
\newcommand{\llangle}[1][]{\savebox{\@brx}{\(\m@th{#1\langle}\)}%
  \mathopen{\copy\@brx\kern-0.5\wd\@brx\usebox{\@brx}}}
\newcommand{\rrangle}[1][]{\savebox{\@brx}{\(\m@th{#1\rangle}\)}%
  \mathclose{\copy\@brx\kern-0.5\wd\@brx\usebox{\@brx}}}
\makeatother
\newcommand{\XX}{\mathbf{x},\mathbf{x}^{\prime}}

\newcommand{\Xp}{\mathbf{x}^{\prime}}
\newcommand{\Xpp}{\mathbf{x}^{\prime\prime}}
\newcommand{\X}{\mathbf{x}}
\newcommand{\aXX}{|\X-\X'|}
\newcommand{\Axx}{|x-x'|}

\newcommand{\VV}{dv^{\prime}\,dv}

\newcommand{\Vp}{dv^{\prime}}
\newcommand{\Vpp}{dv^{\prime\prime}}
\newcommand{\Bo}{\overline{\mathcal{B}}}
\newcommand{\Do}{\overline{\mathcal{D}}}
\newcommand{\Bp}{\partial\mathcal{B}}
\newcommand{\Dp}{\partial\mathcal{D}}
\newcommand{\Bc}{\mathcal{B}}
\newcommand{\Dc}{\mathcal{D}}

\newcommand*{\rpoon}[1]{\overrightharp{\ensuremath{#1}}}
\newcommand*{\lpoon}[1]{\overleftharp{\ensuremath{#1}}}
\newcommand{\RRR}{\mathbb{R}^{3}}
\newcommand{\R}{\mathbb{R}}
\newcommand{\Vhat}{\widehat{\mathbf{V}}}
\begin{document}
%
%
\begin{tcolorbox}[
    colback=gray!5, 
    colframe=gray!70, 
    arc=0mm, 
    width=\textwidth,
    boxrule=0.5pt,
    top=2mm, bottom=2mm, left=2mm, right=2mm
]
\small
\textbf{Please cite this article as:} \\
  Matthew R. Kuhn (2026),
  ``A thermomechanical framework for strongly nonlocal continua,''
  \emph{International Journal of Engineering Science},
  Vol. 229,
  104651,
  https://doi.org/10.1016/j.ijengsci.2026.104651.
\end{tcolorbox}
\vspace{1em} 
\begin{frontmatter}  
\title{\Large A thermomechanical framework for strongly nonlocal continua}
\author[1]{Matthew R. Kuhn}
%
%
%
\affiliation[1]{organization={Emeritus Professor,
                              Donald P. Shiley School of Engr.,
                              Univ. of Portland},
                city={Portland},
                state={Oregon},
                country={U.S.A.}}
%
\begin{abstract}
The paper presents a consistent thermomechanical framework
for strongly nonlocal continua,
a type of generalized media
in which a material's response
at a point depends on deformation and temperature gradients within
its neighborhood.
Such dependence is responsible for localization
phenomena having an intrinsic size, and
nonlocal modeling also serves as a practical regularization
tool to prevent mesh dependence.
Nonlocality is of the integral type,
using a kernel function that gives the
relative influence of neighboring points on a central point.
The paper develops three sources of nonlocality:
1) nonlocal momentum balance equations;
2) nonlocal conservation laws
(including the first and second laws of thermodynamics); and
3) a material's nonlocal constitutive structure.
Balance equations are derived using the principle of virtual power,
permitting non-smooth force and displacement fields
and non-smooth boundary surfaces with indistinct normal directions.
The balance equations differ from those of classical local continua.
Conservation laws also differ from those of local continua,
with the mechanical power and heating at a point being averaged
over its neighborhood.
These laws are developed for processes that are sufficiently slow
to minimize additional meso-scale kinetic energy due to
internal turbulence.
Stress, entropy, and dissipative forces are obtained
as averaged derivatives of the free energy function
with respect to strain, temperature, and internal variables.
An example is presented of a stretched elastoplastic bar
with a small defect, and nonlocality is shown to impart
a characteristic size to the deformation pattern.
\end{abstract}
\end{frontmatter}
%
\section{\large Introduction}
The proposition of local action is a
common assumption in continuum mechanics and forms
the basis of grade-one, simple materials \cite{Noll:1958a}.
The proposition holds that stress at a point depends on the first
gradient of displacement (deformation),
the gradient's rate,
and the gradient's history \emph{at that same point}.
This assumption is relaxed in weakly nonlocal media,
where the stress at a point also depends
on higher gradients of displacement, though still evaluated locally.
The paper concerns a \emph{strongly} nonlocal hypothesis
of material behavior,
asserting that behavior at a point depends on deformation
within the point's neighborhood or even within the entire body.
As defined by Edelen \cite{Edelen:1976a}, a nonlocal theory
is one in which
``the stress at a point ($X$) ... depends on the strain
measures at point ($Z$) of the body other than
the point ($X$).'' 
\par
The generalization of the balance equations of motion for
such nonlocal continua traces back to
the work of Piola \cite{dellIsola:2014a}, 
who applied the principle of virtual work to nonlocal media
that predates its resurgence in the 1970s by over a century.
The first application of a nonlocal theory appeared in the mechanics of
fluid capillarity,
proposed by van der Waals in 1890,
who proposed that the internal energy
at a point depends on the density at neighboring points
\cite{vanderWaals:1979a}. 
He recognized that errors from neglecting nonlocal effects
are significant near abrupt transitions in density but vanish
``completely when the state of equilibrium
is that of a homogeneous distribution of the substance,''
although even then,
``there remains only an error at the limits where the
substance is in contact with the boundaries''
(e.g., at the extremities of a meniscus).
This early work highlights key features of nonlocal descriptions:
namely, that nonlocal effects are most significant
near internal inhomogeneities,
that they increase with increasing
nonuniformity of deformation,
and that they can arise
near a body's boundaries.
For histories of nonlocal theories since the 1960s,
one is referred to those of Bazant
\cite{Bazant:2002a}, Eringen \cite{Eringen:2002a},
and Maugin \cite{Maugin:2013a}.
\par
Nonlocal continuum theories have several motivations.
The most fundamental is the recognition that viewing real
bodies as continua is a conceptual abstraction,
one that conceals at smaller scales~---
granular, crystalline, cluster, dislocation, or molecular~---
behavior arising from discrete interactions of
a body's constituents
\cite{Silling:2010a}. 
Although nonlocal theories also treat bodies
as continua,
they aim to incorporate the effects
of such distant internal interactions.
These interactions are thought to be responsible for meso-scale
localization
and patterning phenomena observed in many materials,
phenomena that are often characterized by intrinsic length scales
(in shear bands, compaction bands, chevron cracking, etc.).
Even when distant interactions are not explicitly modeled,
nonlocal material models (both strongly and weakly nonlocal)
are introduced in numerical solvers as practical regularization tools
to prevent mesh dependence, to avoid numerical ill-posedness,
to avoid an undue
concentration of strain when a material
softens or suffers damage during deformation
\cite{Bazant:1976a}, 
to regularize deformation at a crack tip
\cite{Xia:1993a},
or to improve agreement with other phenomena that arise
at small scales
\cite{Eringen:1981a}.
\par
Similar motivations are given for other generalized continua,
which include
weakly nonlocal media,
such as higher-gradient (weakly nonlocal) models and
augmented continua (e.g., Cosserat and micro-polar media).
Although these alternative continua do not overtly model
distant interactions,
they supply a proxy for their effects.
For example, second-gradient models have been shown to resolve the
capillary effects that van der Waals had addressed with nonlocal
arguments \cite{Auffray:2015a}.
Furthermore,
a Taylor expansion of the strain field around a point leads to
a third-gradient models that approximate
the kernel-averaged strain centered at the point
\cite{Vardoulakis:1989c}.
Other comparisons of nonlocal kernel-averaged strain models
and higher-gradient models are provided in
\cite{Peerlings:2001a}. 
With such weakly nonlocal and augmented continua,
force balance requires introduction
of somewhat abstruse higher-order or generalized stresses
(e.g., couple stress) within a body's interior
and their corresponding generalized tractions on the body's surface.
This difficulty is avoided with strongly nonlocal models,
such as that of this paper.
\par
In summary,
strongly nonlocal models provide a more nuanced description
of material behavior by engendering long-distance internal
interactions and offering advantages in numerical implementation,
but without the need for abstruse stress-like quantities.
\par
The paper develops a consistent thermomechanical framework for
strongly nonlocal media.
Rather than accepting conventional local balance equations
and simply inserting nonlocal constitutive relations,
we instead develop a
consistent coherent framework in which the balance equations,
thermomechanic principles, and constitutive forms
are intrinsically nonlocal.
Starting with the principle of virtual power,
applied to a broad class of virtual velocities~---
both mechanical and thermal~---
we derive the corresponding balance equations.
The balance equations differ from those
of classical, local continua that ensue from a more restricted
space of virtual motions.
Section~\ref{sec:thermomech} then
transitions from virtual to actual velocities,
where the mechanical and thermal balances combine
to yield an energy balance in a nonlocal setting.
From this energy balance,
the first and second laws of thermomechanics are then
developed, following the structure of Green and Naghdi
\cite{Green:1977a,Green:1978a,Green:1991b}, but in
a nonlocal context (Sections~\ref{sec:firstlaw}
and~\ref{sec:secondlaw}).
By using a free energy with a state space that
includes internal variables,
a consistent constitutive framework is proposed in
Section~\ref{sec:constit},
where we caution against the use of alternative constitutive systems
based on spatially averaged state variables.
A simple example is presented in Section~\ref{sec:examples},
and although similar to other examples in the literature,
we demonstrate the application of the paper's consistent framework.
\section{\large Mechanical and thermal balance}\label{sec:statics}
In this section,
we apply the Germain--Maugin formalism
in a nonlocal setting,
invoking the principle of virtual power~---
power induced by
a space of virtual velocities~--- to derive
the momentum balance equations for
both mechanical and thermal forces and momenta.
Section~\ref{sec:thermomech} follows in a natural way
by replacing virtual velocities with actual velocities,
and by deriving an energy balance that
forms the basis of the first and second laws of thermomechanics.
Unlike the Newton--Cauchy (and, more recently, Truesdell)
approach to momentum balance,
in which force is accepted as a primitive object,
we adopt the virtual power approach
of Piola, Hellinger, and Germain
(and the antecedent infinitesimal-displacement, virtual-velocity,
virtual-work, and
variational approaches of Johann Bernoulli,
D'Alembert, L. Carnot, and Lagrange
\cite{Capecchi:2012a,Pisano:2017a}) 
who viewed force as an emergent dual of motion,
the latter being accepted as primitive
\cite{Eugster:2017a,Germain:1973a,Maugin:1980a}.
Hellinger observed that this approach permits the derivation of
local balance conditions both within the interior of a region
and on its boundaries,
an advantage leveraged here.
\par
We also include the rate of thermal displacement (i.e., temperature)
as an additional kinematic variable, also suggested by Hellinger,
which will yield the conditions of thermostatic balance.
As will be shown,
it is only after imposing certain smoothness conditions
on a nonlocal medium that the mechanical and thermal
balance equations become similar to those
of the more restrictive local, conventional media,
even though the underlying strains, temperatures, and forces
have a nonlocal character
(see also \cite{Antman:1979a} for a demonstration of this
equivalence for strictly local media).
\par
Not pursued here, however, are continua that involve
higher-order derivatives (grade-two and higher) of the virtual
velocity fields
or continua with virtual rotation fields that are independent
of a velocity's vorticity
(e.g., \cite{Maugin:2010a}).
As will be seen,
a nonlocal continuum of integral-type avoids
certain difficulties of these higher-order and Cosserat continua,
such as the need for higher-order and moment stresses
(and their constitutive descriptions)
and the complications arising from surface tractions on
point-, line-, and surface-cuts \cite{dellIsola:2012a}. 
\subsection{Virtual power framework}
A material region (body)
$\mathcal{B}$ is a connected open set
in $\RRR$
with boundary $\partial\mathcal{B}$ and closure
$\overline{\mathcal{B}}=\mathcal{B}\cup\partial\mathcal{B}$.
The positions $\X$ of material points in $\Bc$ at time $t$,
constituting the deformed configuration,
are mapped to their reference positions $\mathbf{X}$ at $t=0$
by $\X=\boldsymbol{\chi}(\mathbf{X},t)$,
with $\mathbf{X}=\boldsymbol{\chi}(\mathbf{X},0)$.
We consider a real Hilbert space $\Vhat$
of \emph{virtual} mechanical velocities
(i.e., time rates of infinitesimal displacements),
with elements $\widehat{\mathbf{v}}\in\widehat{V}$
that are functions $\widehat{\mathbf{v}}$
of the positions $\X$ of material points
in $\Bc$ at $t$
and that map these positions to virtual velocities as
$\widehat{\mathbf{v}}\colon\, \X\in\Bo\rightarrow\RRR$.
Appropriate inner
products for $\Vhat$ are described later.
The space $\Vhat$ may be restricted by
boundary conditions or by internal material constraints,
in which case, $\Vhat$ represents the kinematically
admissible virtual velocities,
provided that $\Vhat$ remains a Hilbert space.
\par
The space $\Vhat$ must include the subspace
of virtual rigid-body
velocities, $\Vhat_{\text{r}}\subseteq\Vhat$,
with elements that are functions
$\widehat{\mathbf{v}}_{\text{r}}\in\Vhat_{\text{r}}$,
described later.
Space $\Vhat$
must also include the
\emph{actual} velocities $\mathbf{V}$,
with $\mathbf{V}\in\Vhat$, 
also described later.
We do not consider spaces enriched with
local virtual rotations (as in Cosserat continua),
nor the
additional kinetic energy
that may arise from such rotations.
\par
The virtual mechanical velocity space $\Vhat$ is augmented by
a real Hilbert space of virtual
thermal velocities $\widehat{\Theta}$,
analogous to virtual temperatures,
with elements $\widehat{\theta}\in\widehat{\Theta}$ that
are functions $\widehat{\theta}(\X)$ that
map material points in $\Bc$ to scalars,
$\widehat{\theta}: \X\in\overline{\mathcal{B}}\rightarrow\R$.
Each element $\widehat{\theta}(\X)$  represents an average
rate of thermal (i.e., molecular)
displacements $\alpha(\X)$, in the manner
of Green and Naghdi \cite{Green:1991a}
and as originally introduced in a virtual
power setting by Hellinger
(e.g., \cite{Eugster:2018c}.
Note that the scalar field $\widehat{\theta}(\X)$ is less general
than the entropic vector velocity field of Giorgio~et~al.
\cite{Giorgio:2024a}).
As described later,
the space $\widehat{\Theta}$ includes both
the rigid-body thermal velocity subspace $\widehat{\Theta}_{\text{r}}$
with elements $\widehat{\theta}_{\text{r}}$
and the space of \emph{actual} thermal velocities $\Theta$
having elements $\theta$,
with $\widehat{\Theta}_{\text{r}}\subseteq\widehat{\Theta}$
and $\Theta\in\widehat{\Theta}$.
\par
At this point,
the scalar field $\widehat{\theta}$ is formally analogous to
an additional fourth
component $\widehat{v}_{4}$ of velocity field $\widehat{\mathbf{v}}$~---
differing only in its physical interpretation
(thermal versus mechanical)~---
and one could consider merging the two into a single field
(a similar apparent congruence occurs between heat and work in
\cite{Beretta:2015a}).
The fundamental distinction is introduced
in Section~\ref{sec:thermomech},
arising from the second law and the non-negativity of $\theta$.
\par
The product space of virtual mechanical and thermal velocities is
the Hilbert space
$\widehat{W}=\Vhat\times\widehat{\Theta}$,
and we write its elements $\widehat{\mathbf{w}}\in\widehat{W}$
as a stacked array of functions
$\widehat{\mathbf{w}}(\X)=[\widehat{\mathbf{v}}(\X) \,/\, \widehat{\theta}(\X)]$.
\par
As with the Germain--Maugin formalism
\cite{Germain:1973a,Maugin:1980a},
we identify a linear operator $\mathscr{L}$ that maps
$\widehat{W}$ to a space
$\widehat{Z}=\{\widehat{\mathbf{z}}(\X)\!\!:\,\widehat{\mathbf{z}}(\X)=\mathscr{L}\widehat{\mathbf{w}},\,\widehat{\mathbf{w}}\in\widehat{W}\}$,
with the image $\widehat{Z}$ also being a Hilbert space.
The space $\widehat{Z}$
will induce the essential actors
(forces, stresses, entropy flux, etc.)
appearing in the external and internal virtual powers, introduced below.
For a deformable body with nonhomogeneous temperature, the enhanced
$\widehat{Z}$ may be taken as the product space of virtual velocities
$\widehat{W}$ and their spacial derivatives
(i.e., virtual velocity gradients
and virtual temperature gradients).
For a rigid body with homogeneous temperature, by contrast,
$\mathscr{L}$ is reduced to the identity operator $\mathscr{I}\!$, with
$\mathscr{L}=\mathscr{I}:\widehat{W}\rightarrow\widehat{W}$,
$\widehat{\mathbf{w}}\mapsto\widehat{\mathbf{w}}$.
\par
We consider the dual space $\widehat{Z}^{\ast}$,
with each of its elements
$\mathscr{F}_{\mathcal{B}}\in\widehat{Z}^{\ast}$
being a linear mapping
of $\,\widehat{\mathbf{z}}(\X)=\mathscr{L}\widehat{\mathbf{w}}(\X)$
into $\R$, as
$\mathscr{F}_{\mathcal{B}}\colon\widehat{Z}\rightarrow\R$.
Because $\widehat{Z}$ is a Hilbert space,
each element $\mathscr{F}_{\mathcal{B}}$
of $\widehat{Z}^{\ast}$ is represented
with a unique Riesz inner product $\langle\cdot,\cdot\rangle$,
\begin{equation}\label{eq:inner}
  \text{For each }\mathscr{F}_{\mathcal{B}}\in\widehat{Z}^{\ast},\;
  \exists\,\text{unique }\mathbf{f}\in \widehat{Z}:\;
  \mathscr{F}_{\mathcal{B}}\left[\mathscr{L}\widehat{\mathbf{w}}\right]
  =
  \langle\mathbf{f},\mathscr{L}\widehat{\mathbf{w}}\rangle_{\Bc}
  \in\R,
  \;\;
  \forall\widehat{\mathbf{w}}\in\widehat{W}
\end{equation}
Functions $\mathbf{f}(\X)$ are the generalized forces
on (and inside of) $\Bc$,
both mechanical and thermal, and comprise the dual space
$\widehat{Z}^{\ast}$.
That is,
each functional $\mathscr{F}_{\mathcal{B}}$ in Eq.~(\ref{eq:inner})
defines the virtual power of forces $\widehat{P}_{F}(\Bc)$
produced by virtual velocities $\widehat{\mathbf{w}}$,
where $\widehat{P}_{F}:\widehat{Z}\rightarrow\R$ is identically
\begin{equation}\label{eq:PFB}
  \widehat{P}_{F}(\Bc)
  \equiv
  \mathscr{F}_{\mathcal{B}}\left[\mathscr{L}\widehat{\mathbf{w}}\right]
\end{equation}
\par
We now introduce the dynamic elements of the system.
Following the Germain--Maugin formalization of Hellinger's approach,
we consider an instance of an \emph{actual} velocity field
and its rate (acceleration):
$\mathbf{w}(\X)=[\mathbf{v}(\X)/\theta(\X)]$ and
$\dot{\mathbf{w}}(\X)=[\dot{\mathbf{v}}(\X)/\dot{\theta}(\X)]$,
where
$\mathbf{w}(\X),\dot{\mathbf{w}}(\X)\in W$.
(Note the missing hat ``$\,\widehat{\;\;}\,$''.)
\par
We denote by $\mathbf{m}(\X)$
the actual (stacked)
mechanical and thermal momenta per unit of current volume,
corresponding to actual velocities $\mathbf{w}(\X)$.
The mechanical momentum is given by the product $\rho(\X)\mathbf{v}(\X)$
of the current positive scalar mass density field $\rho(\X)$
and the actual velocity $\mathbf{v}(\X)$, the latter
defined as
$\mathbf{v}=\dot{\X}=\partial\boldsymbol{\chi}(\mathbf{X},t)/\partial t$,
or the mechanical momentum per unit mass.
Analogously, the thermal momentum per unit volume is
the product of $\rho(\X)$ and the specific (per unit current mass)
thermal momentum, $\eta(\X)$ \cite{Podio:2009a},
which is shown below to represent the specific entropy.
The two momenta are assembled into a stacked array
$\mathbf{m}=[\rho\mathbf{v}\big/\rho\eta]$.
\par
The corresponding \emph{inertial forces}
are the actual
rates $\rho\dot{\mathbf{v}}(\X)$
and $\rho\dot{\eta}(\X)$,
written as the stacked array $\dot{\mathbf{m}}$,
\begin{equation}\label{eq:m}
  \dot{\mathbf{m}}(\X) \equiv
  \left[
     \begin{array}{@{\extracolsep{\fill}}c@{\extracolsep{\fill}}}
       \rho\dot{\mathbf{v}}(\X)%
       \\[1pt]\hdashline[1pt/1pt]
       \rule{0pt}{2.0ex}\rho\dot{\eta}(\X)
     \end{array}
   \right]
\end{equation}
where, henceforth, the shorthand $\rho\dot{\mathbf{v}}(\X)$
and $\rho\dot{\eta}(\X)$ is used in place
of $\rho(\X)\dot{\mathbf{v}}(\X)$
and $\rho(\X)\dot{\eta}(\X)$.
We assume that $\rho(\X)$ is sufficiently smooth,
so that $\dot{\mathbf{m}}\in W$.
\par
The virtual power of inertial forces (or acceleration forces),
$\widehat{P}_{A}:\widehat{W}\rightarrow\R$,
is defined as the inner product of the virtual velocities
and their dual, inertial forces,
\begin{equation}\label{eq:PAB}
  \widehat{P}_{A}(\Bc)
    = \langle
         \dot{\mathbf{m}},
         \mathscr{I}
         \widehat{\mathbf{w}}
       \rangle_{\Bc}
\end{equation}
which is separated into
mechanical and thermal parts,
\begin{align}
 &\widehat{P}_{A}(\Bc)
  = \widehat{P}_{A,\text{mech}}(\Bc) + \widehat{P}_{A,\text{therm}}(\Bc)\\[1ex]
 &\begin{aligned}\label{eq:Pamhat}
    &\widehat{P}_{A,\text{mech}}(\Bc) =
      \langle
         \rho\dot{\mathbf{v}},
         \mathscr{I}
         \widehat{\mathbf{v}}
       \rangle_{\Bc} \\
    &\widehat{P}_{A,\text{therm}}(\Bc) =
      \big\langle
         \rho\dot{\eta},
         \mathscr{I}
         \widehat{\theta}\,
       \big\rangle_{\Bc}
  \end{aligned}
\end{align}
in which we have disallowed coupling of the parts,
and where $\mathscr{I}$ is the identity operator,
$\mathscr{I}\widehat{\mathbf{w}} = \widehat{\mathbf{w}}$.
\par
The principle of virtual power is comprised of
two postulates that apply to the virtual power of forces
$\widehat{P}_{F}$,
with each postulate restricting the dual space of forces $\mathbf{f}$
in Eq.~(\ref{eq:inner})
\cite{Germain:1973a,Maugin:1980a}.
Both postulates, stated below, apply to any interior open
connected set $\mathcal{D}\subseteq\mathcal{B}$,
with each region $\mathcal{D}$ being an instance of the
set $\mathscr{D}$
of all open connected subsets of $\mathcal{B}$.
The boundary of a region $\Dc$ is $\Dp$, and its closure is $\Do$.
By considering sub-regions $\mathcal{D}$ within $\mathcal{B}$,
we posit forces $\mathbf{f}$ that act internally
within $\mathcal{B}$~---
internal forces that act between a region $\Dc$
and its surroundings $\Bc\setminus\Do$.
\par
To extend the foregoing 
functionals on $\Bc$ to those on $\Dc$,
a functional
$\widehat{P}_{\bullet}(\Bc)=\mathscr{F}_{\Bc}[\bullet]$,
restricted to $\Dc$, is
written as
$\widehat{P}_{\bullet}(\Dc)\equiv\mathscr{F}_{\Dc}[\bullet]$,
with the associated inner product
$\langle\bullet,\bullet\rangle_{\Dc}$.
For example,
$\widehat{P}_{F}(\Dc)$
is the functional $\widehat{P}_{F}(\Bc)$ restricted to $\Dc$, as
$\widehat{P}_{F}(\Dc):\widehat{Z}\times\mathscr{D}\rightarrow\R$.
\par
The first postulate of virtual power
characterizes both mechanical and thermal balances
of a dynamic system:
\begin{equation}\label{eq:PVP}
  \widehat{P}_{F}(\Dc) 
  = \widehat{P}_{A}(\Dc),
  \quad
  \forall\widehat{\mathbf{w}}\in\widehat{W},\;
  \forall\Dc\in\mathscr{D}
\end{equation}
thus linking the duals of $\mathscr{L}\widehat{\mathbf{w}}$
and $\mathscr{I}\widehat{\mathbf{w}}$,
where the inner products $\langle\cdot,\cdot\rangle_{\mathcal{B}}$
in Eqs~(\ref{eq:inner})--(\ref{eq:Pamhat})
now apply to the sub-domain $\mathcal{D}$,
as $\langle\cdot,\cdot\rangle_{\mathcal{D}}$.
In the special case where $\dot{\mathbf{m}}=\mathbf{0}$,
the system is in static equilibrium,
and Eq.~(\ref{eq:PVP}) implies that
$\langle\mathbf{f},\mathscr{L}\widehat{\mathbf{w}}\rangle_{\Bc}$
is degenerate,
with its kernel (null space) being the space of equilibrium forces.
For such equilibrium systems,
Eq.~(\ref{eq:PVP}) restricts the full dual space
of all $\mathbf{f}\in\widehat{Z}^{\ast}$ to the
equilibrium ``E'' subspace
$\widehat{Z}^{\ast}_{\text{E}}$, or
$\mathbf{f}\in\widehat{Z}^{\ast}_{\text{E}}\subseteq\widehat{Z}^{\ast}$.
\par
The second postulate
involves the subspace of rigid-body virtual
velocities $\widehat{W}_{\text{r}}$, with
$\widehat{W}_{\text{r}}=\Vhat_{\text{r}}
\times\widehat{\Theta}_{\text{r}}\subseteq\widehat{W}$,
defined as
\begin{equation}\label{eq:Ur}
  \begin{aligned}
  \Vhat_{\text{r}}
  &=
  \left\{
  \widehat{\mathbf{v}}_{\text{r}}(\text{x})\in\Vhat
  \colon\;
  \widehat{\mathbf{v}}_{\text{r}}(\X)
  = \widehat{\mathbf{a}} + \widehat{\boldsymbol{\phi}}\times\X,\;
  \forall\X\in\Bc,\;\forall\,
    \widehat{\mathbf{a}},\widehat{\boldsymbol{\phi}}\in\RRR
  \right\} \\
  \widehat{\Theta}_{\text{r}}
  &=
  \left\{
  \widehat{\theta}_{\text{r}}(\text{x})\in\widehat{\Theta}
  \colon\;
  \widehat{\theta}_{\text{r}}(\X)
  = \widehat{\beta},\;
  \forall\X\in\Bc,\;\forall \widehat{\beta}\in\R
  \right\}
  \end{aligned}
\end{equation}
where the thermal displacement rate $\widehat{\beta}$
is an arbitrary scalar,
and velocity $\widehat{\mathbf{a}}$ and spin
$\widehat{\boldsymbol{\phi}}$ are arbitrary 3-vectors
(operator ``$\times$'' in Eq.~(\ref{eq:Ur}\textsubscript{1})
denotes the vector product).
\par
The second postulate applies to internal forces within $\Bc$.
The virtual power of forces $\widehat{P}_{F}(\Dc)$
is additively decomposed into functionals
$\widehat{P}_{F\text{(ext)}}(\Dc)$ and
$\widehat{P}_{F\text{(int)}}(\Dc)$,
identified as the external and internal
parts of $\widehat{P}_{F}(\Dc)$;
the latter part is subject to the additional restriction
imposed by this postulate.
The second postulate requires that
the virtual internal power
$\widehat{P}_{\text{F(int)}}(\Dc)$
vanishes for all rigid-body virtual motions,
regardless of interior domain $\mathcal{D}$:
\begin{equation}\label{eq:equil2}
  \widehat{P}_{F\text{(int)}}(\Dc)
  \equiv
  \mathscr{F}_{\mathcal{D},\text{(int)}}[\mathscr{L}\widehat{\mathbf{w}}_{\text{r}}]=0,
  \quad
  \forall\widehat{\mathbf{w}}_{\text{r}}\in\widehat{W}_{\text{r}},\;
  \forall\mathcal{D}\in\mathscr{D}
\end{equation}
This requirement is Germain's statement of objectivity:
that the virtual power of the internal forces is indifferent to
``the frame in which the motion is observed \cite{Germain:1973a}.'' 
When combined with the first postulate,
it becomes Hellinger's ``rigidifying principle'':
the external and boundary forces on a part $\mathcal{D}$ cut
from a body $\mathcal{B}$ must themselves
be in equilibrium as a rigid body.
\par
Together,
the two postulates
are the \emph{principle of virtual work}:
\begin{equation}\label{eq:master}
  \begin{alignedat}{2}
    \widehat{P}_{F}(\Dc)
    \equiv
    \widehat{P}_{F\text{(ext)}}(\Dc) + \widehat{P}_{F\text{(int)}}(\Dc)
    &
      =\widehat{P}_{A}(\Dc),
      \quad&&
      \forall\widehat{\mathbf{w}}\in\widehat{W},\;
      \forall\mathcal{D}\in\mathscr{D}
  \\
    \widehat{P}_{F\text{(int)}}(\Dc) &\equiv
  \mathscr{F}_{\mathcal{D},\text{(int)}}
    \left[\mathscr{L}\widehat{\mathbf{w}}_{\text{r}}\right]
      =0,
      \quad&&
      \forall\widehat{\mathbf{w}}_{\text{r}}\in\widehat{W}_{\text{r}},\;
      \forall\mathcal{D}\in\mathscr{D}
  \end{alignedat}
\end{equation}
which establish momentum balance
(both mechanical and thermal)
of a system.
The two postulates apply to all open connected subregions
$\Dc\in\mathscr{D}$, and henceforth, this condition will be implied
although not always explicitly stated.
\par
The choice of a continuum~---
classical, generalized, local, nonlocal, etc.~---
is defined by three choices,
thus determining its essential nature:
the choice of Hilbert space
$\widehat{W}$ (from which the rigid-body subspace
$\widehat{W}_{\text{r}}$ is identified);
the choice of inner product $\langle\cdot,\cdot\rangle$
appropriate to $\widehat{W}$; and
the choice of operator $\mathscr{L}$,
which defines the space $\widehat{Z}$.
Although the partition of power $\widehat{P_{F}}$ into
internal and external parts constitutes an additional choice,
it is typically determined by the particular
rigid body subspace $\widehat{W}_{\text{r}}$.
Different Hilbert spaces are selected in
Sections~\ref{sec:localstatics}
and~\ref{sec:nonlocalstatics} for local and non-local media.
\par
Henceforth,
linear operators $\mathscr{L}$ are taken to be integral
operators with kernel functions
$\kappa(\XX):\Bc\,\times\,\RRR\rightarrow\R$ that
apply over the full domain $\Xp\in\RRR$
and that render the space
$\widehat{Z}=\{ \mathscr{L} \widehat{\mathbf{w}}: \widehat{\mathbf{w}}\in\widehat{W}\}$
a Hilbert space.
Kernel $\kappa$ encodes the nonlocal (or possibly local) behavior at $\X$,
capturing the influence~--- if any~---
of neighboring points $\Xp$ on $\X$.
The inner products $\langle\cdot,\cdot\rangle$
in Eqs.~(\ref{eq:inner}), (\ref{eq:PAB}), and~(\ref{eq:Pamhat})
are taken as integrals of function products,
say $f(\X)$ and $g(\X)$, over a specified domain $\mathcal{R}$.
Thus, operator $\mathscr{L}$ and inner product
$\langle\cdot,\cdot\rangle_{\mathcal{R}}$ are
\begin{equation}\label{eq:defL}
  \begin{aligned}
  \mathscr{L}y(\X)
  &=
  \int_{\RRR}
  \kappa(\XX)y(\Xp)\,\Vp
  \\
  \quad
  \langle
    f,g 
  \rangle_{\mathcal{R}}
  &= \int_{\mathcal{R}}
  f(\X) g(\X)\,dv
  \end{aligned}
\end{equation}
where $\Vp$ and $dv$ are measures appropriate to the domain.
\par
In representing various components of
$\langle f,\mathscr{L}y\rangle_{\mathcal{R}}$,
the chosen notation depends
on whether a component is a scalar field $y(\X)$ or
a vector field $\mathbf{y}(\X)$
(for example, $\widehat{\theta}$ or $\widehat{\mathbf{v}}$
respectively),
whether a scalar kernel $\kappa$ or a vector kernel $\mathbf{k}$
is being applied
(for example, scalar $\kappa$ or a gradient
$\mathbf{k}=\kappa_{,i}$ respectively),
and whether the region $\mathcal{R}$
is a volume or a surface
(for example, region $\mathcal{D}$ or boundary $\partial\mathcal{D}$):
\begin{equation}\label{eq:abbrev}
  \begin{aligned}
     \llangle
      f,\kappa\,y
     \rrangle_{\mathcal{R}}
    &\equiv
    \int_{\mathcal{R}}
    f(\X)
    \left(\;
      \int_{\RRR}
      \kappa(\XX)
      y(\Xp)\,\Vp
    \right)
    d\mu \\
    \llangle
      \mathbf{f},\kappa\,\mathbf{y}
    \rrangle_{\mathcal{R}}
    &\equiv
    \int_{\mathcal{R}}
    \mathbf{f}(\X)\cdot
    \left(\;
      \int_{\RRR}
      \kappa(\XX)
      \mathbf{y}(\Xp)\,\Vp
    \right)
    d\mu \\
    \llangle
      \mathbf{f},\mathbf{k}y
    \rrangle_{\mathcal{R}}
    &\equiv
    \int_{\mathcal{R}}
    \mathbf{f}(\X)\cdot
    \left(\;
      \int_{\RRR}
      \mathbf{k}(\XX)
      y(\Xp)\,\Vp
    \right)
    d\mu \\
    \llangle
      \mathbf{F},\mathbf{k}\otimes\mathbf{y}
    \rrangle_{\mathcal{R}}
    &\equiv
    \int_{\mathcal{R}}
    \mathbf{F}(\mathbf{x}) :
    \left(\;
      \int_{\RRR}
      \mathbf{k}(\XX)\otimes
      \mathbf{y}(\Xp)\,\Vp
    \right)
    d\mu
  \end{aligned}
\end{equation}
where $d\mu$ designates the measure of volume $dv$
or surface area $da$, and
the prime ``\,$^{\prime}\,$'' distinguishes the two integration variables
and their measures.
Such operators $\mathscr{L}$ and bilinear forms
$\llangle\cdot,\cdot\rrangle$
are sufficiently general to support both local and nonlocal continua.
\subsection{Momentum balance of local continua}\label{sec:localstatics}
Before addressing nonlocal continua,
we briefly consider a classical, local continuum
(see
\cite{Eugster:2017a,Germain:1973a,Maugin:1980a}).
The space $\widehat{W}$ is here taken as
the pre-Hilbert space of continuous and continuously
differentiable functions on $\Bc$,
or $\widehat{W}=C^{2}(\Bc)$, which is
a smaller space than that used for nonlocal media
(Section~\ref{sec:nonlocalstatics}).
The chosen operator $\mathscr{L}$ maps
an element $\widehat{\mathbf{w}}$
to the product space of $\widehat{\mathbf{w}}$
and its spatial derivatives,
expressed as a stacked array:
\begin{equation}\label{eq:local1}
  \mathscr{L}\colon\widehat{W}\rightarrow\widehat{Z},
  \quad
  \mathscr{L}\widehat{\mathbf{w}}=
  \left[
    \begin{array}{@{\extracolsep{\fill}}c@{\extracolsep{\fill}}}
      \mathscr{L}_{1}\widehat{\mathbf{w}}(\X)\\\hdashline[1pt/1pt]
      \mathscr{L}_{2}\widehat{\mathbf{w}}
    \end{array}
  \right]=
  \left[
    \begin{array}{@{\extracolsep{\fill}}c@{\extracolsep{\fill}}}
      \widehat{\mathbf{w}} \\\hdashline[1pt/1pt]
      \partial\widehat{\mathbf{w}}/\partial\X
    \end{array}
  \right]
  \quad\text{or}\quad
  \left[
    \begin{array}{@{\extracolsep{\fill}}c@{\extracolsep{\fill}}}
      \int_{\RRR}\delta(\XX)
                     \,\widehat{\mathbf{w}}(\Xp)\,\Vp
                  \\[1pt]\hdashline[1pt/1pt]
      \rule{0ex}{2.5ex}
      \int_{\RRR}\widetilde{\boldsymbol{\nabla}}(\XX)
                    \,\widehat{\mathbf{w}}(\Xp)\,\Vp
    \end{array}
  \right]
\end{equation}
where $\widehat{Z}$ is the product (stacked) space of
$\widehat{W}$ and its derivatives.
For the local medium of Eq.~(\ref{eq:local1}),
kernels $\kappa$ and $\boldsymbol{\kappa}$ in
Eq.~(\ref{eq:abbrev}) are
the notional Dirac delta kernel and its gradient analogue,
$\delta(\XX)$ and $\widetilde{\boldsymbol{\nabla}}(\XX)$,
which effect these mappings:
    $\int_{\RRR}\delta(\XX)
       \widehat{\mathbf{w}}(\Xp)\,\Vp
       \equiv \widehat{\mathbf{w}}(\X)$
and $\int_{\RRR}\widetilde{\boldsymbol{\nabla}}(\XX)
       \widehat{\mathbf{w}}(\Xp)\,\Vp
       \equiv
       \left.\boldsymbol{\nabla}\widehat{\mathbf{w}}\right|_{\X}$.
Thus only the values and gradients of
$\widehat{\mathbf{w}}$ at points $\X\in\Bc$ are involved in 
$\mathscr{L}\widehat{\mathbf{w}}$, nullifying the influence of
nearby points $\Xp$.
Finally, the inner product of two functions on $\Dc$
is the integral of their product over $\Dc$
(Eq.~\ref{eq:defL}\textsubscript{2}).
\par
With these choices,
the external and internal
virtual powers,
$\widehat{P}_{F\text{(ext)}}(\Dc)$ and $\widehat{P}_{F\text{(int)}}(\Dc)$,
are associated respectively with the
operators $\mathscr{L}_{1}$ and $\mathscr{L}_{2}$ in
Eq.~(\ref{eq:local1}).
When Eqs.~(\ref{eq:PFB})--(\ref{eq:Pamhat}) are
restricted to the domain $\mathcal{D}$ and its closure,
the virtual powers of a local, classical continuum are as follows
(see Eq.~\ref{eq:master}):
\begin{align}
  &\begin{aligned}\label{eq:local2}
    \widehat{P}_{F}(\Dc) &=
    \widehat{P}_{F\text{(ext)}}(\Dc) + \widehat{P}_{F\text{(int)}}(\Dc)
    \\
      &=
      \langle
        \mathbf{f}_{1},\mathscr{L}_{1}\widehat{\mathbf{w}}
      \rangle_{\Dc} +
      \langle
        \mathbf{f}_{2},\mathscr{L}_{2}\widehat{\mathbf{w}}
      \rangle_{\Dc}
    \\
  \end{aligned}\\[1ex]
  &\widehat{P}_{A}(\Dc) =
  \langle
        \dot{\mathbf{m}},\mathscr{L}_{1}\widehat{\mathbf{w}}
  \rangle_{\Dc} =
  \llangle
        \rho\dot{\mathbf{v}},\delta\widehat{\mathbf{v}}\,
  \rrangle_{\Dc} +
  \llangle
        \rho\dot{\eta},\delta\widehat{\theta}\,
  \rrangle_{\Dc}
  \\[1ex]
  \label{eq:local}
  &\begin{aligned}
     \widehat{P}_{F\text{(ext)}}(\Dc) &=
     \llangle
       \rho\mathbf{b},\delta\,\widehat{\mathbf{v}}\,
     \rrangle_{\Dc} +
     \llangle
       \mathbf{t},\delta\,\widehat{\mathbf{v}}\,
     \rrangle_{\Dp} +
     \llangle
       \rho s,\delta\,\widehat{\theta}\,
     \rrangle_{\Dc} +
     \llangle
       \rho\xi,\delta\,\widehat{\theta}\,
     \rrangle_{\Dc} +
     \llangle
       -k,\delta\,\widehat{\theta}\,
     \rrangle_{\Dp}
     \\
     \widehat{P}_{F\text{(int)}}(\Dc) &=
     \llangle
       -\mathbf{T},\widetilde{\boldsymbol{\nabla}}\widehat{\mathbf{v}}\,
     \rrangle_{\Dc} +
     \llangle
       \mathbf{p},\widetilde{\boldsymbol{\nabla}}\widehat{\theta}\,
     \rrangle_{\Dc}
  \end{aligned}
\end{align}
where $\widehat{P}_{F\text{(int)}}$
is the internal virtual power,
to which postulate~(\ref{eq:master}\textsubscript{2}) is applied,
and the various function $\mathbf{b}$, $\mathbf{T}$, $k$, etc.
apply to separate contributions~---
external and internal, interior and boundary,
and mechanical and thermal~---
to the virtual power of forces.
\par
In Eq.~(\ref{eq:local}),
the dual, work-conjugate mechanical force quantities are
interpreted as follows:
$\mathbf{b}(\X)$ is
the external body force per unit mass;
$\mathbf{t}(\X)$ is
the surface traction per unit area; and
$\mathbf{T}(\X)$ is
the Cauchy stress.
The dual thermal force quantities are interpreted
as supplies or flows of entropy (thermal momentum), a connection
made more definite in Section~\ref{sec:thermomech}:
$-k(\X)$ is
the external entropy flux per unit area, and
$\mathbf{p}(\X)$ is
the internal entropy flow vector.
The external power $\widehat{P}_{F\text{(ext)}}$
includes two sources of entropy per unit of mass:
$s(\X)$ is the specific external entropy supply rate (per unit mass);
whereas
$\xi(\X)$ is the specific rate of net internal entropy production.
Although two sources,
$s(\X)$ and $\xi(\X)$,
appear in the external power, they are treated separately:
$s(\X)$ is a supply that can be directly controlled (or reversed)
by an external agent,
whereas $\xi(\X)$ is generated internally in response
to mechanical and thermal processes.
The latter plays a central role in the second law of
thermomechanics, due to its dissipative and irreversible character.
In a similar manner, one could distinguish between
body forces of external and internal origins, say
$\mathbf{b}_{\text{ext}}$ and $\mathbf{b}_{\text{int}}$,
with the former controlled by an external agent
(e.g., gravity), and the latter arising from internal
attractions and repulsions within $\Bc$.
No advantage is gained by this distinction here,
and $\mathbf{b}$ includes both contributions.
\par
When applying the principle of virtual power to a local medium,
one must acknowledge a limitation of the space $C^{2}(\Bc)$:
as a pre-Hilbert (non-complete) space,
there exist functionals
$\widehat{P}_{F}$ and $\widehat{P}_{A}$ whose dual functions
$\mathbf{b}$, $\mathbf{T}$, $\xi$,\ etc.
do not lie within $C^{2}(\Bc)$.
However, the derivations in this section rely on
the fundamental lemma of variational calculus,
which only requires the smoothness conditions of space $C^{2}(\Bc)$.
\par
The balance equations for the the local continuum are derived by
applying the Germain--Maugin formalism
(see
\cite{Germain:1973a,Lidstrom:2012a,Green:1991a}).
Briefly,
\begin{itemize}
\item
By applying Eqs.~(\ref{eq:Ur}) and~(\ref{eq:master}\textsubscript{2})
with
Eq.~(\ref{eq:local}\textsubscript{2}) and with
$\widehat{\mathbf{a}}=\mathbf{0}$,
$\widehat{\beta}=0$
but arbitrary $\widehat{\boldsymbol{\phi}}$,
the stress tensor $\mathbf{T}$ must be symmetric.
\item
Applying Eqs.~(\ref{eq:Ur}) and~(\ref{eq:master}\textsubscript{1}),
with $\widehat{\mathbf{w}}=\widehat{\mathbf{w}}_{\text{r}}$ and
with
$\widehat{\boldsymbol{\phi}}=\mathbf{0}$, $\widehat{\beta}=0$ but
arbitrary $\widehat{\mathbf{a}}$,
the global force balance equation is obtained
for an arbitrary region $\mathcal{D}$:
$\int_{\mathcal{D}}\rho\mathbf{b}\,dv+\int_{\partial\mathcal{D}}\mathbf{t}\,da=
\int_{\mathcal{D}}\rho\dot{\mathbf{v}}\,dv$.
%
\item
A similar application of
Eq.~(\ref{eq:master}\textsubscript{1})
yields the global moment balance,
with $\widehat{\mathbf{w}}=\widehat{\mathbf{w}}_{\text{r}}$ and
with $\widehat{\mathbf{a}}=\mathbf{0}$,
$\widehat{\beta}=0$, and
arbitrary $\widehat{\boldsymbol{\phi}}$:
$\int_{\mathcal{D}}(\rho\mathbf{b}\times\X)\,dv+\int_{\partial\mathcal{D}}(\mathbf{t}\times\X)\,da=\int_{\mathcal{D}}(\rho\dot{\mathbf{v}}\times\X)\,dv$.
\item
Assuming arbitrary $\widehat{\mathbf{v}}\in\Vhat$ and
$\widehat{\theta}(\X)=0$, and
applying the divergence theorem,
the local balance equations follow:
\begin{equation}\label{eq:equillocal}
  \begin{aligned}
    &\rho\mathbf{b}(\X)+\boldsymbol{\nabla}\cdot\mathbf{T}(\X)
    =\rho\dot{\mathbf{v}}(\X), \quad\forall\X\in\Bc
    \\
    &\mathbf{t}(\X)
    =\mathbf{T}(\X)\cdot\mathbf{n}(\X), \quad\forall\X\in\Bp
  \end{aligned}
\end{equation}
of which static forms were derived by Piola in 1848,
also using the principle of virtual work \cite{Capecchi:2007a}.
\item
Assuming arbitrary $\widehat{\theta}(\X)$ but
$\widehat{\mathbf{v}}(\X)=\mathbf{0}$,
and applying the Gauss theorem,
the local balance equations are derived:
$\rho s+\rho\xi-\text{div}(\mathbf{p})=\rho\dot{\eta}$
for all $\X\in\mathcal{B}$,
and $k=\mathbf{p}\cdot\mathbf{n}$
for all $\X\in\partial\mathcal{B}$.
\end{itemize}
Note that the assumption of
virtual velocities $\widehat{\mathbf{w}}(\X)$ in the pre-Hilbert
$C^{2}(\Bc)$ precludes Heaviside-like jump discontinuities
along singular surfaces, such as
slip surfaces 
and at material and phase interfaces.
Daher and Maugin \cite{Daher:1986a}
addressed this situation within a virtual power framework by partitioning
$\Bc$ into sub-bodies, with $C^{2}$ velocities applied
separately within each sub-body and along their interfaces.
This complication is not needed for nonlocal media,
considered hereafter,
since the velocities $\widehat{\mathbf{w}}(\X)$ are assumed in
$L^{2}$, which admits more general velocity fields.
\subsection{Momentum balance of nonlocal continua}%
\label{sec:nonlocalstatics}
For nonlocal continua,
we start with a space $\widehat{W}$ of virtual velocities,
the product space of $\widehat{V}\times\widehat{\Theta}$,
that is larger than was
assumed for its local counterpart:
the Hilbert space of square-integrable functions
on $\RRR$,
$\widehat{W}=L^{2}(\RRR)$,
with $\widehat{\mathbf{w}}\in\widehat{W}$.
Instead of the Dirac kernels
$\delta$ and $\widetilde{\boldsymbol{\nabla}}$,
we introduce 
a function $\varphi$ as the kernel of a linear
integral transform that serves as a ``mollifier''
or influence function, 
which averages, blurs, or attenuates \cite{Bazant:2002a}
otherwise local quantities, with
$\varphi(\XX)
\colon\Bc\times\RRR\rightarrow\R$.
The linear mapping $\mathscr{L}$ to the Hilbert space $\widehat{Z}$ is
\begin{equation}\label{eq:nonlocal1}
  \mathscr{L}\colon\widehat{W}\rightarrow\widehat{Z},
  \quad
  \mathscr{L}\widehat{\mathbf{w}}=
  \left[
    \begin{array}{@{\extracolsep{\fill}}c@{\extracolsep{\fill}}}
      \mathscr{L}_{1}\widehat{\mathbf{w}}\\\hdashline[1pt/1pt]
      \mathscr{L}_{2}\widehat{\mathbf{w}}
    \end{array}
  \right]=
  \left[
    \begin{array}{@{\extracolsep{\fill}}c@{\extracolsep{\fill}}}
      \varphi\widehat{\mathbf{w}} \\\hdashline[1pt/1pt]
      (\boldsymbol{\nabla}\varphi)\otimes\widehat{\mathbf{w}}
    \end{array}
  \right]
  = 
  \left[
    \begin{array}{@{\extracolsep{\fill}}c@{\extracolsep{\fill}}}
      \int_{\RRR}\varphi(\XX)\,
                        \widehat{\mathbf{w}}(\Xp)\,\Vp \\[1pt]\hdashline[1pt/1pt]
      \rule{0ex}{2.5ex}
      \int_{\RRR}\boldsymbol{\nabla}_{\X}
                                   \varphi(\XX)\otimes
                                   \widehat{\mathbf{w}}(\Xp)\,\Vp
    \end{array}
  \right]
\end{equation} 
in which external and internal
virtual powers,
$\widehat{P}_{F\text{(ext)}}(\Dc)$ and $\widehat{P}_{F\text{(int)}}(\Dc)$,
are associated with the stacked
operators $\mathscr{L}_{1}$ and $\mathscr{L}_{2}$ respectively,
and in which
$(\boldsymbol{\nabla}\varphi)\otimes\widehat{\mathbf{w}}$
is the function of $\X$
given by the stacked array
\begin{equation}\label{eq:nonlocal2}
  \begin{aligned}
    \big[
    (\boldsymbol{\nabla}\varphi)\otimes\widehat{\mathbf{w}}
    \big](\X)
    &=
    \left[
      \begin{array}{@{\extracolsep{\fill}}c@{\extracolsep{\fill}}}
        (\boldsymbol{\nabla}\varphi)\otimes\widehat{\mathbf{v}}\\[1pt]\hdashline[1pt/1pt]
        \rule{0pt}{2.5ex}(\boldsymbol{\nabla}\varphi)\widehat{\theta}
      \end{array}
    \right]
    =
    \left[
      \begin{array}{@{\extracolsep{\fill}}c@{\extracolsep{\fill}}}
        \int_{\RRR}
        \varphi_{,j}(\XX)
        \widehat{v}_{i}\,\Vp\\[1pt]\hdashline[1pt/1pt]
        \rule{0pt}{2.5ex}
        \int_{\RRR}
        \varphi_{,i}(\XX)
        \widehat{\theta}\,\Vp
      \end{array}
    \right]
  \end{aligned}
\end{equation}
where $\varphi_{,j}$ expresses
differentiation with respect to the first argument $\X$,
as $\varphi_{,j}(\XX)=\partial\varphi(\XX)/\partial{x_{j}}$.
Expression~(\ref{eq:nonlocal2})
thus represents the virtual gradients
of the motions $\widehat{\mathbf{v}}$ and $\widehat{\theta}$ at $\X$,
taken as weighted averages
over a neighborhood of $\X$.
\par
To enable further developments,
we limit function $\varphi$ to one that is sufficiently smooth
to ensure that $\mathscr{L}\widehat{\mathbf{w}}$ is also
square-integrable on $\Bc$ and that admits application of
the Gauss divergence theorem to the gradient $\boldsymbol{\nabla}\varphi$.
Specifically, $\varphi$ is assumed bounded and continuous,
with first derivatives
that are almost everywhere bounded and continuous:
$\varphi\in C^{0}(\Bc)\cap W^{1,1}(\Bc)$.%
\footnote{
A broader class of kernels $\varphi$
is certainly available to ensure that
$\mathscr{L}\widehat{\mathbf{w}}\in L^{2}(\Bc)$.
As examples, either a Carleman kernel $\kappa(\X,\Xp)$,
such that $\kappa(\X,\Xp)\in L^{2}(\RRR)$
for almost every fixed $\X$,
or a kernel $\kappa(\X,\Xp)\in L^{2}(\RRR\times\RRR)$
would suffice \cite{Jorgens:1982a}.}
For the same reasons,
the support of $\varphi$ at $\X$, written as
$\text{supp}(\varphi,\X)=\{ 
\Xp\in\RRR:\varphi(\XX)\neq 0
\}$,
is assumed Lipschitz-smooth and compact in $\RRR$
for all $\X\in\Bo$.
This support defines the nonlocal neighborhood that influences
point $\X$.
\par
These assumptions are sufficient to ensure that
$\widehat{Z}=\{ \mathscr{L} \widehat{\mathbf{w}}: \widehat{\mathbf{w}}\in\widehat{W}\}$
is a Hilbert space,
though more general conditions on $\varphi$ could be adopted.
Further restrictions on $\varphi$ will later follow from the
postulate of virtual power.
Specifically, the postulate restricts
$\varphi$ to a mollifier $\varphi(\XX)$ of the more
limited form $\varphi(\XX )=\varphi(\X-\Xp)$;
under this restriction,
$\varphi\in W^{1,1}(\Bc)$ suffices
to ensure that $\mathscr{L}\widehat{\mathbf{w}}\in L^{2}(\Bc)$
(a consequence of Young's convolution inequality
\cite{Bogachev:2007a}).
\par
The mollifier $\varphi$ serves as a characteristic property
of the material,
supplementing its other properties.
It is similar to convolution operators that are
used in strain-driven nonlocal constitutive models
\cite{Eringen:1983a,Bazant:1984c,Polizzotto:2001a}.
Such constitutive models are expedient means
of introducing a length-scale into an otherwise
local continuum;
for example, the support's width
may serve as an intrinsic length.
The framework discussed here, however, does not
merely smooth the strain field
(actually, the stiffness operator);
it also demands
reappraisal of the concepts of body force and traction,
stress and entropy rates,
the governing balance and kinetic equations,
and permissible constitutive forms,
all for intrinsically nonlocal media.
\par
With these choices of
$\widehat{W}$ and $\mathscr{L}$ for a nonlocal continuum,
the virtual powers of the forces and inertias
and of the external and internal virtual powers are as follows:
\begin{align}
  \label{eq:nonlocalPF}
  &\begin{aligned}
    \widehat{P}_{F}(\Dc) &=
    \widehat{P}_{F\text{(ext)}}(\Dc) + \widehat{P}_{F\text{(int)}}(\Dc)
      \\
      &=
      \langle
        \mathbf{f}_{1},\mathscr{L}_{1}\widehat{\mathbf{w}}
      \rangle_{\Dc} +
      \langle
        \mathbf{f}_{2},\mathscr{L}_{2}\widehat{\mathbf{w}}
      \rangle_{\Dc}
  \end{aligned}\\[1ex]
  \label{eq:virtPAD}
  &\widehat{P}_{A}(\Dc) =
  \langle
        \dot{\mathbf{m}},\mathscr{L}_{1}\widehat{\mathbf{w}}
  \rangle_{\Dc} =
  \llangle
        \rho\dot{\mathbf{v}},\varphi\widehat{\mathbf{v}}\,
  \rrangle_{\Dc} +
  \llangle
        \rho\dot{\eta},\varphi\widehat{\theta}\,
  \rrangle_{\Dc}
  \\[1ex]
  \label{eq:nonlocal4}
  &\begin{aligned}
     \widehat{P}_{F\text{(ext)}}(\Dc) &=
     \llangle
       \rho\mathbf{b},\varphi\widehat{\mathbf{v}}\,
     \rrangle_{\Dc} +
     \llangle
       \mathbf{t},\varphi\widehat{\mathbf{v}}\,
     \rrangle_{\Dp} +
     \llangle
       \rho s,\varphi\,\widehat{\theta}\,
     \rrangle_{\Dc} +
     \llangle
       \rho\xi,\varphi\,\widehat{\theta}\,
     \rrangle_{\Dc} +
     \llangle
       -k,\varphi\,\widehat{\theta}\,
     \rrangle_{\Dp}
     \\
     \widehat{P}_{F\text{(int)}}(\Dc) &=
     \llangle
       -\mathbf{T},
       \big(\boldsymbol{\nabla}\varphi\big)\otimes\widehat{\mathbf{v}}\,
     \rrangle_{\Dc} +
     \llangle
       \mathbf{p},
       \big(\boldsymbol{\nabla}\varphi\big)\widehat{\theta}\,
     \rrangle_{\Dc}
  \end{aligned}
\end{align}
where the dual force quantities in
$\widehat{P}_{F\text{(ext)}}$ and $\widehat{P}_{F\text{(int)}}$~---
$\mathbf{T}$, $\mathbf{t}$, $s$, $\xi$, etc.~---
are those that were
characterized following Eq.~(\ref{eq:local}).
The Eq.~(\ref{eq:nonlocal4}) includes all terms that are consistent
with the rigidifying principle~(\ref{eq:master}\textsubscript{2})
in the context of a first-gradient but nonlocal framework, and
as shown below, the principle
also predicates a particular separation of
external and internal virtual powers
that is implicit in Eq.~(\ref{eq:nonlocal4}).
\par
We now apply the principle of virtual power~---
Eqs.~(\ref{eq:master}\textsubscript{1})
and~(\ref{eq:master}\textsubscript{2})~---
to nonlocal continua.
The principle yields restrictions
on the force quantities
in the form of balance equations.
The principle also motivates
restrictions on the kernel function $\varphi$
that simplify these equations.
In general, the more (less) restrictive the conditions that are chosen for
$\varphi$, the less (more) restrictive are the conditions
required on the force quantities.
\begin{enumerate}[listparindent=3ex,parsep=0ex]
\item
\emph{Choice of internal/external partition:}
Although the partition of $\widehat{P}_{F}(\Dc)$ into the external and
internal parts of Eq.~(\ref{eq:nonlocal4}) may seem arbitrary,
it is justified by applying
condition~(\ref{eq:master}\textsubscript{2})
with an arbitrary
vector $\widehat{\mathbf{a}}$
but with zero rotation, $\widehat{\boldsymbol{\phi}}=\mathbf{0}$,
and under static conditions, $\dot{\mathbf{m}}(\X)=\mathbf{0}$.
In particular, if any of the four parts
of Eq.~(\ref{eq:nonlocal4}\textsubscript{1})
were included in the \emph{internal} virtual power, then
the combined internal power
could not be zero unless
the associated dual force was identically zero
for all $\mathbf{x}\in\Bc$.
Such restrictions are immoderate, and
a reasonable conclusion is that 
all terms in Eq.~(\ref{eq:nonlocal4}\textsubscript{1})
are part of the external power,
and that, indeed, the internal virtual power
is correctly associated with the
terms in Eq.~(\ref{eq:nonlocal4}\textsubscript{2}).
\item
\emph{Rigid translational power and restrictions on the mollifier:}
Having established that the
terms in Eq.~(\ref{eq:nonlocal4}\textsubscript{1})
belong to the external power,
we consider the expression for $\widehat{P}_{F\text{(int)}}$
in Eq.~(\ref{eq:nonlocal4}\textsubscript{2}) and again apply
arbitrary rigid virtual velocities
$\widehat{\mathbf{a}}$ and $\widehat{\beta}$ but
zero rotation $\widehat{\boldsymbol{\phi}}=\mathbf{0}$.
To satisfy principle~(\ref{eq:master}\textsubscript{2}),
restrictions must be placed either on the stress $\mathbf{T}(\X)$
and the entropy flux $\mathbf{p}(\X)$, 
on the mollifier function $\varphi$, or on combinations of
these two, such that
\begin{equation}\label{eq:condition2}
  \int_{\Dc}\!\!
  \big(
  -\widehat{a}_{i} T_{ji}(\X)
  + \widehat{\beta} p_{j}(\X)
  \big)
  \left(\:
  \int_{\RRR}
  \varphi_{,j}(\XX )\,\,\Vp
  \right)
  dv = 0
  ,\;\:
  \forall\widehat{\mathbf{a}}\in\RRR,\;
  \forall\widehat{\beta}\in\R,\;
  \forall\mathcal{D}\in\mathscr{D},\;i\in\{1,2,3\}
\end{equation}
in which $\varphi_{,j}$ implies differentiation with respect
to the first argument $\X$,
and summation applies to index $j$.
To avoid forcing $\mathbf{T}$ or $\mathbf{p}$ to vanish identically,
the following restriction on mollifier $\varphi$ is
reasonable and sufficient
for satisfying condition~(\ref{eq:condition2}):
\begin{equation}\label{eq:condition1}
  \int_{\RRR}
  \varphi_{,j}(\XX )\,\Vp = 0
   ,\quad
   \forall\,\X\in\Bo
\end{equation}
Because derivatives $\varphi_{,j}$ are assumed bounded and
continuous,
the strong form of
the Leibniz--Reynolds transport theorem applies,
and since
region $\RRR$ is independent of $\X$,
\begin{equation}\label{eq:unvary}
  \int_{\RRR}
  \frac{\partial}{\partial x_{i}}\varphi(\XX)\,\Vp
  =
  \frac{\partial}{\partial x_{i}}
  \int_{\RRR}
  \varphi(\XX)\,\Vp
  =
  0
\end{equation}
so that the scalar-valued integral $\int_{\RRR}\varphi(\XX)\,\Vp$
is spatially uniform on $\Bc$
(i.e., independent of position $\X$).
Note that the shape of $\varphi$ can vary with $\X$,
but Eq.~(\ref{eq:unvary}) requires its integral is constant.
Although this assumption of macroscale homogeneity is
adopted henceforth,
for bodies in which the integral varies with position
(e.g., in bodies with distinct regions of different materials),
one must revert to Eq.~(\ref{eq:condition1}) and impose
its restriction
on the fields $\mathbf{T}(\X)$ and $\mathbf{p}(\X)$.
\item
\emph{Rigid rotational power and further restrictions on the mollifier:}
Applying condition~(\ref{eq:master}\textsubscript{2})
to the internal virtual power,
with arbitrary rotation $\widehat{\boldsymbol{\phi}}$
but with $\widehat{\mathbf{a}}=\mathbf{0}$ and $\widehat{\beta}=0$,
yields
\begin{equation}\label{eq:condition3}
  \int_{\Dc}\!\!
  T_{\ell i}(\X)
  \left(\:
  \int_{\RRR}
  e_{ijk}\widehat{\phi}_{j}x^{\prime}_{k}
  \varphi_{,\ell}(\XX )
  \,\Vp
  \right)
  dv
  = 0,
  \quad
  \forall\widehat{\boldsymbol{\phi}}\in\RRR,\;
  \forall\mathcal{D}\in\mathscr{D}
\end{equation}
where $\varphi_{,\ell}=\partial\varphi/\partial x_{\ell}$,
and summation applies to all indices.
\par
Equation~(\ref{eq:condition3}) is satisfied
for all $\widehat{\boldsymbol{\phi}}$ by placing
conditions on either $\varphi$ or $\mathbf{T}$, or on  both.
A reasonable condition is requiring $\varphi$
to be a function of the difference $\X-\Xp$,
with $\varphi(\XX )=\varphi(\X-\Xp)$,
so that the integral operator $\mathscr{L}y(\X)$
in Eq.~(\ref{eq:defL}\textsubscript{1}) is, in general,
a negative convolution.
In particular,
\begin{equation}\label{eq:condition30}
  \begin{aligned}
    \int_{\RRR}
    &e_{ijk}\widehat{\phi}_{j}x^{\prime}_{k}
    \frac{\partial}{\partial x_{\ell}}
    \varphi(\X-\Xp)\,\Vp
    =
    -\int_{\RRR}
    e_{ijk}\widehat{\phi}_{j}x^{\prime}_{k}
    \frac{\partial}{\partial x^{\prime}_{\ell}}
    \varphi(\X-\Xp)\,\Vp\\
    &= -\int_{\RRR}
    e_{ijk}\widehat{\phi}_{j}
    \Big[
    \frac{\partial}{\partial x'_{\ell}}
    \Big(
      x^{\prime}_{k}\varphi(\X-\Xp)
    \Big)
    -\frac{\partial x^{\prime}_{k}}{\partial x^{\prime}_{\ell}}
     \varphi(\X-\Xp)
    \Big]\,\Vp\\
    &=
    e_{ij\ell}\widehat{\phi}_{j}
    \int_{\RRR}
    \varphi(\X-\Xp)\,\Vp
  \end{aligned}
\end{equation}
where we have shifted from differentiation with $\X$
to differentiation with $\Xp$ and have applied
the smoothness assumptions of $\varphi$ and
the assumption of compact support $\text{supp}(\varphi,\X)$.
Substituting Eq.~(\ref{eq:condition30}) in~(\ref{eq:condition3}),
\begin{equation}\label{eq:condition31}
  e_{ij\ell}\widehat{\phi}_{j}
  \int_{\Dc}\!\!
  T_{\ell i}(\X)
  \Big(
  \int_{\RRR}
    \varphi(\X-\Xp)\,\Vp
  \Big)\,dv
  =0,
  \quad
  \forall\widehat{\boldsymbol{\phi}}\in\RRR,\;
  \forall\mathcal{D}\in\mathscr{D}
\end{equation}
If Eq.~(\ref{eq:condition31})
is satisfied for all vectors $\widehat{\boldsymbol{\phi}}$
and for all domains $\mathcal{D}$,
then
$T_{\ell i}(\X)$ must equal $T_{i\ell}(\X)$ at all $\X\in\Bc$,
so the local Cauchy stress $\mathbf{T}$ is symmetric
whenever $\varphi(\XX )$ is of the form $\varphi(\X-\X')$.
Hereafter,
we assume the paired restrictions
of $\varphi(\XX)$ having form
$\varphi(\X-\Xp)$
and of $\mathbf{T}$ being symmetric.
If, however, the former is not imposed,
one must revert to the more general balance Eq.~(\ref{eq:condition3}),
which admits non-symmetric stress.
\item
\emph{Global momentum balance:}
The mechanical balance equations for nonlocal media
that follow from Eqs.~(\ref{eq:virtPAD})--(\ref{eq:nonlocal4})
are developed by applying the virtual power
principle~(\ref{eq:master}\textsubscript{2}).
Global balance for a region $\Dc$ is derived by
restricting the virtual velocities to the sub-space
$\widehat{V}_{\text{r}}\subseteq\widehat{V}$
in Eq.~(\ref{eq:master}\textsubscript{1}),
noting that internal power
$\widehat{P}_{F\text{(int)}}=0$ for rigid-body velocities.
With arbitrary translation $\widehat{\mathbf{a}}$ but
zero rotation, $\widehat{\boldsymbol{\phi}}=0$,
\begin{equation}\label{eq:globaltrans}
    \int_{\Dc}
     \big(
     \rho b_{i}(\X)
     - \rho\dot{v}_{i}(\X)
     \big)
     \left(
     \int_{\RRR}
     \!\varphi(\XX )\,\Vp
     \right)
     dv
    +\int_{\Dp}
     \!t_{i}(\X)
     \left(
     \int_{\RRR}
     \!\varphi(\XX )\,\Vp
     \right)
     \,da
     =0
     ,\;
     \forall\mathcal{D}\in\mathscr{D},\;i\in\{1,2,3\}
\end{equation}
where the common factor $\widehat{a}_{i}$ is eliminated.
If $\varphi$ is restricted by Eq.~(\ref{eq:unvary})~---
that is, if the integral $\int_{\R_3}\varphi\,\Vp$
is restricted as spatially unvarying, with a
common value at all points $\X$ in $\Bo$~---
then the integral factors out, with the result
\begin{equation}\label{eq:trans}
  \begin{aligned}
    &\int_{\Dc}
    \rho b_{i}(\X)\,dv
    +
    \int_{\Dp}
    t_{i}(\X)\,da
    =
    \int_{\Dc}
    \rho\dot{v}_{i}(\X)\,dv,\quad
  \forall\mathcal{D}\in\mathscr{D}
  \end{aligned}
\end{equation}
yielding the condition of translational balance for
nonlocal media~---
a condition that is identical to that for local, classical media.
The more general Eq.~(\ref{eq:globaltrans}), however,
applies when the restriction
of Eq.~(\ref{eq:unvary}) is not met.
\item
\emph{Global angular momentum balance and further restrictions on
the mollifier:}
The condition of rotational balance,
which might impose further restrictions on $\varphi$,
is found by applying principle~(\ref{eq:master}\textsubscript{2})
and Eq.~(\ref{eq:Ur}\textsubscript{1})
with arbitrary $\widehat{\boldsymbol{\phi}}$ but
$\widehat{\mathbf{a}}=\mathbf{0}$:
\begin{equation}\label{eq:global1}
   \begin{aligned}
    \int_{\Dc}
      &
      \rho b_{j}(\X)
      \left(
      \int_{\RRR}
      e_{ijk}
      x^{\prime}_{k}\varphi(\XX )\,\Vp
      \right)
      dv
      +
     \int_{\Dp}
     t_{j}(\X)
     \left(
     \int_{\RRR}
     e_{ijk} 
     x^{\prime}_{k}\varphi(\XX )\,\Vp
     \right)
     da\\
    &=
    \int_{\Dc}
    \rho\dot{v}_{j}(\X)
     \left(
     \int_{\RRR}
     e_{ijk}
     x^{\prime}_{k}\varphi(\XX )\,\Vp
     \right)
     dv
     ,\;\;
     \forall\mathcal{D}\in\mathscr{D}
     ,\;
     i\in\{1,2,3\}
  \end{aligned}
\end{equation}
where summations applies over subscripts $j$ and $k$, and
where the common arbitrary factor $\widehat{\phi}_{i}$
has been eliminated.
Note that the ``moment arm'' differs between
a general nonlocal medium and a local medium,
since, in general,
$\int_{\RRR}\Xp\varphi(\XX)\Vp\neq\X$,
nor is $\int_{\RRR}\Xp\varphi(\XX)\Vp$
equal to a common factor times $\X$.
Consequently,
the rotational balance of a general nonlocal continuum
differs from that of local continua.
\par
The condition (\ref{eq:global1}) can be met by placing
restrictions on either $\varphi$,
on the various force quantities, or on both.
An unrestricted $\varphi$ leads to the general
rotational balance Eq.~(\ref{eq:global1})
(i.e., a restriction on forces), which
differs from that of a local medium.
A particular restriction on $\varphi$, however, is sufficient to
make rotational balance similar to that of local media.
Begin by noting the identity
\begin{equation}\label{eq:identity2}
  \int_{\RRR}x'_{k}\varphi(\XX)\,\Vp
  =
  \int_{\RRR}(x'_{k}-x_{k})\varphi(\XX)\,\Vp
  +
  x_{k}\int_{\RRR}\varphi(\XX)\,\Vp
\end{equation}
If we assume, as before, that
$\varphi(\XX )=\varphi(\X-\Xp)$
and if we further assume that $\varphi$ is \emph{symmetric},
such that
$\varphi(\X-\Xp)=\varphi(\Xp-\X)$,
then the integrand of the first term on the right of
Eq.~(\ref{eq:identity2}) is an odd function whose integral
vanishes.
Substituting the remaining term into Eq.~(\ref{eq:global1})
and noting that the integral
$\int_{\RRR}\varphi(\XX)\,\Vp$
is a common term in each of the three parts
of Eq.~(\ref{eq:global1}),
a term that is unvarying throughout $\Bc$
(\emph{cf}. Eq.~\ref{eq:unvary}),
it cancels as a common factor,
with the result
\begin{equation}\label{eq:momentbalance}
    \int_{\Dc}
      e_{ijk}
      x_{k}
      \rho b_{j}(\X)\,
      dv\:
      +
     \int_{\Dp}
     e_{ijk}
     x_{k}
     t_{j}(\X)\,
     da\: =
    \int_{\Dc}
    e_{ijk}
    x_{k}
    \rho\dot{v}_{j}(\X)\,
     dv
     ,\;\;
     \forall\mathcal{D}\in\mathscr{D}
\end{equation}
Therefore,
if the mollifier $\varphi$ is restricted to
a symmetric function,
with $\varphi(\XX)=\varphi(\X-\Xp)=\varphi(\Xp-\X)$,
then global rotational balance for
nonlocal media takes the same form as for local media.
Symmetric forms satisfying these conditions include
\begin{equation}
  \varphi(\aXX)
  \quad\text{and}\quad
  \varphi
  \big(
    G_{ij}(x_{i}-x'_{i})
    (x_{j}-x'_{j})
  \big)
\end{equation}
where the $|\cdot|$ is a general isotropic norm: for example,
$|x_{1}-x_{1}^{\prime}|$,
$\langle \X-\X',\X-\X' \rangle^{1/2}$, etc.
The second form applies an anisotropic norm,
with $G_{ij}$ being a symmetric positive-definite matrix,
appropriate for a material having an anisotropic internal fabric.
Either norm serves as a metric of distances between $\X$
and points in its neighborhood,
assuring the mutual influence of distant points
$\X$ and $\Xp$,
with $\varphi(\XX)=\varphi(\Xp,\X)$.
\item
\emph{Local mechanical balance:}
The rigidifying principle~(\ref{eq:master}\textsubscript{2})
was applied in the preceding items
to obtain global balance conditions and restrictions on $\varphi$.
The \emph{local} mechanical balance follows from
principle~(\ref{eq:master}\textsubscript{1}).
Substituting Eq.~(\ref{eq:nonlocal4}) with $\widehat{\theta}(\Xp)=0$,
%
\begin{equation}\label{eq:equil5}
  \begin{aligned}
    \int_{\Dc}\int_{\RRR}
      \big[
        \rho b_{i}(\X)\varphi(\XX) &- T_{ji}(\X) 
          \varphi_{,j}(\XX)
      \big]\,
      \widehat{v}_{i}(\Xp)\,\VV \\
    +\int_{\Dp}\int_{\RRR}&
       t_{i}(\X)\varphi(\XX)
     \widehat{v}_{i}(\Xp)\,\Vp\,da\\
     &-
     \int_{\Dc}\int_{\RRR}
     \rho\dot{v}_{i}(\X)
     \varphi(\XX)
     \widehat{v}_{i}(\Xp)\,\VV =0,
     \quad\forall\widehat{\mathbf{v}}\in\widehat{V}
     ,\;
     \forall\mathcal{D}\in\mathscr{D}
  \end{aligned}
\end{equation}
%
with summation applying to $i$ and $j$.
In the equation,
$\mathbf{b}$, $\mathbf{t}$, $\dot{\mathbf{v}}$,
and $\mathbf{T}$ are assumed
square-integrable on $\Bc$ and on surfaces $\Dp$;
mollifier $\varphi$ is assumed bounded and continuous; and
derivatives $\varphi_{,j}$ are assumed
almost everywhere bounded and continuous.
Exchanging the order of integration, Eq.~(\ref{eq:equil5})
yields a restriction of the form
\begin{linenomath*}
\begin{align}\label{eq:F}
    \int_{\RRR}
      \mathcal{F}_{i}(\Xp)
      \widehat{v}_{i}(\Xp)\,\Vp =0,
      \quad\forall\widehat{\mathbf{v}}\in\Vhat
\end{align}
\end{linenomath*}
in which summation applies to index $i$.
The two functions,
$\widehat{\mathbf{v}}$ and $\boldsymbol{\mathcal{F}}$,
are in the Hilbert space $L^{2}(\RRR)$,
to which the positive definite property of inner product spaces,
when applied to Eq.~(\ref{eq:F}), requires
\begin{linenomath*}
\begin{align}\label{eq:equil1}
  &\mathcal{F}_{i}(\Xp)
  = 0, \quad \forall\Xp\in\RRR
  ,\;
  \forall\mathcal{D}\in\mathscr{D}
  ,\;
  i\in\{1,2,3\}
  \\ \label{eq:FFF}
  &\begin{aligned}
      \mathcal{F}_{i}(\Xp) \equiv
      &
      \int_{\Dc}
        \big[
          \big(\rho b_{i}(\X) -
               \rho \dot{v}_{i}(\X)
          \big)
          \varphi(\XX)
          - T_{ji}(\X) 
          \varphi_{,j}(\XX)
        \big]
      \,dv \\
      & +
      \int_{\Dp}
        t_{i}(\X)\varphi(\XX)
      \,da
  \end{aligned}
\end{align}
\end{linenomath*}
%
%
with summation applying to $j$.
Together, Eqs.~(\ref{eq:equil1})--(\ref{eq:FFF})
express the condition of local balance at $\Xp\in\RRR$
for a nonlocal medium.
\par
This local balance condition,
which differs from the
Eq.~(\ref{eq:equillocal})
of local continua, has the following implications:
\begin{enumerate}[label=\arabic{enumi}.\alph*.,leftmargin=3em]
\item
Unlike local continua,
where the two conditions~(\ref{eq:equillocal}\textsubscript{1})
and~(\ref{eq:equillocal}\textsubscript{2})
apply separately to the interior and boundary,
a single condition~--- that of Eq.~(\ref{eq:equil1})~---
governs both for nonlocal continua.
\item
Eq.~(\ref{eq:equil1})
must be satisfied at all points $\Xp$,
which in general extend beyond the boundary of a sub-body $\Dc$
and even beyond $\Bc$.
Although, in principle, the equation applies to the infinite region
$\Xp\in\RRR$,
points $\Xp$ beyond the support of $\varphi$ make no
contribution to the equation's integrals,
so that the equation applies to the Minkowski sum
of $\Dc$ and $\text{supp}(\varphi,\X)$, i.e.
$\Xp\in\{\Dc+\text{supp}(\varphi,\X): \X\in\Dc\}$.
\item
Because $\mathbf{b}(\X)$
includes body forces originating at $\X$ and in its neighborhood,
the balance condition encompasses external
gravitational and surface forces at $\X$
as well as internal attractions and repulsions
within $\Bc$, as in peridynamic models
\cite{Silhavy:1983a,Javili:2019a}
(see discussion following Eq.~\ref{eq:local}).
\item
As shown below,
the term $T_{ji}\varphi_{,j}$ in Eq.~(\ref{eq:FFF})
produces an averaging of the derivatives $T_{ji,j}$,
even when these derivatives are discontinuous or ill-defined;
moreover, for points on boundary $\Dp$,
the term $T_{ji}\varphi_{,j}$ yields
an averaging of the traction $T_{ji}n_{j}$,
even when the boundary is non-smooth and its
normal $\mathbf{n}$ is ill-defined.
\item
Equation~(\ref{eq:equil1}) applies to all subdomains
$\Dc\in\mathscr{D}$, and as such,
function $\boldsymbol{\mathcal{F}}$ is, more precisely,
a function of both position $\Xp\in\RRR$ and
domain $\Dc$,
with Eq.~(\ref{eq:equil1}) becoming
\begin{equation}\label{eq:FDx}
  \boldsymbol{\mathcal{F}}:
  \mathscr{D}\times\RRR
  \,\rightarrow\,
  \RRR,
  \quad
  \mathcal{F}_{i}(\mathcal{D},\Xp) = 0,\;\;
  \forall\mathcal{D}\in\mathscr{D},\;
  \forall\Xp\in\RRR,\;
  i\in\{1,2,3\}
\end{equation}
where $\mathcal{F}_{i}(\mathcal{D},\Xp)$ is given by
the right side of Eq.~(\ref{eq:FFF}).
Fosdick and Virga \cite{Fosdick:1989a} analyzed the
mechanical balance of a local medium
and showed that balance for arbitrary subdomains $\Dc$
leads to the familiar Cauchy equation
of Eq.~(\ref{eq:equillocal}\textsubscript{2}),
$\mathbf{t}(\X)=\mathbf{T}(\X)\cdot\mathbf{n}(\X)$,
without appealing to infinitesimal tetrahedra.
Because the two quantities, $\mathbf{T}$ and $\mathbf{t}$,
exist within the interior $\mathcal{B}$
and both are mutually bound by the Cauchy equation,
Fosdick \cite{Fosdick:2011b} argued that one or the
other, $\mathbf{T}$ or $\mathbf{t}$,
should hold primacy as the fundamental measure of internal force,
with the other derived from the Cauchy equation
(Fosdick opted for traction $\mathbf{t}$ as primitive).
\par
\quad
For nonlocal continua,
we depart from these findings in two respects.
First, stress $\mathbf{T}(\X)$
is viewed as the fundamental internal force,
while tractions $\mathbf{t}$ at points \emph{inside} of $\Bc$
are viewed as derived quantities,
resulting as solutions of Eqs.~(\ref{eq:equil1})
and~(\ref{eq:FDx})
(however, we consider tractions applied
on the body's tangible boundary $\Bp$ as being fundamental).
This primacy of stress reflects its role as a thermomechanic
state variable or, alternatively, its
being derived from a free energy that is
present at all points in $\Bc$ (Section~\ref{sec:firstlaw}).
The second departure,
as noted above, is that the
traction $\mathbf{t}$ in the balance
Eqs.~(\ref{eq:equil1})--(\ref{eq:FFF}) applies even at boundary points
where a well-defined normal $\mathbf{n}$ does not exist.
\par
\quad
We must also consider the possibility of $\mathbf{t}(\X)$
depending on the \emph{shape} of boundary $\Dp$
in the neighborhood of $\X$, a matter addressed
in 6.g below.
\item
Newton's reaction principle (third law) \cite{Noll:1959a}
prevails within the interior of a nonlocal continuum $\Bc$.
Following the standard approach for local continua,
consider two disjoint but adjoining regions,
$\Dc_{\text{a}}$ and $\Dc_{\text{b}}$,
whose boundaries partially overlap:
$\Dc_{\text{a}}\cap\Dc_{\text{b}}=\emptyset$ and
$\Dp_{\text{a}}\cap\Dp_{\text{b}}\neq\emptyset$.
The latter condition means that the overlap
has Hausdorff dimension at least 2
(i.e., it excludes line or point overlaps).
The surface tractions
$\mathbf{t}^{\text{a}}$ and
$\mathbf{t}^{\text{b}}$ acting on the opposing regions along
$\Dp_{\text{a}}\cap\Dp_{\text{b}}$,
as derived from Eqs.~(\ref{eq:equil1})--(\ref{eq:FFF}), satisfy
\begin{equation}\label{eq:overlap1}
\begin{aligned}
  \int_{\Dp_{\text{a}}\cap\Dp_{\text{b}}}
    \! t^{\text{a}}_{i}\varphi\,da
  &= -\int_{\Dc_{\text{a}}}\! f_{i}\,dv
    -\int_{\Dp_{\text{a}}\backslash(\Dp_{\text{a}}\cap\Dp_{\text{b}})}
      \! t_{i}\varphi\,da,
    \;\;\forall\Xp\in\RRR,\;
  i\in\{1,2,3\}
  \\
  \int_{\Dp_{\text{a}}\cap\Dp_{\text{b}}}\! t^{\text{b}}_{i}\varphi\,da
  &= -\int_{\Dc_{\text{b}}}\! f_{i}\,dv
    -\int_{\Dp_{\text{b}}\backslash(\Dp_{\text{a}}\cap\Dp_{\text{b}})}\! t_{i}\varphi\,da,
    \;\;\forall\Xp\in\RRR,\;
  i\in\{1,2,3\}
\end{aligned}
\end{equation}
where $f_{i}(\X)$ is the integrand
of the first integral of Eq.~(\ref{eq:FFF}),
and $t_{i}\varphi$
is the integrand of the second.
Likewise, applying Eqs.~(\ref{eq:equil1})--(\ref{eq:FFF})
to the combined region $\Dc_{\text{a}}\cup\Dc_{\text{b}}$ and its
boundary $(\Dp_{\text{a}}\cup\Dp_{\text{b}})\backslash(\Dp_{\text{a}}\cap\Dp_{\text{b}})$,
gives
\begin{equation}\label{eq:overlap2}
  \int_{\Dc_{\text{a}}}f_{i}\,dv + \int_{\Dc_{\text{b}}}f_{i}\,dv
  + \int_{\Dp_{\text{a}}\backslash(\Dp_{\text{a}}\cap\Dp_{\text{b}})}t_{i}\varphi\,da
  + \int_{\Dp_{\text{b}}\backslash(\Dp_{\text{a}}\cap\Dp_{\text{b}})}t_{i}\varphi\,da = 0
\end{equation}
for all $\Xp\in\RRR$.
Adding the Eqs.~(\ref{eq:overlap1})--(\ref{eq:overlap2}) yields
\begin{equation}\label{eq:tandt}
  \int_{\Dp_{\text{a}}\cap\Dp_{\text{b}}}t_{i}^{\text{a}}\,da +
  \int_{\Dp_{\text{a}}\cap\Dp_{\text{b}}}t_{i}^{\text{b}}\,da = 0
\end{equation}
confirming that tractions on opposing boundaries
are equal and opposite.
Note that the assumption
$\Dp_{\text{a}}\cap\Dp_{\text{b}}\neq\emptyset$
excludes integrals in Eq.~(\ref{eq:tandt}) that are zero
due to zero measure.
\par\quad
A corollary results is that
the traction on a surface $\Dp_{\text{a}}\cap\Dp_{\text{b}}$ is unique:
the same traction applies
for any other region $\Dc_{\text{c}}$
whose boundary shares the same overlap with
$\Dp_{\text{a}}$,
such that
$\Dp_{\text{a}}\cap\Dp_{\text{b}}=\Dp_{\text{a}}\cap\Dp_{\text{c}}$.
\item
As a consequence of the reaction principle,
traction $\mathbf{t}(\X)$ at $\X$ on boundary $\Dp$
is independent of the boundary's
shape in the neighborhood of $\X$.
Because the first integral in Eq.~(\ref{eq:tandt})
(i.e., of $\mathbf{t}^{\text{a}}$)
equals the negative of the second integral
along the shared boundary $\Dp_{\text{a}}\cap\Dp_{\text{b}}$,
the first integral is unchanged by the shape of
$\Dc_{\text{a}}$ and boundary $\Dp_{\text{a}}$,
provided that the overlap with $\Dc_{\text{b}}$ is unchanged.
\item
Using Eq.~(\ref{eq:equil1}) to determine balance among the
four functions~---
$\mathbf{T}$, $\mathbf{b}$, $\rho\dot{\mathbf{v}}$, and $\mathbf{t}$~---
might appear to require toilsome verification
across all regions $\Dc\in\mathscr{D}$.
However, balance reduces to verifying that Eq.~(\ref{eq:equil1})
is satisfied for the single region $\Bc$ and for the three functions
$\mathbf{T}$, $\mathbf{b}$, and $\rho\dot{\mathbf{v}}$ in
$\Bc$ and
the traction $\mathbf{t}$ on $\Bp$.
\par\quad
Suppose that Eq.~(\ref{eq:equil1}) is satisfied for $\Bc$.
The terms in the equation
can be derived for any other region $\Dc\in\mathscr{D}$
by beginning with region $\Bc$ and reducing it to $\Dc$,
through a sequence
$\Dc_{0}, \Dc_{1}, \Dc_{2},\ldots,\Dc_{\text{final}}$,
in infinitesimal increments $\Delta\Dc_{i}$, with
$\Dc_{0}=\Bc$, $\Dc_{i+1}=\Dc_{i}\backslash\Delta\Dc_{i}$,
and $\Dc_{\text{final}}=\Dc$.
Each increment alters the first integral
in Eq.~(\ref{eq:FFF}), which is a function of $\Xp\in\RRR$;
since the traction function $\mathbf{t}(\X)$
on the new boundary $\Dp_{i+1}$
is a derived quantity (see item 6.e), the traction
of the previous boundary (i.e., the second integral in Eq.~\ref{eq:FFF})
is altered by changes in the first integral, maintaining
$\boldsymbol{\mathcal{F}}(\Dc_{i+1},\Xp)=\mathbf{0}$.
Thus, treating the traction $\mathbf{t}$ within the interior of $\Bc$
as a derived quantity yields
\begin{equation}\label{eq:mechBalance}
  \boldsymbol{\mathcal{F}}(\Bc,\Xp) = \mathbf{0},\;
  \forall\Xp\in\RRR
  \iff
  \boldsymbol{\mathcal{F}}(\Dc,\Xp) = \mathbf{0},\;
  \forall\Dc\in\mathscr{D},\;\Xp\in\RRR
\end{equation}
so that the $\Dc\in\mathscr{D}$ and $\Dp$ in
criterion~(\ref{eq:FFF}) may be replaced by the single
$\Bc$ and $\Bp$.
\item
In numerical implementations, Eq.~(\ref{eq:FFF}) can be
discretized, with the
statics matrix derived from the integral $\int\varphi_{,j}(\X,\Xp)$
(e.g., Section~\ref{sec:examples} and \ref{app:example}).
\end{enumerate}
\par
We now return to the specialized (and more familiar)
case of local continua.
Although the balance Eqs.~(\ref{eq:equil1})--(\ref{eq:FFF})
of nonlocal continua
differ fundamentally from those of local continua,
the local balance equations
ensue when certain smoothness conditions apply to a
nonlocal continuum.
If one is assured that the dual fields
$\mathbf{b}$ and $\dot{\mathbf{v}}$
are continuous in $\Bc$, that $\mathbf{T}$
is continuous and continuously differentiable
in $\Bc$
(i.e., the pre-Hilbert spaces,
$\mathbf{b},\dot{\mathbf{v}}\in C^{0}(\Bc)$,
$\mathbf{T}\in C^{1}(\Bc)$),
and that boundary $\Bp$ is piecewise smooth,
then the nonlocal balance Eqs.~(\ref{eq:equil1})--(\ref{eq:FFF})
simplify to the same form as those governing a local continuum.
This result follows by observing that
when $\mathbf{T}$ is continuously differentiable,
the product rule applies to
the first term on the right of
Eq.~(\ref{eq:nonlocal4}\textsubscript{2}),
\begin{equation}
  \llangle\!
    -\mathbf{T},
       (\boldsymbol{\nabla}\varphi)\otimes\widehat{\mathbf{v}}
  \rrangle_{\Bc} =
  \llangle
    -T_{ji},\,\varphi_{,j}\widehat{v}_{i}
  \rrangle_{\Bc} =
  \llangle
    -(T_{ji}\varphi)_{,j},\,\widehat{v}_{i}
  \rrangle_{\Bc} +
  \llangle
    T_{ji,j},\varphi\,\widehat{v}_{i}
  \rrangle_{\Bc}
\end{equation}
where Eq.~(\ref{eq:nonlocal4}\textsubscript{2}) is applied
to the full region $\Bc$.
The assumed smoothness of $\mathbf{T}$ permits
the Ostrogradsky--Gauss theorem to be applied to
postulate~(\ref{eq:master}\textsubscript{1}),
\begin{equation}\label{eq:OG}
  \llangle
    (\rho b_{i}+T_{ji,j}),\varphi\,\widehat{v}_{i}
  \rrangle_{\mathcal{B}} +
  \llangle
    (t_{i}-T_{ji}n_{j}),\varphi\,\widehat{v}_{i}
  \rrangle_{\Bp}
  -\llangle
    \rho\dot{v}_{i},\varphi\,\widehat{v}_{i}
  \rrangle_{\mathcal{B}}
  = 0,
  \quad
  \forall\widehat{\mathbf{v}}\in\widehat{V}
\end{equation}
After exchanging the order of integration and invoking the
compact support of $\varphi$,
the fundamental lemma of the calculus of variations
implies the following restrictions on the
force quantities $\mathbf{b}$, $\mathbf{t}$, and $\mathbf{T}$,
\par 
\begin{equation}\label{eq:equil3}
  \begin{aligned}
    &\int_{\mathcal{B}}
      \Big[
        \rho b_{i}(\X)+ T_{ji,j}(\X)
        -\rho\dot{v}_{i}(\X)
        \Big]\varphi(\XX)\,dv
        = 0
        \quad \forall\Xp\in\RRR,\;
  i\in\{1,2,3\}
        \\
    &\int_{\partial\mathcal{B}}
      \Big[
        t_{i}(\X)- T_{ji}(\X)n_{j}(\X)\Big]\varphi(\XX)\,dv
        =0,
        \quad \forall\Xp\in
        \RRR,\;
  i\in\{1,2,3\}
  \end{aligned}
\end{equation}
noting that the standard form of the lemma requires continuity of
$\rho\mathbf{b}$, $\mathbf{t}$, $\rho\dot{\mathbf{v}}$,
$\mathbf{T}$, and $\varphi\widehat{\mathbf{v}}$
\cite{Gelfand:1963a}.
\par
In deriving Eq.~(\ref{eq:OG}),
we have assumed that boundary $\Bp$ is sufficiently smooth
to admit an unambiguous outward normal $\mathbf{n}$
at all boundary points.
A more demanding requirement is that Eq.~(\ref{eq:equil3})
applies at all points
$\Xp\in\RRR$, so that averaging
with $\varphi$ across $\Bc$ gives the same result
(i.e., zero),
regardless of the
point $\Xp$ at which the averaging is centered.
Rather than placing immoderate restrictions on $\varphi$,
the following simpler conditions assure a result of
zero for each $\X'$: 
\begin{equation}\label{eq:internalbalance}
  \begin{aligned}
    &\rho b_{i}(\X) + T_{ji,j}(\X)
     =\rho\dot{v}_{i}(\X),
     \quad \forall\X\in\Bc
    \\
    &t_{i}(\X)
    = T_{ji}(\X)n_{j}(\X),
    \quad \forall\X\in\Bp
  \end{aligned}
\end{equation}
These balance conditions~---
identical in form to those of a local medium~---
hold for a nonlocal medium,
but one with a presumed continuity of the force quantities
$\mathbf{b}$, $\mathbf{t}$, and $\mathbf{T}$ and
a regular boundary $\Bp$.
\item
\emph{Global thermal balance:}
The principle~(\ref{eq:master}\textsubscript{1}) is applied
to the nonlocal medium of
Eqs.~(\ref{eq:virtPAD})--(\ref{eq:nonlocal4}) by
restricting the virtual motions to the rigid-motion sub-space
$\widehat{\Theta}_{\text{r}}\subseteq\widehat{\Theta}$
with an arbitrary scalar 
$\widehat{\beta}$ (in Eq.~\ref{eq:Ur}\textsubscript{2})
but with $\widehat{\mathbf{v}}=\mathbf{0}$.
Because the internal power
$\widehat{P}_{F\text{(int)}}$ is zero for rigid motions,
\begin{equation}
  \begin{aligned}
    &\int_{\Dc}\big(\rho s(\X) + \rho\xi(\X) \big)
    \left(
    \int_{\RRR}
    \!\varphi(\XX )\,\Vp
    \right)
    dv
    -
    \int_{\Dp}
    \rho k(\X)
    \left(
    \int_{\RRR}
    \!\varphi(\XX )\,\Vp
    \right)
    da
    \\
    &\quad
    =
    \int_{\Dc}\rho \dot{\eta}(\X)
    \left(
    \int_{\RRR}
    \!\varphi(\XX )\,\Vp
    \right)
    dv,\quad
    \forall\Dc\in\mathscr{D}
  \end{aligned}
\end{equation}
As with Eq.~(\ref{eq:globaltrans}),
the $\int_{\R_3}\varphi\,\Vp$ integrations
are spatially unvarying scalars,
so can be factored, such that
\begin{equation}\label{eq:globaltherm}
    \int_{\Dc}
    \big(\rho s(\X) + \rho\xi(\X)\big)\,
    dv\;
    -
    \int_{\Dp}
    \rho k(\X)\,
    da\;
    =
    \int_{\Dc}\rho \dot{\eta}(\X)\,
    dv,\quad
    \forall\Dc\in\mathscr{D}
\end{equation}
This equation of global thermal (entropy) balance
for a nonlocal medium is
identical to that of a local medium.
\item
\emph{Local thermal balance:}
The local balance of thermal fields is derived by
applying condition~(\ref{eq:master}\textsubscript{1})
and Eq.~(\ref{eq:nonlocal4}\textsubscript{1}),
with mechanical velocity $\widehat{\mathbf{v}}=\mathbf{0}$:
\begin{equation}\label{eq:thermal1}
  \begin{aligned}
    &\int_{\Dc}\int_{\RRR}
      \Big\{
        \big[\rho s(\X)+\rho\xi(\X)\big]
        \varphi(\XX) + p_{i}(\X)
          \varphi_{,i}(\XX)
      \Big\}\,
      \widehat{\theta}(\Xp)\,\VV \\
    &\quad\quad
    +\int_{\Dp}\int_{\RRR}
     \big[
       -k(\X)\varphi(\XX)\big]\,
     \widehat{\theta}(\Xp)\,\Vp\,da\\
     &\quad\quad\quad\quad
     -\int_{\Dc}\int_{\RRR}
     \big[
       \rho\dot{\eta}(\X)\,
       \varphi(\XX)\,
       \widehat{\theta}(\Xp)\,dv
     \big]\,\Vp= 0,
     \quad\forall\,\widehat{\theta}\in\widehat{\Theta},\;
    \forall\Dc\in\mathscr{D}
  \end{aligned}
\end{equation}
in which summation applies to index $i$.
As with local mechanical balance,
the order of integration can be exchanged,
and by applying algebraic properties of inner product spaces,
\begin{linenomath*}
\begin{align}\label{eq:thermal4}
  &\mathcal{G}(\Xp)
  = 0, \quad \forall\Xp\in\RRR,\;
    \forall\Dc\in\mathscr{D}
  \\
  \label{eq:thermal5}
  &\begin{aligned}
    \mathcal{G}(\Xp)
    \equiv
    &\int_{\Dc}
      \Big\{
        \big[\rho s(\X) + \rho\xi(\X) - \rho\dot{\eta}(\X)\big]
        \varphi(\XX) + p_{i}(\X)
          \varphi_{,i}(\XX)
      \Big\}\,dv
    \\
    &-
     \int_{\Dp}
     k(\X)\varphi(\XX)
     \,da
  \end{aligned}
\end{align}
\end{linenomath*}
for all domains $\Dc\in\mathscr{D}$ and
for each direction $i\in\{1,2,3\}$. Eq.~(\ref{eq:thermal4})
is the condition of local thermal (entropy) balance at $\Xp$ for
a nonlocal continuum,
and Eq.~(\ref{eq:thermal5}) combines conditions
at both the boundary and interior of $\Dc$.
As with the discussion of Eq.~(\ref{eq:mechBalance}),
if Eq.~(\ref{eq:thermal5}) is met for the single region $\Bc$,
then balance is assured for all sub-regions $\Dc\in\mathscr{D}$.
\par
In a manner complementary to mechanical balance,
if the fields
$s$, $\xi$, and $\dot{\eta}$ are known to be continuous
and $\mathbf{p}$ is continuously differentiable,
the condition of Eqs.~(\ref{eq:thermal4})--(\ref{eq:thermal5})
can be replaced with the more familiar conditions
\begin{equation}\label{eq:inttherm}
  \begin{aligned}
    &\rho s(\X)+\rho\xi(\X)
    - \text{div}\big(\mathbf{p}(\X)\big)
    = \rho\dot{\eta}(\X),
    \quad\forall\X\in\Bc
    \\
    &k(\X)
    = \mathbf{p}(\X)\cdot\mathbf{n}(\X),
    \quad\forall\X\in\Bp
  \end{aligned}
\end{equation}
which are the same as those of
a local medium.
\end{enumerate}
\section{\large Nonlocal thermomechanics}\label{sec:thermomech}
We now transition from virtual to actual velocities
and develop a consistent thermomechanics of nonlocal media.
In the previous section,
restrictions were developed for
the mollifier $\varphi$ and for the various
dual force quantities;
henceforth, we adopt the following restrictions:
\begin{itemize}[itemsep=0.7ex,parsep=0ex,topsep=0.7ex]
\item
The following quantities are square-integrable on $\Bc$:
forces
$\mathbf{b}$, $s$, and $\xi$;
momenta $\rho\dot{\mathbf{v}}$ and $\rho\dot{\eta}$;
stress $\mathbf{T}$ and entropy flow $\mathbf{p}$;
velocities $\mathbf{v}$ and $\theta$; and density $\rho$.
Likewise, surface quantities $\mathbf{t}$ and $k$ are square-integrable
on $\Bp$ and on $\Dp$, for all $\Dc\in\mathscr{D}$.
Note that differentiability of $\mathbf{T}$ and $\mathbf{p}$
are not assumed.
\item
Mollifier $\varphi(\XX)$
is bounded and continuous,
and its first derivatives with respect to $\X$ and $\Xp$
are assumed to be almost everywhere bounded and continuous,
$\varphi\in C^{0}\cap W^{1,1}$.
The support of $\varphi$, $\text{supp}(\varphi,\X)$,
is assumed Lipschitz-smooth and compact in $\RRR$
for all $\X\in\Bc$.
\item
The integral $\int_{\RRR}\varphi(\XX)\,\Vp$
is spatially uniform for all $\X\in\Bc$, as in Eq.~(\ref{eq:unvary}).
\item
Kernel $\varphi(\XX)$ is a
function of the difference $\X-\Xp$, i.e.
$\varphi(\XX )=\varphi(\X-\Xp)$,
with the collateral restriction that stress $\mathbf{T}$ is symmetric
(see Eq.~\ref{eq:condition31}).
\item
Kernel $\varphi(\XX)$ is an even function of $\X-\X'$,
with $\varphi(\X-\Xp)=\varphi(\Xp-\X)$.
Although more general forms are permitted, we assume that
$\varphi(\XX )=\varphi(\aXX)$, where
$|\cdot|$ is a norm on $\RRR$.
By assuming that $\varphi$ is an even function of $\X-\X'$,
the global rotational balance equation
coincides with that of a local medium
(see Eq.~\ref{eq:global1} and its discussion).
\item
Global mechanical balance is established by
Eqs.~(\ref{eq:trans}) and (\ref{eq:momentbalance}),
and global thermal balance by
Eq.~(\ref{eq:globaltherm}).
\item
Local mechanical and thermal balance requires that
Eqs.~(\ref{eq:equil1})
and~(\ref{eq:thermal4}) are satisfied respectively,
with $\mathcal{F}_{i}$ and $\mathcal{G}$ defined
by Eqs.~(\ref{eq:FFF}) and~(\ref{eq:thermal5}).
\end{itemize}
Note that the assumption of
$\int_{\RRR}\varphi(\XX)\,\Vp$
being spatially uniform for $\X\in\Bc$ does not imply that
the same mollifier $\varphi(\aXX)$ applies throughout $\Bc$,
but only that
the integral $\int_{\RRR}\varphi(\aXX)\,\Vp$ is uniform on $\Bc$.
The function $\varphi(\aXX)$ should therefore be
properly written as $\varphi(\X,\aXX)$,
with the understanding that $\int_{\RRR}\varphi(\XX)\,\Vp$
is constant.
(Note that Eq.~\ref{eq:condition3} must also apply, and because the
identity of Eq.~\ref{eq:condition30} does not hold for
more general $\varphi$, the stress $\mathbf{T}$ may be non-symmetric
in transitional regions where
$\varphi(\aXX)$ changes with position $\X$.)
With this understanding, we continue with
the notation $\varphi(\aXX)$.
\par
The requirement of finite support $\text{supp}(\varphi,\X)$
is consistent with
Eringen's \emph{axiom of neighborhood}
(or the smooth neighborhood hypothesis)
\cite{Eringen:2002a,Fosdick:1998a}.
Although a further restriction is natural,
we do not require that $\varphi$ is a
decreasing function of distance $\aXX$
(i.e., Eringen's \emph{attenuating neighborhood hypothesis}
is not assumed).
\par
To this point, $\varphi_{,i}(\XX)$ has denoted
differentiation with respect to the first argument,
$\partial\varphi/\partial x_{i}$.
Differentiation with respect to the second argument
is at times more natural,
so the following notation is adopted:
\begin{equation}
  \lpoon{\boldsymbol{\nabla}}\varphi\;\;\text{or}\;\;
  \lpoon{\varphi}_{,i}
  =
  \varphi_{,i}(\XX)
  =
  \partial\varphi(\XX)/\partial x_{i}
  \quad\text{and}\quad
  \rpoon{\boldsymbol{\nabla}}\varphi\;\;\text{or}\;\;
  \rpoon{\varphi}_{,i}
  =
  \partial\varphi(\XX)/\partial x'_{i}
\end{equation}
When a function $g(\Xp)$ is continuously differentiable,
the following identity holds:
\begin{equation}\label{eq:identity}
  \begin{aligned}
    \int_{\RRR}
      \lpoon{\varphi}_{,i}(\aXX) g(\Xp)\,\Vp
    &=-\int_{\RRR}
      \rpoon{\varphi}_{,i}(\aXX) g(\Xp)\,\Vp
    \\
    \mkern-18mu
    &=-\int_{\RRR}
        \frac{\partial}{\partial x'_{i}}
        \Big(
          \varphi(\aXX)g(x')
        \Big)\,\Vp
      +\int_{\RRR}
       \varphi(\aXX)
       \frac{\partial}{\partial x'_{i}}
       g(\Xp)\,\Vp
       \\
    \mkern-18mu 
    &=\int_{\RRR}
       \varphi(\aXX)
       g_{,i}(\Xp)\,\Vp
  \end{aligned}
\end{equation}
where $g_{,i}=\partial g/\partial x'_{i}$ on the last line.
(Because
$\text{supp}(\varphi,\X)$ is assumed continuous and compact
with respect to $\Xp$ in $\RRR$
for all $\X\in\Bc$, the first integral on
the second line vanishes.)
Equation~(\ref{eq:identity}) states that the
inner product of $g$ and the
\emph{derivative-average} of $f$ is equal
to the inner product of $f$ and the
\emph{averaged derivative} of $g$.
Note, however, that this application of the divergence
theorem applies only when $g(\Xp)$ is continuously differentiable;
when it is not, one must
revert to
$\int_{\RRR}\lpoon{\varphi}_{,i}(\aXX) g(\Xp)\,\Vp$
in place of
$\int_{\RRR}\varphi(\aXX)g_{,i}(\Xp)\,\Vp$.
\par
In addition to the itemized restrictions, we introduce three
further restrictions:
(a)~a normalization of $\varphi$ (Eq.~\ref{eq:unity});
(b)~positivity of $\varphi$ within its support,
enabling localization of the second law
(Eq.~\ref{eq:xilocal}); and
(c)~non-negative temperature $\theta$,
also associated with the second law.
Moreover,
when transitioning from virtual to actual velocities,
the regularity requirements for $\mathbf{v}$ and $\theta$ will
at times be tightened, requiring smoother fields,
as noted when applicable.
\par
The normalizing restriction on $\varphi$ is
\begin{equation}\label{eq:unity}
  \int_{\RRR}
  \varphi(\XX )\,\Vp = 1
   ,\quad
   \forall\,\X\in\Bc
\end{equation}
although values other than 1 are admissible,
provided that the force quantities are scaled accordingly.
\par
Having adopted the above restrictions,
we henceforth write $\varphi(\aXX)$ in place
of $\varphi(\XX)$, where $|\cdot|$ is a general
(possibly anisotropic) norm.
\subsection{Energy balance}\label{sec:energybalance}
By transitioning to actual velocities $\mathbf{v}$,
the kinetic energy $K$ of an open connected
region $\Dc$
and its time-rate $\dot{K}$ are
\par
\begin{alignat}{2}
  K(\Dc) =\; &
  \frac{1}{2}
  \langle\rho\mathbf{v},\delta\mathbf{v}\rangle_{\Dc}
                 \;&=\; & \frac{1}{2}
                   \int_{\Dc} \rho(\X)\mathbf{v}
                   (\X)
                   \cdot\mathbf{v}(\X)\,dv\\
  \dot{K}(\Dc) =\;  & 
  \langle\rho\dot{\mathbf{v}},\delta\mathbf{v}\rangle_{\Dc}
                  \;&=\; &
                   \int_{\Dc} \rho(\X)\dot{\mathbf{v}}(\X)
                                        \cdot\mathbf{v}(\X)\,dv
\end{alignat}
where $\delta$ is the Dirac operator defined following
Eq.~(\ref{eq:local1}).
\par
Analogously to their virtual counterparts in Eq.~(\ref{eq:virtPAD}),
the actual inertial powers,
$P_{A,\text{mech}}(\Dc)$ and $P_{A,\text{therm}}(\Dc)$,
are defined as
\begin{equation}\label{eq:PA}
  P_{A,\text{mech}}(\Dc) =
    \llangle\rho\dot{\mathbf{v}},\varphi\mathbf{v}\rrangle_{\Dc},
  \quad
  P_{A,\text{therm}}(\Dc) =
    \llangle\rho\dot{\eta},\varphi\theta\rrangle_{\Dc}
\end{equation}
using the notation of Eq.~(\ref{eq:abbrev})
but with the virtual hat ($\widehat{\rule{2ex}{0ex}}$)
of Eq.~(\ref{eq:Pamhat}\textsubscript{1})
removed.
\par
Mollifier $\varphi$ can be a manifestation of distant internal
interactions among material points or, alternatively, of an
inability to discern local micro-scale velocities,
accelerations, or temperatures, other than in a general,
averaged sense.
In the absence of long-range interactions
and when acuity is faultless,
function $\varphi$ is reduced to the Dirac $\delta$
of local media,
as in Section~\ref{sec:localstatics},
and the actual rate $\dot{K}$ equals
the inertial power $P_{A,\text{mech}}$.
However,
for a nonlocal medium,
$\dot{K}$ is not necessarily equal to $P_{A,\text{mech}}$,
and we define a residual
$\dot{R}_{K}$ as the difference
\begin{equation}\label{eq:residualR}
  \dot{R}_{K}(\Dc) =
  \dot{K}(\Dc) - P_{A,\text{mech}}(\Dc)
\end{equation}
where $\dot{R}_{K}$ represents
a ``hidden'' rate of kinetic energy that is not discerned
with the meso-scale measure $P_{A,\text{mech}}(\Dc)$,
analogous to heat flux being a measure of additional
kinetic energy at the molecular scale.
For example, with granular materials, the full kinetic energy
within a region $\Dc$ that has a similar size as
support $\text{supp}(\varphi,\X)$
is partitioned among the average translational
kinetic energy $\dot{K}(\Dc)$,
the thermal energy (measured by the region's hotness),
and the grain-scale \emph{velocity fluctuations} captured by
$\dot{R}_{K}(\Dc)$.
An intermediate energy $\dot{R}_{K}(\Dc)$ can also arise when
atomic models are used to describe solids at high temperatures,
accounting for the second moment of thermal
fluctuations \cite{Parthasarathy:2018a}.
Note that
as the rate of a process decreases,
both $\dot{K}(\mathcal{D})$ and $P_{A,\text{mech}}(\Dc)$
are reduced by the square of the rate,
and the residual $\dot{R}_{K}(\Dc)$ is likewise reduced,
so that under quasi-static conditions, $\dot{R}_{K}(\Dc)$ vanishes.
\par
Recalling that the space of virtual velocities $\widehat{W}$
includes the actual velocity field $\mathbf{w}(\X)$
as a particular instance,
$\mathbf{w}(\mathbf{x})\in W\subset\widehat{W}$,
we substitute Eqs.~(\ref{eq:nonlocalPF})--(\ref{eq:nonlocal4}),
(\ref{eq:PA}), and~(\ref{eq:residualR})
into the principle of virtual power,
Eq.~(\ref{eq:master}\textsubscript{1}),
but now evaluated with the
\emph{actual} mechanical and thermal velocity fields
$\mathbf{v}(\X)$ and $\theta(\X)$,
\begin{equation}\label{eq:balance5}
  \begin{aligned}
    \Big(
      \llangle
        \rho\mathbf{b}, & \varphi\mathbf{v}
      \rrangle_{\Dc} +
      \llangle
        \mathbf{t},\varphi\mathbf{v}
      \rrangle_{\Dp}
    \Big)
    +
    \Big(
      \llangle
        \rho s, \varphi\,\theta
      \rrangle_{\Dc} -
      \llangle
        k,\varphi\,\theta
      \rrangle_{\Dp}
    \Big)\\
    =\;
    &\frac{d}{dt}
    \Big(
      K(\Dc) +
      \llangle
        \rho\eta,\varphi\,\theta
      \rrangle_{\Dc}
    \Big)\\
    &+
    \Big(
      \llangle[\big]
        \mathbf{T},\big(\lpoon{\boldsymbol{\nabla}}\varphi\otimes\mathbf{v}\big)
      \rrangle[\big]_{\Dc} -
      \llangle[\big]
        \mathbf{p},\big(\lpoon{\boldsymbol{\nabla}}\varphi\big)\,\theta
      \rrangle[\big]_{\Dc} -
      \llangle
        \rho\eta,\varphi\,\dot{\theta}
      \rrangle_{\Dc} -
      \llangle
        \rho\xi,\varphi\,\theta
      \rrangle_{\Dc}
      - \dot{R}_{K}(\Dc)
    \Big)
  \end{aligned}
\end{equation}
where we have assumed that $\eta(\X,t)$ and $\theta(\X,t)$
are differentiable in time $t$,
so that the identity
$\llangle\rho\dot{\eta},\varphi\theta\,\rrangle_{\Dc}
= \frac{d}{dt}\llangle\rho\eta,\varphi\theta\,\rrangle_{\Dc}-
\llangle\rho\eta,\varphi\dot{\theta}\,\rrangle_{\Dc}$
applies for a fixed region $\Dc$.
Within the terms on the left,
one recognizes the external mechanical power
(terms with $\mathbf{b}$ and $\mathbf{t}$)
and the net entropy supply
(the term with $s$ for the external supply and the term with $k$
for the entropy flux).
Note that
the mollifier gradient $\lpoon{\boldsymbol{\nabla}}\varphi$
in the third line of Eq.~(\ref{eq:balance5})
smooths the gradients of $\mathbf{v}$ and $\theta$ in a manner
similar to that of
the smoothing kernel in strain-rate averaging constitutive models
(e.g., a neighborhood-average strain rate
$\int\!\varphi(\aXX)\dot{\boldsymbol{\varepsilon}}(\Xp)\,\Vp$
in \cite{Kroner:1967a,Polizzotto:2001a}).
\par
Although Eq.~(\ref{eq:balance5})
expresses a balance of combined power and entropy, it
does not yet represent an energy balance~---
nor a law of thermodynamics~---
until a further assumption and a further principle are introduced.
An energy balance ensues if the external per-volume
supply of energy $r(\X)$
and the external boundary flux of energy $h(\X)$,
both associated with molecular-scale thermal energy (heat),
are taken to equal the nonlocal products of their corresponding
entropy quantities at $\X$ and the \emph{average temperature} $\theta$
near $\X$:
\begin{equation}\label{eq:heatquants}
  \begin{alignedat}{2}
  \rho r(\X) &=
  \int_{\RRR}
  \rho s(\X)\varphi(\aXX)\theta(\Xp)\,\Vp
  &&=\langle\rho s,\varphi\theta\rangle_{\RRR}
  \\ 
  h(\X) &=
  \int_{\RRR}
  k(\X)\varphi(\aXX)\theta(\Xp)\,\Vp
  &&=\langle k,\varphi\theta\rangle_{\RRR}
  \end{alignedat}
\end{equation}
After
substituting Eq.~(\ref{eq:heatquants}) into Eq.~(\ref{eq:balance5})
and applying Eq.~(\ref{eq:identity}),
one recognizes the result as an
energy balance, but one for a nonlocal medium:
\begin{equation}\label{eq:energyconserve}
  \begin{aligned}
    \Big(
      \llangle
        &\rho\mathbf{b},\varphi\mathbf{v}
      \rrangle_{\Dc} +
      \llangle
        \mathbf{t},\varphi\mathbf{v}
      \rrangle_{\Dp}
    \Big)
    +
    \Big(
      \int_{\Dc}\rho r(\X)\,dv -
      \int_{\Dp}h(\X)\,da
    \Big)\\
    &=\,\frac{d}{dt}
    \Big(
      K(\Dc) +
      \llangle
        \rho\eta,\varphi\,\theta
      \rrangle_{\Dc}
    \Big) \\
    &\;\;\;\;+
    \Big(
      \llangle[\big]
        \mathbf{T},\big(\lpoon{\boldsymbol{\nabla}}\varphi\otimes\mathbf{v}\big)
      \rrangle[\big]_{\Dc} +
      \llangle[\big]
        \mathbf{q}/\theta
        ,
        \varphi
        \theta^{2}
        \boldsymbol{\nabla}(1/\theta)
      \rrangle[\big]_{\Dc}
      -
      \llangle
        \rho\eta,\varphi\,\dot{\theta}
      \rrangle_{\Dc} -
      \llangle
        \rho\xi,\varphi\,\theta
      \rrangle_{\Bc}
      - \dot{R}_{K}(\Dc)
    \Big)
  \end{aligned}
\end{equation}
where $\mathbf{q}(\X)$ is the internal heat flow,
with entropy flow $\mathbf{p}(\X)=\mathbf{q}(\X)/\theta(\X)$.
The combined rate of mechanical work and heating equals
the rate of change of kinetic energy
and of internal heating plus
the rate at which internal energy (mechanical and thermal) is
generated, plus the residual rate of kinetic energy $\dot{R}_{K}$.
The thermal momentum $\eta(\X)$ is identified as the local entropy
(as in the discussion following Eqs.~\ref{eq:PFB} and~\ref{eq:local}),
and the
thermal motion $\theta(\X)$
is the corresponding local temperature,
a measure of hotness at $\X$.
Henceforth, the (actual)
temperature $\theta(\X)$ is assumed positive:
\begin{equation}
  \theta(\X) > 0,\quad\forall\X\in\Bo
\end{equation}
Note that $\rho\mathbf{r}$ and $h$
in Eq.~(\ref{eq:energyconserve}),
although local to $\X$, are derived from an
average, nonlocal temperature (i.e., thermal velocity),
similar to mechanical work being derived from
an average, nonlocal velocity $\mathbf{v}$.
\subsection{First Law for nonlocal media}\label{sec:firstlaw}
In the approach of Serrin \cite{Serrin:1986a},
\v{S}ilhav\'{y} \cite{Silhavy:1983a},
and Green and Naghdi \cite{Green:1978a,Green:1991b},
the first law (in its strong form) asserts the existence of
an internal energy~---
a state variable~---
such that the combined external supply
of mechanical work and heat
to a thermodynamic
system vanishes for any cyclic process
returning the system to its original condition.
In our context, a system is defined by a region $\Dc\in\mathscr{D}$ 
in $\Bc$ and by the collection of ordered paths
that are available to the region's \emph{condition}.
The system's condition is characterized by its
kinetic energy $K(\Dc)$,
its configuration $\boldsymbol{\chi}(\mathbf{X},t)$,
its temperature field $\theta(\X)$,
and other state parameters,
although redistributions
of the internal quantities $\mathbf{T}$, $\mathbf{p}$, $\eta$, and $\xi$
(i.e., the original dual force quantities)
are possible during a cycle.
For a non-local medium,
the external supply of work and heat along a path
is the combined time-integral $\int(\cdot)\,dt$
of the four scalar rates on the left of
Eq.~(\ref{eq:energyconserve}).
Recalling that $\mathbf{v}(\X)$ and $\theta(\X)$
are rates of displacement,
the time-integral of the rate on the second line
of Eq.~(\ref{eq:energyconserve}) vanishes for a cyclic path.
After canceling these terms in the first and
second lines of the equation,
the remaining terms on the right~---
those on the third line of
Eq.~(\ref{eq:energyconserve})---
when integrated from the beginning to the end of a cycle,
must sum to zero.
Because this sum is zero regardless of the cyclic path,
the fundamental theorem of line integrals (the gradient theorem)
requires that these
remaining terms are the time rate of a potential, say $\Psi_{\Dc}$
\cite{Green:1978a,Green:1991b},
\begin{equation}\label{eq:psidot1}
  \frac{d}{dt}\Psi_{\Dc} =
      \llangle
        \mathbf{T},(\lpoon{\boldsymbol{\nabla}}\varphi\otimes\mathbf{v})
      \rrangle_{\Dc} -
      \llangle
        \mathbf{p},(\lpoon{\boldsymbol{\nabla}}\varphi)\,\theta
      \rrangle_{\Dc} -
      \llangle
        \rho\eta,\varphi\,\dot{\theta}
      \rrangle_{\Dc} -
      \llangle
        \rho\xi,\varphi\,\theta
      \rrangle_{\Dc}
      - \dot{R}_{K}(\Dc)
\end{equation}
where $\Psi_{\Dc}$ is identified as
the Helmholtz free energy of a nonlocal continuum.
\par
Henceforth, we assume that processes are sufficiently slow
(i.e. quasi-static)
that any cycle also returns the hidden
kinetic energy $R_{K}(\Dc)$ to initial value (taken as negligibly small),
so that $R_{K}(\Dc)$ is henceforth neglected in rate $d\Psi_{\Dc}/dt$.
\par
Because rate $d\Psi_{\Dc}/dt$ is
associated with a region $\Dc$ but
applies to all such regions $\Dc\in\mathscr{D}$ in $\Bc$,
a local rate of the specific free energy, $\dot{\psi}(\X)$,
must exist at all points in $\Bc$:
\begin{equation}\label{eq:dHelmholtz}
  \begin{aligned}
  \rho\dot{\psi}(\X) =
    &\; T_{ji}(\X)\!\int_{\RRR}
      \lpoon{\varphi}_{,j}(\aXX)
      v_{i}(\Xp)\,\Vp
    - \rho\eta(\X)\!\int_{\RRR}
         \varphi(\aXX) \dot{\theta}(\Xp)\,\Vp\\
    &- p_{i}(\X)\!\int_{\RRR}
      \lpoon{\varphi}_{,i}(\aXX)
      \theta(\Xp)\,\Vp  
     - \rho\xi(\X)\!\int_{\RRR}
         \varphi(\aXX) \theta(\Xp)\,\Vp
     ,\quad\forall\X\in\Bc
  \end{aligned}
\end{equation}
in which summations apply over $i$ and $j$, and where,
in accordance with the inner products
defined in Eq.~(\ref{eq:abbrev}),
the rate $\dot{\Psi}_{\Dc}$ in Eq.~(\ref{eq:psidot1}) is
\begin{equation}\label{eq:dHelmholtz2}
  \dot{\Psi}_{\Dc}
  =
  \int_{\Dc}\rho\dot{\psi}(\X)\,dv
\end{equation}
\par
After substituting Eq.~(\ref{eq:dHelmholtz})
in Eqs.~(\ref{eq:energyconserve}),
the first law of thermodynamics for a nonlocal medium is
\begin{equation}\label{eq:engergies}
  \begin{aligned}
  &
  \frac{d}{dt}
    \bigg(
      K(\Dc)
      +
      \int_{\Dc}
      \rho e(\X)\,dx
    \bigg)
  \\
  &\quad =
  \Big(
      \llangle
        \mathbf{b}, \varphi\mathbf{v}
      \rrangle_{\Dc} +
      \llangle
        \mathbf{t},\varphi\mathbf{v}
      \rrangle_{\Dp}
    \Big)
    +
    \Big(
      \int_{\Dc}\rho r(\X)\,dv -
      \int_{\Dp}h(\X)\,da
    \Big),
    \quad
    \forall\mathcal{D}\in\mathscr{D}
  \end{aligned}
\end{equation}
where the
rate of the internal energy potential per unit mass,
$\dot{e}(\X)$, is present as
\begin{equation}\label{eq:edot}
  \begin{aligned}
    \rho\dot{e}(\X)
    =&\;
    \rho\dot{\psi}(\X)
    +
    \frac{d}{dt}
    \bigg(
    \int_{\RRR}\rho\eta(\X)\varphi(\aXX)\theta(\Xp)\,\Vp
    \bigg)
    \\
    =
    &\;T_{ji}(\X)\!\int_{\RRR}
      \lpoon{\varphi}_{,j}(\aXX)
      v_{i}(\Xp)\,\Vp
     + \rho\dot{\eta}(\X)\!\int_{\RRR}
        \varphi(\aXX) \theta(\Xp)\,\Vp\\
    &
     - p_{i}(\X)\!\int_{\RRR}
       \lpoon{\varphi}_{,i}(\aXX)\theta(\Xp)\,\Vp
     - \rho\xi(\X)\!\int_{\RRR}
        \varphi(\aXX) \theta(\Xp)\,\Vp
     ,\quad\forall\X\in\Bc
  \end{aligned}
\end{equation}
\par
With Eq.~(\ref{eq:engergies}), one can attribute the combined rates
of kinetic and internal energies to
the causative effects of the external supply of mechanical power
$\dot{\mathcal{W}}(\Dc)$ and of heating $\dot{\mathcal{H}}(\Dc)$
to region $\Dc$ (as in \cite{Green:1977a}),
\begin{equation}
  \frac{d}{dt}
    \Big(
      K(\Dc)
      +
      \int_{\Dc}
      \rho e(\X)\,dx
    \Big)
  =
  \dot{\mathcal{W}}(\Dc)
  +
  \dot{\mathcal{H}}(\Dc)
\end{equation}
where, after substituting Eqs.~(\ref{eq:heatquants}),
\begin{align}
  \label{eq:Wdot}
  &\begin{aligned}
    \dot{\mathcal{W}}(\Dc)
      &=
        \llangle
        \mathbf{b}, \varphi\mathbf{v}
        \rrangle_{\Dc} +
        \llangle
          \mathbf{t},\varphi\mathbf{v}
        \rrangle_{\Dp}
      \\
      &=
      \frac{d}{dt}
      \Big(
        K(\Dc)
        +
        \int_{\Dc}
        \rho e(\X)\,dx
      \Big)
      -
      \llangle
        \rho\dot{\eta},\varphi\theta
      \rrangle_{\Dc}
      +
      \int_{\Dc}\rho\dot{\omega}(\X)\,dv
  \end{aligned}
  \\
  \label{eq:Hdot}
  &\begin{aligned}
    \dot{\mathcal{H}}(\Dc)
      &=
        \llangle
          \rho s, \varphi\,\theta
        \rrangle_{\Dc} -
        \llangle
          k,\varphi\,\theta
        \rrangle_{\Dp}
       \\
       &=
        \llangle
          \rho\dot{\eta},\varphi\theta
        \rrangle_{\Dc}
        -
        \int_{\Dc}\rho\dot{\omega}(\X)\,dv
  \end{aligned}
\end{align}
with rate $\dot{\omega}$ defined by
\begin{equation}\label{eq:defrhow}
  \begin{aligned}
    \rho\dot{\omega}(\X)
    &=
    - \rho\dot{\psi}(\X)
    + T_{ji}(\X)\!\int_{\RRR}
        \lpoon{\varphi}_{,j}(\aXX)
      v_{i}(\Xp)\,\Vp
    - \rho\eta(\X)\!\int_{\RRR}
        \varphi(\aXX) \dot{\theta}(\Xp)\,\Vp
    \\
    &=
    p_{i}(\X)\!\int_{\RRR}
      \lpoon{\varphi}_{,i}(\aXX)
      \theta(\Xp)\,\Vp
    +
    \rho\xi(\X)\!\int_{\RRR}
      \varphi(\aXX) \theta(\Xp)\,\Vp
  \end{aligned}
\end{equation}
a quantity that takes a central role in the next section.
\subsection{Second Law for nonlocal media}\label{sec:secondlaw}
Although the first law for nonlocal media~---
Eq.~(\ref{eq:engergies})~---
proceeds from
balance conditions derived from the principle
of virtual power when applied to cyclic processes
(thus leading to Eqs.~\ref{eq:balance5}
and~\ref{eq:energyconserve}),
the second law of thermodynamics is an assertion
avowed from observed evidence:
that heat flows from warmer regions toward cooler ones,
that thermal energy converted from mechanical work
cannot be fully restored as mechanical work, etc.
Several similar statements of the second law have been proposed
for continua,
primarily in the context of local media
(see 
\cite{Truesdell:1960a}\S258 and
\cite{Coleman:1963a,Muller:1967a,Green:1977a,Germain:1983a},
etc.).
\par
We adopt the approach of Green and Naghdi \cite{Green:1977a},
in which the second law is expressed as three independent inequalities~---
each grounded in observation~---
here derived in a nonlocal context.
These inequalities impose restrictions on the constitutive
description of a body $\Bc$.
The first observation
is the impossibility of fully reversing a process that transforms
mechanical power into heat:
in every process, a part of the power $\dot{\mathcal{W}}$
(which can be positive, negative, or zero)
that is induced by velocities $\mathbf{v}$
(on the right of Eq.~\ref{eq:Wdot}\textsubscript{1})
produces non-negative heating
(a positive contribution to $\dot{\mathcal{H}}$)
that cannot be extracted (i.e. recovered)
upon reversal of the velocities.
An example is the non-negative heating produced by internal friction.
Therefore,
if the work $\dot{\mathcal{W}}$ in
Eq.~(\ref{eq:Wdot}\textsubscript{2}) is composed of
a part that can be positive, negative, or zero and whose
sign can be reversed by reversing the process,
then any remaining part must be non-negative.
Because the $d(\cdot)/dt$ term in Eq.~(\ref{eq:Wdot}\textsubscript{2})
and the
$\llangle\rho\dot{\eta},\varphi\theta\rrangle_{\Dc}$ term
in Eqs.~(\ref{eq:Wdot}\textsubscript{2})
and~(\ref{eq:Hdot}\textsubscript{2})
can each be positive, negative, or zero, the remaining term~---
collected into $\int_{\Dc}\rho\dot{\omega}(\X)\,dv$~---
must be non-negative:
\begin{equation}\label{eq:rhowle0}
  \int_{\Dc}\rho\dot{\omega}(\X)\,dv \ge 0
\end{equation}
From the thermal balance of Eq.~(\ref{eq:thermal1}) and the
definition of $\rho\dot{\omega}$
in Eq.~(\ref{eq:defrhow}\textsubscript{2}),
inequality~(\ref{eq:rhowle0}) is equivalent to
\begin{equation}\label{eq:CD}
  \llangle
    \rho\dot{\eta},\varphi\theta
  \rrangle_{\Dc}
 -\llangle
    \rho s,\varphi\theta
  \rrangle_{\Dc}
 +\llangle
    k,\varphi\theta
  \rrangle_{\Dp}
  \ge 0
\end{equation}
where $\theta$ is non-negative,
which asserts that the rate of entropy change in a
region $\Dc\in\Bc$ must always
equal or exceed the rate of external entropy supply
(also, \cite{Coleman:1963a,Muller:1967a}).
Substituting the full Eq.~(\ref{eq:thermal1})
and noting that inequality~(\ref{eq:CD}) applies
to all regions $\Dc\in\mathscr{D}$ and that the
regularity conditions imposed at the outset of this section
ensure that the
integrands in~(\ref{eq:CD}) are integrable,
the inequality takes a local form
at point $\X$:
\begin{equation}\label{eq:2ndlaw}
  p_{i}(\X)\!\int_{\RRR}
      \lpoon{\varphi}_{,i}(\aXX)
      \theta(\Xp)\,\Vp
  + \rho\xi(\X)\!\int_{\RRR}
      \varphi(\aXX) \theta(\Xp)\,\Vp
  \ge 0
     ,\quad\forall\X\in\Bc
\end{equation}
where $\mathbf{p}(\X)$ is the entropy flux,
$\xi(\X)$ is the internal entropy production,
and summation applies to index $i$.
This inequality is further simplified by considering
the next observation.
\par
The second observation is that entropy flux
(i.e., heat flux $\mathbf{q}$ divided by the positive $\theta$)
acts in the direction of decreasing temperature,
as expressed by the conduction inequality
\begin{equation}\label{eq:conduction}
  -p_{i}(\X)\!\int_{\RRR}
      \lpoon{\varphi}_{,i}(\aXX)
      \theta(\Xp)\,\Vp
  \ge 0,
  \quad
  \forall\X\in\Bc
\end{equation}
This inequality is more transparent after
applying identity~(\ref{eq:identity}) and assuming that
$\theta(\Xp)$ is continuously differentiable,
\begin{equation}\label{eq:pii}
    -p_{i}(\X)\!\int_{\RRR}
      \lpoon{\varphi}_{,i}(\aXX)
      \theta(\Xp)\,\Vp
    =
    -p_{i}(\X)\!\int_{\RRR}
      \varphi(\aXX)
      \frac{\partial}{\partial x^{\prime}_{i}}
      \theta(\Xp)\,\Vp
      \ge
      0
\end{equation}
confirming that entropy flux,
in a nonlocal continuum,
is in the direction
of decreasing \emph{average} temperature.
\par
Taking note of inequality~(\ref{eq:pii}),
the previous inequality~(\ref{eq:2ndlaw})
can be restated as
the following
dissipation inequality~---
for the dissipation rate $\Phi$ per unit volume~---
that governs systems in thermal equilibrium
or approaching it:
\begin{equation}\label{eq:2ndlaw2}
  \Phi(\X)\equiv
  \rho\xi(\X)\!\int_{\RRR}
    \varphi(\aXX) \theta(\Xp)\,\Vp
  \ge 0,
  \quad
  \forall\X\in\Bc
\end{equation}
requiring the \emph{average} internal entropy production
to be non-negative at every point.
This condition is further localized by noting that both
$\rho$ and $\theta$ are positive, so that if
$\varphi$ is non-negative within its support
$\text{supp}(\varphi,\X)$, the first inequality of the second law
is equivalent to the localized inequality
\begin{equation}\label{eq:xilocal}
  \xi(\X) \ge 0,
  \quad
  \forall\X\in\Bc
\end{equation}
a condition that is the same for both local and nonlocal media.
(Note, however, that if $\varphi$ is negative within part of
its support but still meets the normalizing condition
(\ref{eq:unity}), one must revert to the more general
statement (\ref{eq:2ndlaw2}).)
Combining Eqs.~(\ref{eq:dHelmholtz}) 
and~(\ref{eq:2ndlaw2})
leads to an alternative statement of the second law,
in the manner of the Clausius--Duhem inequality:
\begin{equation}\label{eq:psidot}
  \begin{aligned}
    -&\rho(\X)\dot{\psi}(\X)
    +T_{ji}(\X)\!\int_{\RRR}
      \lpoon{\varphi}_{,j}(\aXX)
      v_{i}(\Xp)\,\Vp
    \\
    &\quad
    -p_{i}(\X)\!\int_{\RRR}
      \lpoon{\varphi}_{,i}(\aXX)
      \theta(\Xp)\,\Vp
    - \rho\eta(\X)\!\int_{\RRR}
        \varphi(\aXX) \dot{\theta}(\Xp)\,\Vp
    \;\ge 0,
    \quad\forall\X\in\Bc
  \end{aligned}
\end{equation}
which is applied in the next section.
\par
The third observation concerns the direction of
temperature change $\dot{\theta}$ during heating,
which is not captured by the inequalities above:
if a region at uniform temperature is heated but remains
rigid during heating,
with $\dot{\mathcal{H}}(\Dc)>0$ and $\dot{\mathcal{W}}(\Dc)=0$,
its temperature increases.
Noting that this observation applies to all regions
$\Dc\in\mathscr{D}$ and that the integrands of
$\dot{\mathcal{H}}$ in Eq.~(\ref{eq:Hdot}) are Lebesgue integrable,
the third condition is the implication
\begin{equation}\label{eq:3rd}
  \dot{\mathcal{H}}(\X)>0\text{ and }\mathbf{v}(\X)\in\Vhat_{\text{r}}
  \quad\Rightarrow\quad
  \dot{\theta}(\X)>0,
  \quad
  \forall\mathbf{x}\in\Bc
\end{equation}
where the rigidity condition,
$\mathbf{v}(\X)\in\Vhat_{\text{r}}$,
applies to any neighborhood of $\X$.
Substituting Eqs.~(\ref{eq:Hdot}\textsubscript{2}),
(\ref{eq:defrhow}\textsubscript{1}),
and~(\ref{eq:edot}\textsubscript{1})
gives the following restriction, also in the form of an implication:
\begin{equation}\label{eq:thetadot}
  \dot{e}(\X)>0
  \text{ and }\mathbf{v}(\X)\in\Vhat_{\text{r}}
  \;\Rightarrow\;
  \dot{\theta}(\X)>0,
  \quad
  \forall\mathbf{x}\in\Bc
\end{equation}
\par
In summary,
the three observations are expressed as three restrictions that,
together with the positivity of temperature, constitute
the second law:
inequality~(\ref{eq:xilocal})
(or any of~\ref{eq:rhowle0}--\ref{eq:2ndlaw} or~\ref{eq:psidot});
inequality~(\ref{eq:pii});
and the implication~(\ref{eq:3rd}) or~(\ref{eq:thetadot}).
\subsection{Nonlocal constitutive framework}\label{sec:constit}
As a general approach,
we adopt the constitutive framework
of Coleman and Gurtin \cite{Coleman:1967a} and
assume that the internal Helmholtz energy $\psi$
is a function of the following state variables:
the displacement gradient
$\mathbf{F}=\boldsymbol{\nabla}_{\!_\mathbf{X}}$\raisebox{1.5pt}{$\boldsymbol{\chi}$} and the temperature $\theta$~---
both of which can be observed, measured, and controlled~---
and a finite set $\boldsymbol{\alpha}$ of ``hidden'' internal variables
$\boldsymbol{\alpha}=\alpha_{1},
\alpha_{2},\ldots\alpha_{n}$,
which individually can be scalars or tensors
(also, \cite{Rice:1971a,Germain:1983a,Maugin:1994a}).
These variables are assumed sufficient in character and number
to describe thermodynamic equilibrium states and
to describe non-equilibrium states
sufficiently close to equilibrium \cite{Bataille:1979a}.
Common internal variables~--- all associated with
irreversible processes~--- include plastic and viscous deformation,
dislocation or crack density, damage measures,
fabric parameters, etc.
Evolution of the internal variables is induced
by deformation and by temperature gradients,
with the evolution rate being a function $\mathbf{g}$
of the state variables, with
$\dot{\boldsymbol{\alpha}}=\mathbf{g}(\mathbf{F},\theta,\boldsymbol{\alpha})$.
The directions of the rates, however, are constrained
by the second law.
As shown below,
the internal variables are not conjugate with either the applied
forces or entropy flux,
so that the variables cannot be directly controlled
by these external quantities
\cite{Maugin:1994a}
(i.e., the variables are
``parameter[s] of state which can be measured but not controlled''
\cite{Bridgman:1953a}).
\par
As implied in Eq.~(\ref{eq:dHelmholtz}),
the free energy density $\psi(\X)$
is taken as the spatial average $\overline{\Psi}$ of a
state function $\Psi$ of the local
quantities
$\mathbf{F}$, $\theta$, and~$\boldsymbol{\alpha}$,
\begin{equation}\label{eq:psiIII}
  \psi(\X)
  =
  \overline{\Psi}
  \big(
    \X,
    \mathbf{F}(\X),
    \theta(\X),
    \boldsymbol{\alpha}(\X)
  \big)
  =
  \int_{\RRR}
  \varphi(\aXX)
  \Psi
  \big(
    \Xp,
    \mathbf{F}(\Xp),
    \theta(\Xp),
    \boldsymbol{\alpha}(\Xp)
  \big)\,\Vp
\end{equation}
where this form of $\psi$ precludes a dependence
on the temperature gradient $\nabla\theta$.
The integrated function $\Psi$ depends on position
$\Xp$ (e.g., from a stiffness modulus, heat capacity, etc. that
can vary with $\Xp$)
and on $\mathbf{F}$, $\theta$, and $\boldsymbol{\alpha}$
at $\Xp$.
Although this form may appear to be discretionary,
we show that it leads to unambiguous relationships
among stress, entropy, free energy, and dissipation.
\par
Since mollifier $\varphi$ is assumed constant in time, the rate
$\dot{\psi}$ in Eq.~(\ref{eq:psidot})
equals the nonlocal, averaged rate
$\dot{\overline{\Psi}}$,
\begin{equation}\label{eq:psiIIa}
  \dot{\psi}(\X)
  =
  \dot{\overline{\Psi}}
  \big(
    \X,
    \mathbf{F}(\X),
    \theta(\X),
    \boldsymbol{\alpha}(\X)
  \big)
  =
  \int_{\RRR}
  \varphi(\aXX)
  \frac{d}{dt}\Psi
  \big(
    \Xp,
    \mathbf{F}(\Xp),
    \theta(\Xp),
    \boldsymbol{\alpha}(\Xp)
  \big)
  \,\Vp
\end{equation}
We proceed in the manner of
Coleman and Noll~\cite{Coleman:1963a}
and Green and Naghdi \cite{Green:1965a},
but now in a nonlocal context,
to develop the constitutive consequences of
energy function $\Psi$, which is assumed
continuously differentiable in its
arguments
(also \cite{Coleman:1967a,Ziegler:1983a}).
Expanding the rate $d{\Psi}/dt$
of Eq.~(\ref{eq:psiIIa}) and substituting $\dot{\psi}$
into the inequality~(\ref{eq:psidot}),
\begin{equation}\label{eq:ineqIII}
  \begin{aligned}
  &\bigg(
    T_{ji}(\X)\!\!\int_{\RRR}\!\!
    \lpoon{\varphi}_{,j}(\aXX)
    v_{i}(\Xp)\,\Vp
    - \rho(\X)\!\!
    \int_{\RRR}\!
    \varphi(\aXX)
    \,\frac{\partial\Psi}{\partial F_{ij}}
    (\Xp)
    L_{ik}(\Xp)F_{kj}(\Xp)\,\Vp
  \bigg)\\
  &\quad
  +
  \bigg(
    - \rho\eta(\X)\!\int_{\RRR}
        \varphi(\aXX) \dot{\theta}(\Xp)\,\Vp
    -\rho(\X)\!\!
    \int_{\RRR}
    \varphi(\aXX)
    \,\frac{\partial\Psi}{\partial \theta}(\Xp)\,
    \dot{\theta}(\Xp)\,\Vp
  \bigg)\\
  &\quad
  +
  \bigg(
  -p_{i}(\X)\!\int_{\RRR}\!
  \lpoon{\varphi}_{,i}(\aXX)
  \theta(\Xp)\,\Vp
  -\rho(\X)\!\!
  \int_{\RRR}\!
  \varphi(\aXX)
  \,\frac{\partial\Psi}{\partial \alpha_{i}}(\Xp)
  \dot{\alpha}_{i}(\Xp)\,\Vp
  \bigg)
  \;\ge
  0
  \end{aligned}
\end{equation}
in which mollifier 
the summation of indices applies,
and $\mathbf{L}=\dot{\mathbf{F}}\cdot\mathbf{F}^{-1}$
is introduced as
the velocity gradient in the
current configuration, $L_{ij}=v_{i,j}$.
The current density $\rho$ relates to the reference density
$\rho_{\text{o}}$ by
$\rho(\mathbf{x})=J^{-1}\rho_{\text{o}}(\mathbf{X})$,
where $J=\text{det}(\mathbf{F})$.
\par
Inequality~(\ref{eq:ineqIII}) applies to arbitrary fields
$\mathbf{v}(\Xp)$ and $\theta(\Xp)$,
which we now assume are
continuously differentiable,
allowing application of identity~(\ref{eq:identity}).
To provide further clarity,
$\mathbf{T}$, $\rho$, $\eta$, and $\mathbf{p}$,
which are independent of $\Xp$,
are moved inside their integrals:
\begin{equation}\label{eq:ineqIIIb}
  \begin{aligned}
  &
    \bigg(
    \int_{\RRR}\!\!
    T_{ji}(\X)
    \varphi(\aXX)
    L_{ij}(\Xp)\,\Vp
    -
    \int_{\RRR}\!
    \rho(\X)
    \varphi(\aXX)
    \,\frac{\partial\Psi}{\partial F_{ij}}
    (\Xp)
    L_{ik}(\Xp)F_{kj}(\Xp)\,\Vp
  \bigg)\\
  &\quad
  +
  \bigg(
    -\int_{\RRR}\!
      \rho\eta(\X)
      \varphi(\aXX) \dot{\theta}(\Xp)\,\Vp
    -
    \int_{\RRR}\!
    \rho(\X)
    \varphi(\aXX)
    \,\frac{\partial\Psi}{\partial \theta}(\Xp)\,
    \dot{\theta}(\Xp)\,\Vp
  \bigg)\\
  &\quad
  +
  \bigg(
  -\int_{\RRR}\!
    p_{i}(\X)
    \varphi(\aXX)
    \theta_{,i}(\Xp)\,\Vp
  -
  \int_{\RRR}\!
    \rho(\X)
    \varphi(\aXX)
  \,\frac{\partial\Psi}{\partial\alpha_{i}}(\Xp)\,
  \dot{\alpha}_{i}(\Xp)\,\Vp
  \bigg)
  \;\ge
  0
  \end{aligned}
\end{equation}
After
collecting common factors in the integrands,
\begin{equation}\label{eq:ineqIIIc}
  \begin{aligned}
  &
    \int_{\RRR}\!
    \varphi(\aXX)
    \left[
    T_{ji}(\X)\delta_{kj}
    -
    \rho(\X)
    \,\frac{\partial\Psi}{\partial F_{ij}}
    (\Xp)
    F_{kj}(\Xp)
    \right]
    L_{ik}(\Xp)
    \,\Vp
  \\
  &\quad
  -
  \int_{\RRR}\!
    \varphi(\aXX)
    \left[
    \rho\eta(\X)
    +
    \rho(\X)
    \,\frac{\partial\Psi}{\partial \theta}(\Xp)
    \right]
    \dot{\theta}(\Xp)\,\Vp
 \\
  &\quad
  +
  \bigg(
  -\int_{\RRR}\!
    p_{i}(\X)
    \varphi(\aXX)
    \theta_{,i}(\Xp)\,\Vp
  -
  \int_{\RRR}\!
    \rho(\X)
    \varphi(\aXX)
  \,\frac{\partial\Psi}{\partial\alpha_{i}}(\Xp)\,
  \dot{\alpha}_{i}(\Xp)\,\Vp
  \bigg)
  \;\ge
  0
  \end{aligned}
\end{equation}
where $\delta_{kj}$ is the Kronecker identity.
At a given time,
functions $\mathbf{L}(\Xp)$ and $\dot{\theta}(\Xp)$
can be arbitrarily assigned (or set to zero),
so that the fundamental lemma implies that
\begin{subequations}
\label{eq:psiIII3}
  \begin{align}
  &
    T_{ji}(\X)
    =
    \rho(\X)
    \int_{\RRR}
    \varphi(\aXX)
    \,\frac{\partial\Psi}{\partial F_{ik}}
    (\Xp)
    F_{jk}(\Xp)
    \,\Vp
  \\
  &
    \eta(\X)
    =
    -
    \int_{\RRR}
    \varphi(\aXX)
    \,\frac{\partial\Psi}{\partial \theta}(\Xp)
    \,\Vp
 \\
  &
  -
  p_{i}(\X)\!
  \int_{\RRR}
    \varphi(\aXX)
    \theta_{,i}(\Xp)\,\Vp
  -
  \rho(\X)\!
  \int_{\RRR}
    \varphi(\aXX)
    \,\frac{\partial\Psi}{\partial\alpha_{i}}(\Xp)\,
    \dot{\alpha}_{i}(\Xp)\,\Vp
  \;\ge
  0
  \end{align}
\end{subequations}
in which we have moved
$\mathbf{T}$, $\rho$, $\eta$, and $\mathbf{p}$
from their integrands (again, they are independent of $\Xp$)
and have replaced the fixed value of $\int\varphi(\aXX)\,\Vp$,
taken as unity in Eq.~(\ref{eq:unity}).
A comparison of Eqs.~(\ref{eq:2ndlaw}) and (\ref{eq:psiIII3}c) shows
that rate $(\partial\Psi/\partial\alpha_{i})/\dot{\alpha}_{i}$
is associated with internal entropy production $\xi$.
If variables $\boldsymbol{\alpha}$ have units of strain,
then $\rho\nabla_{\alpha}\Psi$ has units of stress~---
the dissipation stress of
Ziegler \cite{Ziegler:1987a}.
Note that the results in Eq.~(\ref{eq:psiIII3}) differ from those of
Nilsson \cite{Nilsson:1997a}, who assumed that $\Psi$
depends on an averaged strain
$\overline{\mathbf{F}}$ and on averaged internal variables
$\overline{\boldsymbol{\alpha}}$,
although such compound averaging
does not follow from our derivation of
the second law (discussed below).
\par
With Eqs.~(\ref{eq:psiIII3}),
stress $\mathbf{T}(\X)$
depends on a nonlocal averaging of $\Psi$ and its derivatives
in the neighborhood of $\X$,
which, when compared with a local evaluation at $\X$ alone,
may substantially alter the stress at $\X$.
For example, in the context of elasto-plasticity,
if inhomogeneity near $\X$ produces
loading in part of the point's neighborhood but
unloading in the remainder,
an averaging of stiffnesses over the full neighborhood
yields a stiffness much different from
either part considered separately.
\par
The first term on the left of
inequality~(\ref{eq:psiIII3}c) also
appears in the conduction inequality~(\ref{eq:conduction})
(i.e., the second law's second restriction),
which must be non-negative.
The full inequality~(\ref{eq:psiIII3}c)
shows that when a neighborhood of $\X$ has uniform temperature,
so that the first integral is zero, the sum
$-\rho (\partial\Psi/\partial\alpha_{i})\dot{\alpha}_{i}$
must be non-negative, producing internal dissipation $\Phi$,
as in Eqs.~(\ref{eq:2ndlaw2})--(\ref{eq:xilocal}).
Likewise, inequality~(\ref{eq:psiIII3}c) is
consistent with the conduction inequality of a rigid
body undergoing no dissipation
(i.e., with the second term of~\ref{eq:psiIII3}c
being zero).
This comparison of Eqs.~%
(\ref{eq:2ndlaw})--(\ref{eq:conduction}),
(\ref{eq:2ndlaw2})--(\ref{eq:xilocal}),
and~(\ref{eq:psiIII3}c) implies
a positive nonlocal dissipation rate $\overline{\Phi}$,
\begin{equation}\label{eq:dissip2}
  \overline{\Phi}(\X) \:=\:
  \rho\xi(\X)\!\int_{\RRR}\!
    \varphi(\aXX) \theta(\Xp)\,\Vp
  \:=\:
  - \rho(\X)\!
  \int_{\RRR}\!
  \varphi(\aXX)
  \,
  \frac{\partial\Psi}{\partial \alpha_{i}}(\Xp)
  \dot{\alpha}_{i}(\Xp)\,\Vp
    \ge 0
    ,\quad
    \forall\X\in\Bc
\end{equation}
%
Note that the dissipation rate has a nonlocal character,
since the rates of internal variables, $\dot{\boldsymbol{\alpha}}$
and their stress-like thermodynamic forces
$\partial\Psi/\partial\alpha_{i}$
\emph{in the vicinity of} $\X$ govern the dissipation at $\X$.
The difference in dissipation between local and nonlocal media
will be greater when there is large
inhomogeneity during mechanical or thermal processes.
Continuing the example of elasto-plasticity,
the rate $\dot{\alpha}$ may be zero in the part of a point's
neighborhood that is unloading, while non-zero in the part that is
loading, with an average that differs substantially from
either separate rate.
\par
Proceeding in a manner
similar to Rice \cite{Rice:1971a}
but for nonlocal media,
the stress rate 
follows from Eq.~(\ref{eq:psiIII3}):
\begin{equation}
\label{eq:rate1}
\begin{aligned}
  \dot{T}_{ji}(\X) =\:
  & \frac{\dot{\rho}(\X)}{\rho(\X)}\,T_{ji}(\X)
  \:+\:
  \rho(\X)\!
    \int_{\RRR}\!
    \varphi(\aXX)
    \,\frac{\partial\Psi}{\partial F_{ik}}
    (\Xp)
    \dot{F}_{jk}(\Xp)
    \,\Vp
  \\
  &+ \rho(\X)\!
  \int_{\RRR}\!
  \varphi(\XX)
  \frac{\partial^{2}\Psi}{\partial F_{ik}\partial F_{\ell m}}(\Xp)
  F_{jk}(\Xp)\,
  \dot{F}_{\ell m}(\Xp)
  \,\Vp
  \\
  &+ \rho(\X)\!
  \int_{\RRR}\!
  \varphi(\XX)
  \frac{\partial^{2}\Psi}{\partial F_{ik}\partial\theta}(\Xp)
  F_{jk}(\Xp)\,
  \dot{\theta}(\Xp)
  \,\Vp
  \\
  &+ \rho(\X)\!
  \int_{\RRR}\!
  \varphi(\XX)
  \frac{\partial^{2}\Psi}{\partial F_{ik}\partial\alpha_{\ell}}(\Xp)
  F_{jk}(\Xp)\,
  \dot{\alpha}_{\ell}(\Xp)
  \,\Vp
\end{aligned}
\end{equation}
where $\dot{\rho}/\rho=L_{kk}$.
The first and second integrals are recognized as reversible
stress rates, and the stiffness modulus
$\partial^{2}\Psi/\partial F_{ik}\partial F_{\ell m}$ in the
second allows for possible damage and coupled dependence on the internal
variables when
$\partial^{3}\Psi/\partial F_{ik}\partial F_{\ell m}\partial\alpha_{n}$
is nonzero \cite{Collins:1997a}.
The first integral and the
leading term $\dot{\rho}T_{ji}/\rho=L_{kk}T_{ji}$
are negligible relative to the second integral when the
moduli are much larger than the current stress.

The third integral is the stress rate due to temperature change
at constant configuration
(the product of the coefficient of volumetric thermal expansion
and the isothermal bulk modulus).
In the case of small strains,
the stress rate $\dot{T}_{ij}(\X)$ is the sum of the
spatial averages, $\int_{\RRR}\varphi(\X,\Xp)\ldots\Vp$,
of the products of moduli
$\partial^{2}\Psi/\partial\ldots\partial\ldots$
and the rates of strain, temperature, and the internal variables.
Because of the directional restriction on
$\dot{\boldsymbol{\alpha}}$ in Eq.~(\ref{eq:dissip2}),
the last integral in Eq.~(\ref{eq:rate1})
is an irreversible stress
rate, associated with plastic and viscous effects.
\par
The constitutive function
$\overline{\Psi}(\mathbf{F},\theta,\boldsymbol{\alpha})$
is general in form,
provided that $\Psi$ preserves material frame-indifference
\cite{Noll:1959a},
respects a material's internal symmetries and constraints,
satisfies the second law of Eq.~(\ref{eq:psiIII3}c),
and satisfies the second law in
Eq.~(\ref{eq:3rd}) or~(\ref{eq:thetadot})
(not enforced by Eq.~\ref{eq:psiIII3}c).
Note that although $\overline{\Psi}$ appears as a nonlocal average
in Eq.~(\ref{eq:psiIII}),
it may, in principle, be \emph{local} in any of
$\mathbf{F}$, $\theta$, and $\boldsymbol{\alpha}$.
For example,
a material in which the response depends on the local value
of $\theta$ but is nonlocal with respect to
$\mathbf{F}$ and $\boldsymbol{\alpha}$
follows from the following form of $\overline{\Psi}$.
To focus attention on $\theta$, ellipses ``$\ldots$'' are inserted for
the $\mathbf{F}$ and $\boldsymbol{\alpha}$ terms in the first line
of this example:
\begin{equation}
\label{eq:localtheta}
\begin{aligned}
  \overline{\Psi}
  \big(
    \ldots,
    \theta(\X),
    \ldots
  \big)
  &=
  \int_{\RRR}
  \varphi(\aXX)\,
  \Psi
  \Big(
    \ldots,
    \int_{\RRR}
      \varphi^{-1}(|\Xp-\Xpp|)\,
      \theta(\Xpp)\,\Vpp,
    \ldots
  \Big)\,\Vp
  \\
  &=
  \int_{\RRR}
  \varphi(\aXX)
  \Psi
  \big(
    \mathbf{F}(\Xp),
    \theta(\X),
    \boldsymbol{\alpha}(\Xp)
  \big)\,\Vp
\end{aligned}
\end{equation}
in which $\varphi^{-1}$ is the kernel of the
right-inverse integral operator that recovers locality at $\X$,
such that
$\theta(\Xp)=\int\varphi^{-1}(|\Xp-\Xpp|)\theta(\Xpp)\,\Vpp$
and
$\theta(\X)=\int\varphi(\aXX)\int\varphi^{-1}(|\Xp-\Xpp|)\theta(\Xpp)\,\Vpp\,\Vp$.
In this manner,
the general form of $\overline{\Psi}$ admits models in which
damage and plasticity depend on an averaging of $\Psi$'s
dependence on $\boldsymbol{\alpha}$ and $\theta$
but upon a non-averaged
dependence on temperature or on the strain measure $\mathbf{F}$
(e.g., \cite{PijaudierCabot:1987a}).
The form also includes models in which the dependence
of $\overline{\Psi}$ on some
internal variables is local but on others is nonlocal
\cite{Benvenuti:2002a}.
\par
This section concludes with a cautionary note.
Although the constitutive form of $\psi(\X)$ described above
is quite general,
discretion is warranted with a simpler form in which free energy
$\psi(\X)$ is not the nonlocal average
$\overline{\Psi}(\mathbf{F},\theta,\boldsymbol{\alpha})$
of function $\Psi$ but is instead a function of the averaged strains,
temperatures, or internal variables.
This problematic function form $\Psi_{\text{opt}}$ is
\begin{equation}\label{eq:Psiopt}
  \psi(\X) = \Psi_{\text{opt}}
  \big(
    \overline{\mathbf{F}}(\X),
    \overline{\theta}(\X),
    \overline{\boldsymbol{\alpha}}(\X)
  \big)
\end{equation}
in which the overlines denote spatially averaged quantities:
$\overline{z}(\X)=\int_{\RRR}\varphi(\aXX)z(\Xp)\Vp$.
Caution also applies to constitutive models in which
the rate function $\mathbf{g}$,
yield function, flow function, dissipation potential,
damage activation function, etc.
depend on the averaged strain or other averaged
state variables.
Continuing the previous example of
an elasto-plastic body in which
separate parts of the neighborhood of a point $\mathbf{x}$
undergo loading and unloading,
Eq.~(\ref{eq:Psiopt}) claims that the stress at $\X$ is determined
by the average strain in the neighborhood;
whereas Eq.~(\ref{eq:psiIII}) uses the average
of $\Psi(\X)$ in the neighborhood.
Although form~(\ref{eq:Psiopt}) may seem
more direct, its shortcomings are revealed by
proceeding as with
Eqs.~(\ref{eq:psiIIa})--(\ref{eq:ineqIIIc}),
resulting in the inequality
\begin{equation}\label{eq:ineqOpt}
  \begin{aligned}
  &
    \int_{\RRR}\!
    \varphi(\aXX)
    \left[
    T_{ji}(\X)\delta_{kj}
    L_{ik}(\Xp)
    -
    \rho 
    \,\frac{\partial\Psi_{\text{opt}}}{\partial \overline{F}_{ij}}
    (\X)
    \Big(
    L_{ik}(\Xp)F_{kj}(\Xp)
    +
    v_{i}(\Xp)F_{kj,k}(\Xp)
    \Big)
    \right]
    \,\Vp
  \\
  &
  -
  \int_{\RRR}\!
    \varphi(\aXX)
    \left[
    \rho(\X)\eta(\X)
    +
    \rho(\X)
    \,\frac{\partial\Psi_{\text{opt}}}{\partial \overline{\theta}}(\X)
    \right]
    \dot{\theta}(\Xp)\,\Vp
 \\
  &
  +
  \bigg(
  -
    p_{i}(\X)\int_{\RRR}\!
    \varphi(\aXX)
    \theta_{,i}(\Xp)\,\Vp
  -
  \rho(\X)
  \frac{\partial\Psi_{\text{opt}}}{\partial\overline{\alpha}_{i}}(\X)
  \int_{\RRR}\!
    \varphi(\aXX)
    \dot{\alpha}_{i}(\Xp)\,\Vp
  \bigg)
  \;\ge
  0
  \end{aligned}
\end{equation}
Unlike Eq.~(\ref{eq:ineqIIIc}), the term $L_{ik}(\Xp)$ does
not appear as a common factor in the first line,
because of the stray term $v_{i}(\Xp)F_{kj,k}(\Xp)$.
As a result, stress $\mathbf{T}$ cannot be properly defined
as a derivative of $\Psi_{\text{opt}}$,
unless the stray term is included in the dissipation
inequality.
Moreover, if the second line in Eq.~(\ref{eq:ineqOpt})
must equal zero for arbitrary distributions $\dot{\theta}(\Xp)$,
entropy is simply
$\eta(\X)=-\partial\Psi_{\text{opt}}(\X)/\partial\overline{\theta}$:
that is, $\eta(\X)$ reverts to a \emph{local} derivative 
of $\Psi_{\text{opt}}(\X)$.
\par
The difference in the Helmholtz energies of
Eqs.~(\ref{eq:psiIII}) and~(\ref{eq:Psiopt}) is subtle,
and for simple constitutive forms,
the difference may be inconsequential,
but care must be exercised when adopting constitutive forms
for stress and entropy that involve a simple averaging
of strains, temperatures, or internal variables.
\par
In summary, the complete set of field equations for stress and motion
is comprised of the free energy function $\psi(\X)$
and its average $\overline{\Psi}(\X)$
in Eq.~(\ref{eq:psiIII}),
the constitutive stress rate of Eq.~(\ref{eq:rate1}),
the balance condition of
Eqs.~(\ref{eq:equil1})--(\ref{eq:FFF}),
the compatibility condition
$\mathbf{F}=\boldsymbol{\nabla}_{\!_\mathbf{X}}$\raisebox{1.5pt}{$\boldsymbol{\chi}$},
the second-law restriction of
Eq.~(\ref{eq:dissip2}),
and any imposed displacement constraints.
The solution of an example problem is given
in the next section.
\section{\large Example}\label{sec:examples}
\begin{figure}
\centering
\includegraphics{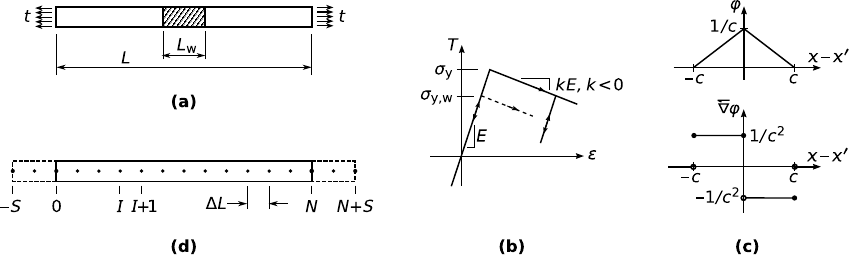}
\caption{One-dimensional bar loaded by stretching: 
(a)~length $L$ with a central
weakened section of length $L_{\text{w}}$,
(b)~linear elasto-plastic material with softening,
(c)~mollifier $\varphi({\aXX})$ and its derivative
$\boldsymbol{\nabla}\varphi$, and
(d)~discretization for the problem's finite-difference solution as
a linear complementarity problem.
\label{fig:Bar}}
\end{figure}

We analyze a stretched prismatic rod having a
weakened central section (Fig.~\ref{fig:Bar}a).
Similar analyses have been reported
for elasto-plastic \cite{Bazant:1988a,Nilsson:1997a,Jirasek:2003a}
and damaging \cite{PijaudierCabot:1987a,Bazant:2002a}
materials,
in which nonlocal averaging is applied to internal variables.
We instead introduce nonlocality by applying the nonlocal
balance conditions (Section~\ref{sec:nonlocalstatics}),
a nonlocal free energy (Section~\ref{sec:energybalance}),
and a nonlocal constitutive operator (Section~\ref{sec:constit}).
\par
The material model is rate-independent
elasto-plasticity
with softening in the form of a decaying yield limit
(Fig.~\ref{fig:Bar}b).
The full and weakened section lengths are $L$ and $L_{\text{w}}$,
with the weakened section having a
yield limit $\sigma_{\text{y, w}}$ that is smaller than the limit
 $\sigma_{\text{y}}$ elsewhere along the rod.
We apply a simple (almost crude)
nonlocal mollifier $\varphi(\aXX)$ having radius $c$ with a
linear decay from its central density of $1/c$,
such that its volume $\int\!\varphi(\aXX)\,dv'=1$,
as required by Eq.~(\ref{eq:unity}) (Fig.~\ref{fig:Bar}c).
\par
Isothermal conditions, small strains,
and slow loading are assumed,
so that the material's free energy $\Psi$ is a function
of strain $\varepsilon=F_{11}-1$ and a single internal variable
$\alpha$, taken as the plastic strain.
Consistent with the form in Eq.~(\ref{eq:psiIII}),
$\Psi$ and its spatial average are
\begin{equation}\label{eq:PsiEx}
\begin{aligned}
  \rho\Psi\big(x',\varepsilon(x'),\alpha(x')\big) &= 
  \frac{1}{2} E(x') \Big(\varepsilon(x') - \alpha(x')\Big)^{2}
  +
  \frac{1}{2} a H(x')\big(\alpha(x')\big)^{2}
  \\
  \rho\overline{\Psi}\big(x,\varepsilon(x),\alpha(x)\big) \,&=
  \int_{\R}
  \varphi(\Axx)
  \left[
  \frac{1}{2} E(x') \Big(\varepsilon(x') - \alpha(x')\Big)^{2}
  +
  \frac{1}{2} a H(x')\big(\alpha(x')\big)^{2}
  \right]
  \,dx'
\end{aligned}
\end{equation}
in which $E$ and $H$ are the elastic and plastic moduli,
$\alpha$ is the plastic strain, and parameter $a\in[0,1]$
is~0 for purely isotropic hardening and~1
for purely kinematic hardening.
Note that Eq.~(\ref{eq:PsiEx}\textsubscript{2})
averages the full energy function $\Psi$ rather
than simply averaging the strain $\varepsilon$ or variable $\alpha$.
The stress and stress rate, from
Eqs.~(\ref{eq:psiIII3}a) and~(\ref{eq:rate1}), are
\begin{equation}
\label{eq:TTdot}
\begin{aligned}
  T(x) &=
  \int_{\R}
  \varphi(\Axx)
  E(x') \Big(\varepsilon(x') - \alpha(x')\Big)\,dx'
  \\
    \dot{T}(x)
    &=
    \int_{\R}
    \varphi(\Axx)
    E(x')\,\dot{\varepsilon}(x')\,dx'
    -
    \int_{\R}
    \varphi(\Axx)
    E(x')\,\dot{\alpha}(x')\,dx'
\end{aligned}
\end{equation}
We use a nonlocal model,
as proposed by Ba{\v{z}}ant and Lin \cite{Bazant:1988a},
in which the
nonlocal kinematic terms $\varphi(\Axx)\varepsilon(x')$
and $\varphi(\Axx)\dot{\varepsilon}(x')$
in this equation are replaced
with the local values $\varepsilon(x)$ and  $\dot{\varepsilon}(x)$.
This alteration is consistent with the treatment described 
in Eq.~(\ref{eq:localtheta}),
enlisting an inverse mollifier $\varphi^{-1}$
and replacing $\varepsilon(x')$ with
$\int\varphi^{-1}(x',x'')\varepsilon(x'')\,dx''$).
Variable $\alpha$ retains its nonlocal form,
as in Eq.~(\ref{eq:TTdot}).
\par
To complete the constitutive description,
the free energy $\psi$ in Eq.~(\ref{eq:Psiopt})
is augmented with a dissipation function $\Phi$.
For the rod,
the local rate $\Phi$ at $x$
depends only on $\alpha$
and its rate $\dot{\alpha}$, or
$\Phi(x,\alpha,\dot{\alpha})$.
To broaden the discussion,
we briefly consider a three-dimensional solid in which
$\boldsymbol{\alpha}$ is a \emph{list} of variables.
The nonlocal, averaged function $\overline{\Phi}(\X)$ is
\begin{equation}
  \overline{\Phi}(\X,\boldsymbol{\alpha},\dot{\boldsymbol{\alpha}})
  =
  \int_{\RRR}\varphi(\aXX)
  \Phi(\Xp,\boldsymbol{\alpha},\dot{\boldsymbol{\alpha}})
  \,\Vp
\end{equation}
Proceeding in the manner of Ziegler \cite{Ziegler:1987a}
and Collins and Houlsby \cite{Collins:1997a},
the material is assumed rate-independent, so that $\overline{\Phi}$
is positive homogeneous of degree one in rate $\dot{\boldsymbol{\alpha}}$,
such that
$\overline{\Phi}(\X;\boldsymbol{\alpha},\lambda\dot{\boldsymbol{\alpha}})
=\lambda\overline{\Phi}(\X;\boldsymbol{\alpha},\dot{\boldsymbol{\alpha}})$,
$\forall\lambda\in\R^{+}$.
Euler's theorem then requires that
\begin{equation}\label{eq:dissip4}
  \overline{\Phi}(\X,\boldsymbol{\alpha},\dot{\boldsymbol{\alpha}})
  =
  \frac{\partial
    \overline{\Phi}(\X,\boldsymbol{\alpha},\dot{\boldsymbol{\alpha}})}
  {\partial\dot{\alpha}_{i}}
  \:
  \dot{\alpha}_{i}
  =
  \int_{\RRR}\varphi(\aXX)\,
  \frac{\partial\Phi}
  {\partial\dot{\alpha}_{i}}(\Xp)
  \,\dot{\alpha}_{i}(\Xp)
  \,\Vp
\end{equation}
Subtracting Eq.~(\ref{eq:dissip2})
from Eq.~(\ref{eq:dissip4}\textsubscript{2}),
\begin{equation}\label{eq:dissip5}
  \int_{\RRR}
    \varphi(\aXX)
    \left(
      \rho(\X)
      \frac{-\partial\Psi}{\partial\alpha_{i}}(\Xp)
      -
      \frac{\partial\Phi}{\partial\dot{\alpha}_{i}}(\Xp)
    \right)
  \,\dot{\alpha}_{i}(\Xp)\,\Vp
  = 0
  ,\quad
  \forall\X\in\Bc
\end{equation}
This condition enforces consistency
of the functions $\Psi$ and $\Phi$ by requiring an
$\varphi$-averaged orthogonality of $\dot{\boldsymbol{\alpha}}$
and the difference in the equation's integrand.
The stronger Ziegler orthogonality principle \cite{Ziegler:1987a}
asserts, in its nonlocal form, an averaged value of zero
for each separate $i$-th variable:
\begin{equation}\label{eq:phipsi}
  \int_{\RRR}
    \varphi(\aXX)
    \left(
      \rho(\X)
      \frac{-\partial\Psi}{\partial\alpha_{i}}(\Xp)
      -
      \frac{\partial\Phi}{\partial\dot{\alpha}_{i}}(\Xp)
    \right)
  \,\Vp
  = 0
  ,\quad
  \forall\X\in\Bc,\; i\in\{1,2,\ldots,m\}
\end{equation}
where $m$ is the length of the variable list $\boldsymbol{\alpha}$.
\par
We now return to the one-dimensional example and introduce
a dissipation function $\Phi$
and its nonlocal counterpart $\overline{\Phi}$
to supplement the free energy in Eq.~(\ref{eq:Psiopt}):
\begin{equation}
\begin{aligned}
  \Phi(x';\alpha,\dot{\alpha})
  &=
  \Big(
  \sigma_{\text{y}}(x')
  +
  (1-a)H(x')\,|\alpha(x')|
  \Big)
  \,
  |\dot{\alpha}(x')|
  \\
  \overline{\Phi}(x;\alpha,\dot{\alpha})
  &=
  \int_{\R}
  \varphi(|x-x'|)\,\Phi(x')\,\Vp
\end{aligned}
\end{equation}
which completes the constitutive description.
In keeping with the consistency condition of Eq.~(\ref{eq:phipsi}),
the evolution of variable $\alpha$
is restricted by a nonlocal yield function
$\overline{f}(\varepsilon,\alpha)$
which must remain non-positive:
\begin{align}\label{eq:yield}
  &\bar{f}(x;\varepsilon,\alpha)
  =
  \int_{\R}\varphi(\Axx)f(x';\varepsilon,\alpha)\,dx'
  \;\le 0
  \\
  &
  \begin{aligned}
  f(x';\varepsilon,\alpha)
  &=
  -\rho(x)\,
  (\partial\Psi(x')/\partial\alpha) - 
  \rho(x)\,(\partial\Phi(x')/\partial\dot{\alpha})
  \\
  &=
  \Big|E(\varepsilon'-\alpha')-aH\alpha'\Big|
  -\Big(\sigma_{\text{y}}+(1-a)H\,|\alpha'|\Big)\,\text{sgn}(\dot{\alpha})
  \end{aligned}
\end{align}
noting that the non-positive condition is placed on the
full, averaged yield function rather than its local counterpart.
In these equations, $\sigma_{\text{y}}$ is the initial yield stress,
parameter $a$ is between 0 (pure isotropic hardening)
and 1 (pure kinematic hardening),
and $H=\partial f/\partial\alpha=kE/(1-k)$ is the plastic modulus,
where $k$ is the slope of the plastic loading branch
(Fig.~\ref{fig:Bar}b) (e.g., see \cite{Lubliner:1990a}).
Both $H$ and $k$ are negative for the strain softening of this example.
We introduce the plastic flow direction $h_{1}$
and the isotropic yield direction $h_{2}$,
both limited to $h_{1},h_{2}\in\{-1,0,1\}$,
\begin{equation}\label{eq:hs}
  h_{1} = \text{sgn}\!
  \left(
    E(\varepsilon - \alpha') - a H \alpha'
  \right)
  ,\quad
  h_{2} = \text{sgn}\!
  \left(\alpha'\right)
\end{equation}
and both are zero when $\bar{f}<0$.
When at the yield limit, $\bar{f}=0$, and compatibility with 
Eq.~(\ref{eq:yield}) requires a non-positive rate $d\bar{f}/dt$:
\begin{equation}\label{eq:fdot}
\begin{aligned}
  \bar{f}(x)=0 \;\Rightarrow\;
  \dot{\bar{f}}(x)
  &=
  \int_{\R}\varphi(\Axx)
  \left[
    \frac{\partial f}{\partial\varepsilon}\dot{\varepsilon}
    +
    \frac{\partial f}{\partial\alpha}\dot{\alpha}
  \right]_{x'}
  \,dx'
  \;\le 0
  \\
  &=
  \int_{\R}
  \varphi(\Axx)
  \Big[
  Eh_{1}\dot{\varepsilon}
  -
  \big(
  (E+aH)h_{1}
  +
  (1-a)Hh_{2}
  \big)\dot{\alpha}
  \Big]_{x'}\,dx'
  \;\le 0
\end{aligned}
\end{equation}
in which the terms in brackets are functions of $x'$.
The equation provides the evolution $\dot{\alpha}=g(\varepsilon,\alpha)$
described in the first paragraph of Section~\ref{sec:constit}.
The signs of $h_{1}$ and $h_{2}$ in Eq.~(\ref{eq:hs})
ensure non-negative dissipation,
as required by Eq.~(\ref{eq:dissip2}).
\par
Viewing variable $\alpha$ as the the plastic strain and
$h_{1}$ as the flow direction of $\dot{\alpha}$,
the advance of variable $\alpha$ in Eq.~(\ref{eq:fdot})
is restricted by the complementarity condition
\begin{equation}
  \bar{f}(x)<=0
  \quad\Rightarrow\quad
  h_{1}\dot{\alpha} \ge 0,
  \quad
  \dot{\bar{f}}(x) \le 0,\quad
  h_{1}\dot{\alpha}\,\dot{\bar{f}}(x) = 0
\end{equation}
whenever the plastic limit $\bar{f}(x)=0$ is reached.
\par
Note that the free energy in Eq.~(\ref{eq:PsiEx})
and the yield restrictions of Eqs.~(\ref{eq:yield})
and~(\ref{eq:fdot})
are consistent with
the constitutive form in Eq.~(\ref{eq:psiIII}),
rather than the problematic Eq.~(\ref{eq:Psiopt}).
With the latter,
the stress rate $\dot{T}(\X)$ in
Eq.~(\ref{eq:TTdot}\textsubscript{2})
would be
$\int E(x)\varphi(\Axx)\dot{\alpha}(x')\,dx'$,
and instead of averaging the full yield function,
as in Eq.~(\ref{eq:yield}),
the function's individual arguments, $\varepsilon$ and $\alpha$,
would be averaged as in the flawed form of Eq.~(\ref{eq:Psiopt}).
Although the differences are subtle,
we comply with the energy-consistent form~(\ref{eq:psiIII}).
\par
The following values are used:
$L = 1$, $\sigma_{\text{y}}=1$, $E=1000$, $k=-0.05$
(softening), $a=0$ (isotropic softening), and $N=200$
(number of elements in the bar's length $L$,
Fig.~\ref{fig:Bar}d).
The central weakened section is assigned length
$L_{\text{w}}=0.02L$,
and its reduced yield strength is 90\%
of the surrounding bar's yield strength:
$\sigma_{\text{y, w}}= 0.9\sigma_{\text{y}}$.
Mollifier $\varphi$ is given radius $c=0.05L$.
\par
The rod is stretched in increments $dL=L\,\dot{\varepsilon}\,dt$,
and the resulting displacements
along the bar's length are solved as a
linear complementarity problem (LCP)
\cite{Maier:1970a}.
To demonstrate resilience of the localization pattern,
bars of lengths $0.5L$ and $2L$
are also analyzed, with
all three sharing the same widths
$L_{\text{w}}$ and $c$
(the grid discretization number $N$ is scaled in proportion to length).
Details of the numerical method are given in
\ref{app:example}.
\par
Although nonlocal and purely local models give the same
result when elongation is within the bar's elastic
range,
a local model is ill-posed
when stress reaches the yield limit and softening
commences, resulting in spurious solutions as localization
becomes concentrated in a zone of vanishing length
\cite{Bazant:1976a,Bazant:2002a}
(for the LCP,
the coefficient matrix $[\mathbf{M}]$ becomes neither a P-matrix
nor an N-matrix, yielding no solution, \ref{app:example}).
\par
\begin{figure}
  \centering
  \raisebox{5ex}{%
  \includegraphics{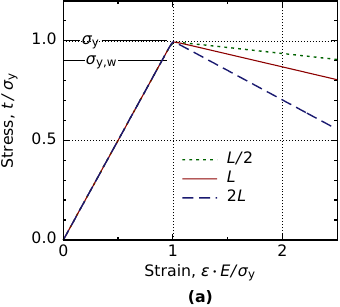}}\quad\quad\ 
  \includegraphics{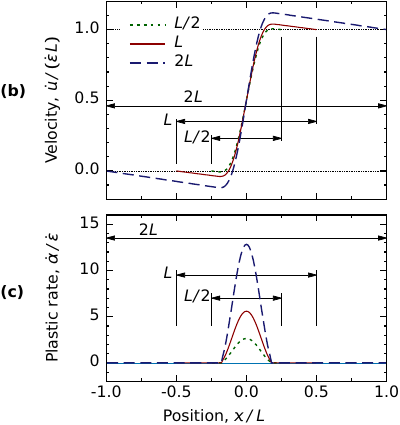}
  \caption{%
     Results for bars of the three lengths,
     $L/2$, $L$, and $2L$:
     (a)~stress and strain during controlled elongation,
     (b)~displacement rates
        $\protect\dot{u}/(\protect\dot{\varepsilon}L)$
        along lengths of the bars
        at strain $\varepsilon=2\cdot\sigma_{\text{y}}/E$,
     and (c)~rates of the internal variable
       $\protect\dot{\alpha}/\protect\dot{\varepsilon}$
       (plastic strain rate)
       along length of bar.
       \label{fig:Results}}
\end{figure}

Stress--strain results are shown
in Fig.~\ref{fig:Results}a for bars of lengths
$0.5L$, $L$, and $2L$.
The figure shows that the bars
soften after reaching a peak strength.
Although softening commences within the weakened section,
the peak occurs well
above the weakened yield strength $\sigma_{\text{y,w}}=0.90$
of the central $0.02L$ section, with the peak
only slightly below the yield strength $\sigma_{\text{y}}$
of the remaining $0.98L$ of the bar.
The width of the mollifier, being $2c=0.10L$, causes the larger
surrounding yield strength to predominate within the weakened
central section.
For stretching past the peak,
the rate of stress reduction increases with the bar's length.
\par
Figure~\ref{fig:Results}b shows displacement rates along lengths 
of the three bars at a strain well past the peak stress,
$\varepsilon=2\sigma_{\text{y}}/E$.
Rates at the fixed left end are zero and at the moving right end
are $1\cdot \dot{\varepsilon}L$.
With each bar, stretching is concentrated around the weakened center;
whereas elastic unloading occurs nearer to the bars' ends.
The width of the central localization is nearly the same
for the three bars,
as this characteristic length is established by the mollifier's width.
The localized zone is about $0.36L$,
roughly four times the mollifier's width of $0.10L$.
Strain localization is also apparent in the rate profiles of the
internal variable, $\dot{\alpha}$
(the plastic strain rate, Fig.~\ref{fig:Results}c),
with all three bars exhibiting the same width of plastic softening.
We also checked sensitivity to grid fineness
by using discretizations with $N$ of 40, 100, and 200,
and we found that
results were insensitive to the bar's discretization,
as Fig.~\ref{fig:Results} is nearly the same for all three $N$ values.
\par
These results conform to other examples of elasto-plastic rods
\cite{Bazant:1988a,Nilsson:1997a,Jirasek:2003a},
which also took care to average the full constitutive operator rather than
merely the strain (a distinction discussed with
Eqs.~\ref{eq:psiIII} and~\ref{eq:Psiopt}).
In the examples, however,
no body force $b(x)$ is applied along the rod's length,
so that the proper nonlocal balance condition of
Eqs.~(\ref{eq:FFF}) and~(\ref{eq:mechBalance}), although applied
in our example (see Eqs.~\ref{eq:Tb}--\ref{eq:Enon}),
yields the same result
as the local condition (Eq.~\ref{eq:equillocal}\textsubscript{1})
used in the previous examples.
\section{Conclusions}
Beginning with a formal application of the principle of virtual power,
a consistent thermomechanics is developed for nonlocal continua,
with a particular focus on processes that are sufficiently slow
to minimize additional meso-scale kinetic energy.
The resulting mechanics admits non-smooth, square-integrable force
and displacement fields as well as non-smooth boundary surfaces whose normal directions might be discontinuous or difficult to discern.
Central to this framework is an averaging kernel function that
conveys the influence on a given point
of its surrounding neighborhood.
The kernel introduces an inherent length scale that
enables the modeling of phenomena exhibiting
a characteristic size or thickness,
reduces sensitivity to a discretization's fineness, and
smooths the effects of abrupt material discontinuities or geometric 
irregularities.
\par
Nonlocality enters the field equations
in three distinct but coherent ways:
in the balance laws, in the conservation laws
(including the first and second laws of thermodynamics),
and in a material's constitutive structure.
The principle of virtual power establishes the nonlocal balance
constraints on force fields~--- both mechanical and thermal~---
and imposes limitations on the averaging kernel function.
When more restrictive conditions are placed on the kernel function,
more relaxed conditions apply to the force fields, and vice versa.
The global, external
balance equations for an entire body reduce to those of
classical local media when modest restrictions are applied to
the kernel:
when the kernel's volume is constant
throughout the body and the kernel is an even function
of its vector argument.
The local, internal balance equations, however,
differ for nonlocal and classical continua.
Notably, the balance equations of stress and traction
and of stress and body force~---
conditions that are separate for local continua~---
collapse into a single condition for a nonlocal continuum.
Consequently, the traction at a surface point is balanced
by both the stress and body force in the neighborhood of the point.
In a similar manner,
entropy flux at a surface point depends on entropy flow
and entropy supply in the point's neighborhood.
\par
Second, energy conservation and the first and second laws
of thermodynamics are derived
in a manner similar to that for local continua,
but the result is a
nonlocal form of the Helmholtz energy
together with nonlocal restrictions imposed by the second law.
The derivation includes the kinetic energies of all motions~---
both macroscopic and molecular~---
within a point's neighborhood, and this energy includes the averaged, 
prevailing motion within the reach (support)
of the kernel function plus any meso-scale kinetic energy
arising from residual motions not captured by this average.
In developing the thermomechanics of Section 3, we assumed
that this residual kinetic energy $\dot{R}_{K}$ is insignificant.
This assumption is reasonable for slow processes
but might not apply 
to rapid turbulent flows, where the kinetic energy
of the prevailing motion, $P_{A,\text{mech}}$,
might neglect kinetic energy
at scales finer than the kernel's support but larger
than the molecular (thermal) scale accounted by the
entropy rate. 
\par
As a constitutive framework,
we adopted an internal state variable approach,
but one endowed with nonlocal character by averaging
the local free energy over the reach of the kernel function.
In this formulation, the stress, entropy, and dissipative forces
are obtained as the averaged derivatives
of the Helmholtz energy function with respect to strain,
temperature, and internal variables respectively.
We cautioned against other approaches that simply average
the state variables (i.e., strain, temperature, and internal variables) 
rather than the thermodynamically consistent approach
of averaging the full energy function.
Other than limitations placed by the second law,
general forms of the energy function are admissible,
including forms that explicitly exempt some of the state variables
from the averaging of the energy function.
An example is given in Section~\ref{sec:examples},
in which an internal variable but not the strain are averaged
with the free energy.
The results illustrate that the averaging kernel
imparts a characteristic length that controls the width
of a localization zone.
\appendix
\section{Example numerics}\label{app:example}
The bar of Section~\ref{sec:examples}
is divided into $N$ elements of length
$\Delta L = L/N$ (Fig.~\ref{fig:Bar}d).
An additional $S$
elements are appended
at each end, encompassing the radius $c$ of mollifier $\varphi$ when
it is centered at the bar's ends, $S=\text{ceil}(Nc/L)$.
Nodes at locations $x_{I}$ are labeled from $-S$ to $N+2S$,
and elements within the bar are labeled $1$ to $N$, with element
$i$ between nodes $I$ and $I+1$.
\par
The local strain rate $\dot{\varepsilon}_{I}$ 
of element $I$ is estimated from
the nodal velocities $\dot{u}_{I}$ at the element's ends,
$\dot{\varepsilon}_{I}=(\dot{u}_{I}-\dot{u}_{I-1})/\Delta L$, 
and the $N$ rates are collected
as the product of an $N\times (N+1+2S)$
local kinematics matrix
$[\mathbf{B}]$ and the vector of nodal velocities
$[\dot{\mathbf{u}}]$:
\begin{equation}\label{eq:epsBu}
  [\dot{\boldsymbol{\varepsilon}}] = [\mathbf{B}][\dot{\mathbf{u}}]
\end{equation}
noting that strain rates $[\dot{\boldsymbol{\varepsilon}}]$
are local to the elements.
The corresponding stress rates $\dot{T}_{I}$ in the elements,
detailed below, must satisfy the balance Eqs.~(\ref{eq:FFF})
and~(\ref{eq:mechBalance}), discretized as
\begin{equation}\label{eq:Tb}
  [\lpoon{\boldsymbol{\nabla}}\boldsymbol{\varphi}]
  [\dot{\mathbf{T}}]
  =
  [\boldsymbol{\varphi}_{1}][\rho\mathbf{b}]
\end{equation}
where the $(N+1+2S)\times N$ nonlocal statics matrix
$[\lpoon{\boldsymbol{\nabla}}\boldsymbol{\varphi}]$
effects the kernel $\varphi_{,j}(\XX)$
in Eq.~(\ref{eq:FFF}),
and the $(N+1+2S)\times(N+1+2S)$
matrix $[\boldsymbol{\varphi}_{1}]$ averages
body forces $\rho b$ by applying the kernel
$\varphi(\XX)$ in Eq.~(\ref{eq:FFF}).
Elements of the two matrices are assigned the values
\begin{equation}
  \lpoon{\boldsymbol{\nabla}}\varphi_{IJ}
  =
  \int_{x^{\prime}_{J}-\Delta L/2}^{x^{\prime}_{J}+\Delta L/2}\!
  \frac{d}{dx}\varphi(|x_{I}-x^{\prime}_J|)\,dx^{\prime}
  \quad
  \text{and}
  \quad
  \varphi_{1,\:IJ}
  =
  \int_{x^{\prime}_{J}-\Delta L/2}^{x^{\prime}_{J}+\Delta L/2}\!
  \varphi(|x_{I}-x^{\prime}_J|)\,dx^{\prime}
\end{equation}
Moduli $E$ are collected in an $N\times N$
diagonal matrix $[\mathbf{E}]$,
and a second averaging matrix $[\boldsymbol{\varphi}_{2}]$ is
applied to $[\mathbf{E}]$, as in Eq.~(\ref{eq:TTdot}):
\begin{equation}\label{eq:Enon}
  [\mathbf{E}]\rightarrow [\boldsymbol{\varphi}_{2}][\mathbf{E}]
  ,\quad
  \varphi_{2,\:IJ}
  =
  \int_{x^{\prime}_{J}}^{x^{\prime}_{J}+\Delta L}\!
  \varphi(|x_{I}-x^{\prime}_J|)\,dx^{\prime}
\end{equation}
in which $\varphi_{2,\:IJ}$ is centered on element $I$,
whereas $\varphi_{1,\:IJ}$ is centered on node $I$.
\par
To accommodate the standard form of a linear complementarity
problem (LCP),
we introduce a variable $\dot{\lambda}$ for each element
that has reached the yield limit, $\hat{f}=0$.
We designate $M$ as the number of the $N$ elements that
have reached the limit.
By defining
$\dot{\lambda}$ as the product
$\dot{\lambda} = h_{1}\dot{\alpha}$, this multiplication
by the yielding direction $h_{1}$ assures a non-negative rate,
$\dot{\lambda}\ge 0$
(note that $\dot{\alpha}$ can be positive, negative,
or zero).
The $\dot{\lambda}$ of those $M$ elements that are
at the yield limit are arranged
as an $M\times 1$ vector $[\dot{\boldsymbol{\lambda}}]$,
which is related to
the $N\times 1$ vector of rates $[\dot{\boldsymbol{\alpha}}]$
as the product
\begin{equation}\label{eq:alphalambda}
  [\dot{\boldsymbol{\alpha}}]
  =
  [h_{1}][\mathbf{Q}][\dot{\boldsymbol{\lambda}}]
\end{equation}
where $[h_{1}]$ is an $N\times N$ diagonal matrix filled with
the elements' $h_{1}$ flow directions.
Those rows in the $N\times M$ matrix
$[\mathbf{Q}]$ that correspond to non-yielding elements
are filled with zeros; whereas,
in the $I$ row of an element at the yield limit for $\lambda_{J}$,
element $Q_{IJ}=1$, with all other elements being zero.
\par
Displacement conditions on boundary nodes
are enforced with an $R\times(N+1+2S)$
constraint matrix $[\mathbf{C}]$:
\begin{equation}\label{eq:Cdotc}
  [\mathbf{C}][\dot{\mathbf{u}}] = [\dot{\mathbf{c}}]
\end{equation}
in which each of the $R$ rows of
$[\mathbf{C}]$ and $[\dot{\mathbf{c}}]$
effects a velocity constraint.
In the example, $R=2+2S$, with zero velocity at node
0 and at the $S$ supplemental nodes on the left end,
and with velocity
$\dot{c} = L\,\dot{\varepsilon}\,dt$ at node $N$ and at the
$S$ supplemental nodes on the right end.
\par
The problem is solved as a linear complementarity problem (LCP)
of the form
\begin{subequations}
\label{eq:LCP}
  \begin{align}
    [\mathbf{M}][\dot{\boldsymbol{\lambda}}]
    + [\dot{\mathbf{q}}]
    &\;\ge\; 0
    \\
    [\dot{\boldsymbol{\lambda}}]
    &\;\ge\; 0
    \\
    [\dot{\boldsymbol{\lambda}}]^{\text{T}}
    \left(
      [\mathbf{M}][\dot{\boldsymbol{\lambda}}]
      + [\dot{\mathbf{q}}]
    \right)
    &\;=\;
    0
  \end{align}
\end{subequations}
As an intermediate step in finding $[\mathbf{M}]$
and $[\dot{\boldsymbol{\lambda}}]$,
consider the mixed LCP,
\begin{subequations}\label{eq:form2b}
\begin{align}
  \renewcommand*{\arraystretch}{1.2}
  \left[
      \begin{array}{@{\extracolsep{0ex}}c;{1pt/1pt}c@{\extracolsep{\fill}}}
        \mathbf{S}_{\mathbf{xx}}
        & \mathbf{S}_{\mathbf{x}\boldsymbol{\lambda}}
        \\\hdashline[1pt/1pt]
        \mathbf{S}_{\boldsymbol{\lambda}\mathbf{x}}
        & \mathbf{S}_{\boldsymbol{\lambda}\boldsymbol{\lambda}}
        \\\hdashline[1pt/1pt]
        \mathbf{C} & \mathbf{0}
        \\\hdashline[1pt/1pt]
        \mathbf{0}
        & \mathbf{I}
      \end{array}
  \right]
  \left[
    \begin{array}{l@{\extracolsep{\fill}}}
        \dot{\mathbf{u}}\,
        \\\hdashline[1pt/1pt]
        \rule{0pt}{12pt}\dot{\boldsymbol{\lambda}}\,
    \end{array}
  \right]
  &
  \begin{array}{c}
    =
    \\
    \ge
    \\
    =
    \\
    \ge
  \end{array}
  \left[
    \begin{array}{c@{\extracolsep{\fill}}}
     \rule[-4pt]{0pt}{16pt}\rho\dot{\mathbf{b}}
     \\\hdashline[1pt/1pt]
     \rule[-4pt]{0pt}{14pt}\mathbf{0}
     \\\hdashline[1pt/1pt]
     \rule[-4pt]{0pt}{14pt}\dot{\mathbf{c}}
     \\\hdashline[1pt/1pt]
     \rule[-4pt]{0pt}{14pt}\mathbf{0}
    \end{array}
  \right]
  \\
  \left[\dot{\boldsymbol{\lambda}}\right]^{\text{T}}
  \Big[
      \begin{array}{@{\extracolsep{0ex}}c;{1pt/1pt}c@{\extracolsep{\fill}}}
        \mathbf{S}_{\boldsymbol{\lambda}\mathbf{x}}
        & \mathbf{S}_{\boldsymbol{\lambda}\boldsymbol{\lambda}}
      \end{array}
  \Big]
  \left[
    \begin{array}{l@{\extracolsep{\fill}}}
        \dot{\mathbf{u}}\,
        \\\hdashline[1pt/1pt]
        \rule{0pt}{11pt}\dot{\boldsymbol{\lambda}}\,
    \end{array}
  \right]
  &\, = \; 0
\end{align}
\end{subequations}
with the four stiffnesses
\begin{subequations}
\label{eq:subeqs}
  \begin{align}
    [\mathbf{S}_{\mathbf{xx}}]
    &=
    [\boldsymbol{\varphi}_{1}]^{-1}
    [\lpoon{\boldsymbol{\nabla}}\boldsymbol{\varphi}]
    \Big(
      [\boldsymbol{\varphi}_{2}]
      \text{ or }
      [\,\mathbf{I}\,]
    \Big)
    [\mathbf{E}]
    [\mathbf{B}]
    \\
    [\mathbf{S}_{\mathbf{x}\boldsymbol{\lambda}}]
    &=
    -
    [\boldsymbol{\varphi}_{1}]^{-1}
    [\lpoon{\boldsymbol{\nabla}}\boldsymbol{\varphi}]
    [\boldsymbol{\varphi}_{2}]
    [\mathbf{E}]
    [h_{1}]
    [\mathbf{Q}]
    \\
    [\mathbf{S}_{\boldsymbol{\lambda}\mathbf{x}}]
    &=
    -
    [\mathbf{Q}]^{\text{T}}
    [\boldsymbol{\varphi}_{2}]
    [\mathbf{E}]^{\text{T}}
    [h_{1}]^{\text{T}}
    [\mathbf{B}]
    \\
    [\mathbf{S}_{\boldsymbol{\lambda}\boldsymbol{\lambda}}]
    &=
    [\mathbf{Q}]^{\text{T}}
    [\boldsymbol{\varphi}_{2}]
    \Big(
      [\mathbf{E}]
      [h_{1}]
      +
      [a H h_{1}h_{2}]
      +
      [(1-a) H h_{2}]
    \Big)
    [\mathbf{Q}]
  \end{align}
\end{subequations}
where $N\times N$ diagonal matrices
$[\mathbf{E}]$,
$[h_{1}]$,
$[aHh_{1}]$, and
$[(1-a)Hh_{2}]$
contain the elements' respective values along
the diagonals.
The first row in Eq.~(\ref{eq:form2b}a) is
a substitution of Eqs.~(\ref{eq:TTdot}\textsubscript{2}),
(\ref{eq:epsBu}), and~(\ref{eq:alphalambda})
in Eq.~(\ref{eq:Tb}).
The second row in Eq.~(\ref{eq:form2b}a) is
a substitution of
Eqs.~(\ref{eq:epsBu}) and~(\ref{eq:alphalambda})
in Eq.~(\ref{eq:fdot}).
\par
The third row in Eq.~(\ref{eq:form2b}a)
(and in Eq.~\ref{eq:FFF}) is the displacement
boundary conditions of Eq.~(\ref{eq:Cdotc}).
If the  $[\rho\dot{\mathbf{b}}]$ are taken as applied forces,
any constraints on displacements induce
reaction forces that supplement these applied forces.
If $[\dot{\mathbf{r}}]$ are the reaction forces,
the first row of Eq.~(\ref{eq:form2b}a) is replaced with
\begin{equation}\label{eq:br}
  [\mathbf{S}_{\mathbf{xx}}]
  [\dot{\mathbf{u}}]
  +
  [\mathbf{S}_{\mathbf{x}\boldsymbol{\lambda}}]
  [\dot{\boldsymbol{\lambda}}]
  =
  [\rho\dot{\boldsymbol{b}}]
  +
  [\dot{\mathbf{r}}]
\end{equation}
and the end tractions $t$ in Fig.~\ref{fig:Bar}a are simply
the values of reactions $[\mathbf{r}]$ at the bar's ends.
The displacement constraints are
enforced with the following generalized inverses and
projection matrices:
\begin{equation}
\begin{gathered}
  [\mathbf{C}]^{\dagger}=
  [\mathbf{C}]^{\text{T}}([\mathbf{C}][\mathbf{C}]^{\text{T}})^{-1}
  ,\quad
  [\mathbf{P}_{\mathcal{L}^{\perp}}]=
  [\mathbf{C}]^{\dagger}[\mathbf{C}]
  ,\quad
  [\mathbf{P}_{\mathcal{L}}]=
  [\,\mathbf{I}\,] -
  [\mathbf{P}_{\mathcal{L}^{\perp}}]
  \\
  [\mathbf{S}_{\mathbf{xx}}]^{(-1)}_{(\mathcal{L})}=
  [\mathbf{P}_{\mathcal{L}}]
  \Big(
  [\mathbf{S}_{\mathbf{xx}}]
  [\mathbf{P}_{\mathcal{L}}]
  +
  [\mathbf{P}_{\mathcal{L}^{\perp}}]
  \Big)^{-1}
\end{gathered}
\end{equation}
where $[\mathbf{P}_{\mathcal{L}}]$ is the projection matrix
onto the null space $\mathcal{L}$ of $[\mathbf{C}]$;
$[\mathbf{P}_{\mathcal{L}^{\perp}}]$ in the projection matrix
onto the column space $\mathcal{L}^{\perp}$
of $[\mathbf{C}]^{\text{T}}$;
$[\mathbf{C}]^{\dagger}$ is the Moore--Penrose inverse
of $[\mathbf{C}]$; and
$[\mathbf{S}_{\mathbf{xx}}]^{(-1)}_{(\mathcal{L})}$
is the Bott--Duffin inverse of $[\mathbf{S}_{\mathbf{xx}}]$
constrained to $\mathcal{L}$.
Applying the inverses to Eqs.~(\ref{eq:Cdotc}) and~(\ref{eq:br})
gives the bar's velocities and reactions,
\begin{equation}
\label{eq:solveur}
\begin{aligned}
[\dot{\mathbf{u}}]
&=
[\mathbf{C}]^{\dagger}[\dot{\mathbf{c}}]
+
[\mathbf{S}_{\mathbf{xx}}]^{(-1)}_{(\mathcal{L})}
\Big(
  [\rho\dot{\mathbf{b}}]
  -
  [\mathbf{S}_{\mathbf{x}\boldsymbol{\lambda}}]
  [\dot{\boldsymbol{\lambda}}]
  -
  [\mathbf{S}_{\mathbf{xx}}]
  [\mathbf{C}]^{\dagger}[\dot{\mathbf{c}}]
\Big)
\\
[\dot{\mathbf{r}}]
&=
-[\rho\dot{\mathbf{b}}]
+
[\mathbf{S}_{\mathbf{xx}}]
[\dot{\mathbf{u}}]
+
[\mathbf{S}_{\mathbf{x}\boldsymbol{\lambda}}]
  [\dot{\boldsymbol{\lambda}}]
\end{aligned}
\end{equation}
Substituting Eq.~(\ref{eq:solveur}\textsubscript{1})
into the second row of Eq.~(\ref{eq:form2b}a)
reveals the $[\mathbf{M}]$ and $[\dot{\boldsymbol{\lambda}}]$
matrices of the LCP of Eqs.~(\ref{eq:LCP}):
\begin{equation}
\begin{aligned}
  [\mathbf{M}]
  &=
  [\mathbf{S}_{\boldsymbol{\lambda}\boldsymbol{\lambda}}]
  -
  [\mathbf{S}_{\boldsymbol{\lambda}\mathbf{x}}]
  [\mathbf{S}_{\mathbf{xx}}]^{(-1)}_{(\mathcal{L})}
  [\mathbf{S}_{\mathbf{x}\boldsymbol{\lambda}}]
  \\
  [\dot{\mathbf{q}}]
  &=
  [\mathbf{S}_{\boldsymbol{\lambda}\mathbf{x}}]
  [\mathbf{S}_{\mathbf{xx}}]^{(-1)}_{(\mathcal{L})}
  [\rho\dot{\boldsymbol{b}}]
  +
  [\mathbf{S}_{\boldsymbol{\lambda}\mathbf{x}}]
  \Big(
  [\:\mathbf{I}\:]
  -
  [\mathbf{S}_{\mathbf{xx}}]^{(-1)}_{(\mathcal{L})}
  [\mathbf{S}_{\mathbf{xx}}]
  \Big)
  [\mathbf{C}]^{\dagger}[\dot{\mathbf{c}}]
\end{aligned}
\end{equation}
The LCP
is solved as a sequential quadratic program,
and the solution values $[\dot{\boldsymbol{\lambda}}]$
are used to compute the rates $[\dot{\mathbf{u}}]$ and
$[\dot{\mathbf{r}}]$ in Eq.~(\ref{eq:solveur}).
The rates are explicitly integrated in time increments $dt$
with simple forward difference advancement of the displacements
$[\mathbf{u}]$, variables $[\boldsymbol{\alpha}]$,
and end tractions $[\mathbf{t}]$.
Results are reported in Section~\ref{sec:examples}.
\section*{\normalsize Data availability}
\small
Data used in the paper is available as a compressed file
that contains
Octave/Matlab files for analyzing the example of
Section~\ref{sec:examples}
and plotting files
for creating Fig.~\ref{fig:Results}.\ 
\texttt{https://faculty.up.edu/kuhn/misc/misc.html}.
\section*{\normalsize Acknowledgment}
\small
The author is grateful for the invaluable conversations
and suggestions of Dr. Anil Misra during the work's early development.
%
%
%
\small
\bibliographystyle{unsrt}
%

\end{document}